\documentclass[aps,prd,twocolumn,groupedaddress,floatfix]{revtex4}

\usepackage{graphicx}
\usepackage{bm}
\usepackage{hyperref}
\usepackage{amsfonts}
\usepackage{amsmath,amssymb,amsopn}

\newcommand{\be}{\begin{equation}}
\newcommand{\ee}{\end{equation}}
\newcommand{\ba}{\begin{eqnarray}}
\newcommand{\ea}{\end{eqnarray}}
\newcommand{\ban}{\begin{eqnarray*}}
\newcommand{\ean}{\end{eqnarray*}}
\newcommand{\nn}{\nonumber}

\begin{document}

\title{Entropy production and equilibration in Yang-Mills quantum mechanics}

\author{Hung-Ming Tsai}
\author{Berndt M\"uller}
\affiliation{Department of Physics, Duke University, Durham, NC 27708, USA}

\date{\today}

\begin{abstract}
The Husimi distribution provides for a coarse grained representation of the phase space distribution of a quantum system, which may be used to track the growth of entropy of the system. We present a general and systematic method of solving the Husimi equation of motion for an isolated quantum system, and we construct a coarse grained Hamiltonian whose expectation value is exactly conserved. As an application, we numerically solve the Husimi equation of motion for two-dimensional Yang-Mills quantum mechanics (the {\em x-y} model) and calculate the time evolution of the coarse grained entropy of a highly excited state. We show that the coarse grained entropy saturates to a value that coincides with the microcanonical entropy corresponding to the energy of the system.
\end{abstract}

\maketitle

\section{Introduction}

Entropy production in isolated quantum systems is an interesting and important research problem with many applications. Due to the unitarity of time evolution in quantum mechanics, the von Neumann entropy of an isolated quantum system remains fixed. A proper definition of the concept of entropy growth for an isolated quantum system thus requires coarse graining which, in turn, must be grounded on a correspondence between quantum and classical physics. Such a correspondence can be constructed from one of the phase-space representations of quantum theories found since the classical works of Wigner and Moyal \cite{Wigner:1932eb, Moyal:1949sk}. Recently, it was suggested by Kunihiro {\em et al.} \cite{Kunihiro:2008gv} that the Husimi representation of the density operator \cite{Husimi:1940, Hillery:1983ms, Lee:1995pr} is suitable for describing the entropy production in an isolated quantum system, because the long-term growth rate of the entropy defined by the Husimi distribution approaches the classical limit for long times. Applications of this formalism include the entropy production  in relativistic heavy-ion collisions and inflationary cosmology.

The process of entropy production in relativistic heavy-ion collisions has been studied extensively. The final entropy per unit rapidity produced in high-energy nuclear collisions at the Relativistic Heavy-Ion Collider (RHIC) is well known experimentally. The final entropy produced per unit rapidity produced in central Au+Au collisions at the top RHIC energy of 200 GeV per nucleon pair in the center-of-mass frame is $5600 \pm 500$ at mid-rapidity \cite{Muller:2005en}. Theoretical studies  suggest that at least half of the final entropy is produced during a rapid equilibration and thermalization period during the initial phase of the nuclear collision, with a
thermalization time about $1.5$~fm/c or less \cite{Fries:2009wh, Kunihiro:2010tg}. Furthermore, it has been pointed out that the nuclear matter is transformed in this rapid equilibration stage from saturated gluonic matter in a universal quantum state, called the color-glass condensate, into a thermally equilibrated quark-gluon plasma \cite{McLerran:2005kk,Muller:2006ee}.  It is an important theoretical  challenge to construct a formalism capable of describing the entropy production during this equilibration and thermalization process.

Another interesting exploration relevant to entropy production of quantum systems is reheating of the universe after inflation \cite{Bassett:2005xm}. The reheating process starts  from a preheating phase \cite{Kofman:1994rk}, where the inflation field is coupled to the  matter fields and it transfers energy to the matter fields. These matter fields then undergo further decays into other particles until the decay products will eventually reach a thermal equilibrium.  Through these stages, the reheating process of the universe after inflation produces a gigantic amount of entropy.

To deal with applications to such a wide range of physical systems, it is desirable to construct a general formalism describing the coarse grained entropy production in an isolated quantum system from the growth of complexity of the quantum system. In this work, we apply the formalism developed in \cite{Kunihiro:2008gv} to study the coarse grained entropy production of a specific non-integrable quantum system and its approach to thermal equilibrium. As an example, we choose a simple quantum system that possesses chaotic dynamical behavior. It is well-known that chaotic dynamical behavior requires that an isolated, conservative dynamical system must have at least four degrees of freedom (two position and two momentum variables) \cite{Biro:1994bi}. The two-dimensional quantum system we have chosen, often called the $xy$-model or two-dimensional Yang-Mills quantum mechanics, has well known chaotic properties \cite{Matinyan:1981dj}. We find that the coarse grained entropy production of this quantum system saturates, and we obtain a characteristic time after which the complexity of the system no longer increases.

This article is structured as follows. In Section II, we briefly introduce the Husimi representation of the density operator and explain how it is applied to a definition of the coarse grained entropy of a quantum system, also known as the Wehrl-Husimi entropy. On the way, we propose an novel method to derive the coarse grained Hamiltonian whose expectation value serves as a constant of motion for time evolution of the Husimi distribution. In Section III, we discuss the equation of motion of the Husimi distribution and introduce the test-particle method for obtaining the numerical solutions to this equation. After transforming the Husimi equation of motion into a system of equations of motion for test particles, we solve these equations to obtain the Husimi distribution and the Wehrl-Husimi entropy as a function of time in Section IV. We analyze the time dependence of the Wehrl-Husimi entropy and obtain a characteristic time scale, after which the entropy is saturated. Besides, we propose a method to investigate the value of the saturated Wehrl-Husimi entropy for an infinitely large test-particle number, which is independent of the test-particle approximation scheme. Furthermore, we compare the saturation value of the Wehrl-Husimi entropy to equilibrium based definitions of the entropy of the same quantum system. The difference between the microcanonical and the Wehrl-Husimi entropy serves as a probe for when and whether the quantum system equilibrates.

\section{General formalism}

\subsection{Husimi distribution and coarse grained entropy}

The main goal of this paper is to study the entropy production of a quantum system as a function of time. To define a coarse grained entropy, it is necessary to construct a mapping which not only creates a correspondence between the dynamics of the quantum system and that of the classical system, but also ensures that  the resulting coarse grained distribution is non-negative and thus can be used for the definition of the coarse grained entropy \cite{Kunihiro:2008gv}. A minimal coarse graining of a quantum system is achieved by projecting its density operator on a coherent state \cite{Husimi:1940}. The resultant distribution function is known as the Husimi distribution $\rho_H( t; \mathbf{q}, \mathbf{p} )$, which is positive semi-definite function on the phase space. We note that the Husimi distribution is not unique, but depends on the choice of the canonical variables $(\mathbf{q}, \mathbf{p})$. Even for a specific choice of $(\mathbf{q}, \mathbf{p})$ it depends on the smearing parameter $\alpha$, as discussed below. For a two-dimensional quantum
system, the Husimi distribution is defined as
\ba
 \rho_H ( t; q_1,q_2,p_1,p_2 ) = \langle
z_1,z_2;\alpha | \hat \rho (t) | z_1, z_2;\alpha  \rangle,
\label{eq:Husimi_def}
\ea
where $\hat \rho (t)$ denotes the density operator, $\alpha$ is a
parameter and the coherent state $| z_1, z_2;\alpha  \rangle$
satisfies,
\ba
\hat a_{1,\alpha} | z_1,z_2;\alpha  \rangle &=& z_{1,\alpha} |
z_1,z_2;\alpha \rangle, \nn \\
\hat a_{2,\alpha} | z_1,z_2;\alpha \rangle &=& z_{2,\alpha} |
z_1,z_2;\alpha \rangle , \nn
\ea
with
\ba
\hat a_{1,\alpha} &=&  \frac{1}{\sqrt{2\alpha}} \left( \hat q_1+ i
\frac{\alpha}{\hbar}  \; \hat p_1 \right),
\\
\hat a_{2,\alpha} &=& \frac{1}{\sqrt{2\alpha}} \left( \hat q_2+ i
\frac{\alpha}{\hbar}  \; \hat p_2 \right).
\ea
Note that $\alpha$ is related to the smearing parameter $\Delta$ in \cite{Kunihiro:2008gv, Fries:2009wh} by $\alpha=\hbar/\Delta$. The definition (\ref{eq:Husimi_def}) ensures that the Husimi distribution is non-negative within all of phase space. Throughout this paper,  the notion of $\rho_H( t; \mathbf{q}, \mathbf{p} )$ always implies a dependence on $\alpha$, as indicated in (\ref{eq:Husimi_def}). The Husimi distribution can also be obtained by Gauss smearing of the Wigner function. Let $W$ be the Wigner function defined by:
\ba
W(t; \mathbf{q}, \mathbf{p})=\int_{-\infty}^{\infty} d^2 \mathbf{x} \; \langle\mathbf{q}-\frac{\mathbf{x}}{2}|\hat \rho(t) |\mathbf{q}+\frac{\mathbf{x}}{2}\rangle e^{\frac{i}{\hbar} \mathbf{p} \cdot \mathbf{x}} .
\label{eq:Wigner}
\ea
The Husimi distribution is obtained by convolution of the Wigner distribution with a Gaussian:
\ba
\rho_{H} (t; \mathbf{q},\mathbf{p})=\frac{1}{\pi^2 \hbar^2}
\int_{-\infty}^{\infty} d^{2}\mathbf{q}'d^{2}\mathbf{p}' \; W(t; \mathbf{q}',
\mathbf{p}')
\nn\\
\times e^{-(\mathbf{q}'-\mathbf{q})^2/\alpha -\alpha(\mathbf{p}'-\mathbf{p})^2 /\hbar^2}.
\label{eq:Husimi-def-2}
\ea
Since the Husimi distribution is a minimally (in the sense of the uncertainty principle) smeared Wigner function, it was proposed in \cite{Kunihiro:2008gv} that the Husimi distribution can be applied to the definition of a minimally coarse grained entropy, the Wehrl-Husimi entropy. In two dimensions
\ba
{S_H }(t) =  - \int \frac{d^2\mathbf{q}\, d^2\mathbf{p}}{(2\pi \hbar )^2}\,
{\rho_H }(t; \mathbf{q},\mathbf{p}) \ln {\rho_H }(t; \mathbf{q},\mathbf{p}) .
\label{eq:entropy}
\ea
The properties of the Wehrl-Husimi entropy are reviewed in \cite{Wehrl:1978zz}. Besides, Wehrl conjectured that $S_H(t) \ge 1$ for any one dimensional system, where the equality holds for a minimum uncertainty distribution \cite{Wehrl:1979rmp}. Lieb  proved this conjecture in \cite{Lieb:1978cmp}. We here generalize Wehrl's conjecture to that of a two-dimensional system:
\ba
S_H(t) \ge 2,
\label{eq:conjecture}
\ea
where the equality holds for a minimum-uncertainty Husimi distribution. We confirm  in Sect.~\ref{sec:Husimi_plots} that our numerical results satisfy the bound (\ref{eq:conjecture}).  To investigate the time dependence of the coarse grained entropy, we now derive the equation of motion for the Husimi distribution.

\subsection{Time evolution of Husimi distribution \label{sec:Husimi_EOM}}

In quantum mechanics, the Liouville equation
\ba
i\hbar \frac{\partial \hat \rho (t) }{\partial t} = [\hat{\mathcal H},\hat \rho (t) ],
\label{eq:Liouville}
\ea
where $\hat{\mathcal H}$ denotes the Hamiltonian operator, describes the time evolution of the density operator. One can study the time evolution of a quantum  system by mapping the equation of motion of the density operator in the Hilbert space onto that of the corresponding density distribution in the phase space.  The Husimi equation of motion is obtained by subjecting both sides of eq.~(\ref{eq:Liouville}) to the Husimi transform (\ref{eq:Husimi_def}). For a one-dimensional quantum system, the Husimi equation of motion was first derived by O'Connell and Wigner \cite{OConnell:1981pla}. Here, we derive the the Husimi equation of motion for two-dimensional quantum system. For a single particle in two dimensions, the classical counterpart of the Hamiltonian  $\hat{\mathcal H}$ reads,
\ba
{\mathcal H} = \frac{1}{2m}\left( {p_1^2 + p_2^2 } \right) +V (q_1,q_2),
\label{eq:Hamiltonian}
\ea
where $m$ is the mass of the particle and $V(q_1,q_2)$ is the potential energy. For  the Hamiltonian system whose potential energy $V(q_1, q_2)$ is a $\mathcal{C}^{\infty}$-differentiable function of $(q_1, q_2)$,
we apply (\ref{eq:Wigner}, \ref{eq:Husimi-def-2}) to (\ref{eq:Liouville}), perform a series expansion of $V$ in powers of $q_1$ and $q_2$, and finally obtain the equation of motion for the Husimi distribution:
\begin{widetext}
\ba
 \frac{\partial \rho_H }{\partial t} &=& -\frac{1}{m} \sum\limits_{j = 1}^2 {\left(
{p_j + \frac{\hbar ^2}{2\alpha }\frac{\partial }{\partial p_j }}
\right)} \frac{\partial \rho_H }{\partial q_j }
+ \sum\limits_{\lambda _i, \mu _i, \kappa _i}
 \left[ {\frac{\left( {i\hbar } \right)^{\lambda _1 + \lambda _2- 1}}
 {2^{\lambda _1 + \lambda _2 + \mu _1 + \mu _2 - 1}}}  \,
 {\frac{\alpha ^{\mu _1 + \mu _2 - \kappa _1 -
\kappa _2 }}{\lambda _1 !\lambda _2 !\kappa _1 !\kappa _2 !\left(
{\mu _1 - 2\kappa _1 } \right)!\left( {\mu _2 -
2\kappa _2 } \right)!}}  \right.
\nn \\
&& \times  \left.  \frac{\partial ^{\lambda _1 + \mu _1 }}{\partial
q_1^{\lambda _1 + \mu _1 } }\frac{\partial ^{\lambda _2 + \mu _2
}}{\partial q_2^{\lambda _2 + \mu _2 } }V\left( {q_1 ,q_2 } \right) \,
 \frac{\partial ^{\lambda _1 }}{\partial
p_1^{\lambda _1 } }\frac{\partial ^{\lambda _2 }}{\partial
p_2^{\lambda _2 } }\frac{\partial ^{\mu _1 - 2\kappa _1 }}{\partial
q_1^{\mu _1 - 2\kappa _1 } }\frac{\partial ^{\mu _2 - 2\kappa _2}}
{\partial q_2^{\mu _2 - 2\kappa _2 } }\rho_H   \right] ,
\label{eq:Husimi_general}
\ea
\end{widetext}
where $\lambda _i$, $\mu _i$ and $\kappa _i$ are summed over all non-negative integers, with the constraints that $\left( {\lambda _1+ \lambda _2 } \right)$ is odd, $\left( {\mu _1 - 2\kappa _1 } \right) \ge 0$ and $\left( {\mu _2 - 2\kappa _2 } \right) \ge 0$. When the potential energy is of polynomial form:
\be
V({q_1 ,q_2 } ) = \sum_{i = 0}^{n_1 } {\sum_{j = 0}^{n_2 }{a_{ij} q_1^i q_2^j } } ,
\ee
with the coefficients $a_{ij}$ and non-negative integers $n_1$ and $n_2$, one finds that the additional constraints $\left( {\lambda _1 + \mu _1 } \right) \leq n_1$ and $\left( {\lambda _2 + \mu _2 } \right) \leq n_2$ should be applied to the sum in (\ref{eq:Husimi_general}).

We now specialize our investigation to the Hamiltonian:
\ba
{\mathcal H} = \frac{1}{2m}\left( {p_1^2 + p_2^2 } \right) + \frac{1}{2}g^2q_1^2 q_2^2,
\label{eq:YMH}
\ea
which describes a dynamical system known as Yang-Mills quantum mechanics \cite{Matinyan:1981dj}.   The Hamiltonian in (\ref{eq:YMH}) is almost globally chaotic, except for a tiny portion of the phase space in which stable orbits has been discovered \cite{Carnegie:1984jpa, Dahlqvist:1990prl}. For the potential energy  in the last term of (\ref{eq:YMH}), the order of the derivatives of $V(q_1,q_2)$ in (\ref{eq:Husimi_general}) is restricted by the relations $( {\lambda _1 + \mu _1 } ) \leq 2$ and $( {\lambda _2 + \mu _2 } ) \leq 2$. Therefore, we can rewrite the Husimi equation of motion (\ref{eq:Husimi_general}) as:
\begin{widetext}
\ba
 \frac{\partial \rho_H }{\partial t}
 &=& - \sum\limits_{j =1}^2 \left[  \frac{p_j}{m}\frac{\partial \rho_H }{\partial q_j }
+\left( \frac{\hbar ^2}{2m\alpha }
-\frac{\alpha^2}{8}  \frac{\partial ^4 V}{\partial q_1^2 \partial
q_2^2 } \right) \frac{\partial \rho_H }{\partial p_j \partial q_j } \right]
 +\sum\limits_{j = 1}^2 \left( {\frac{\partial V}{\partial q_j }\frac{\partial \rho_H}{\partial p_j }
+ \frac{\alpha }{2}\frac{\partial ^2V}{\partial q_j^2}   \frac{\partial \rho_H
}{\partial p_j \partial q_j } } \right)
\nn \\
&&  + \frac{\alpha }{4} \left(\frac{\partial ^3 V}{\partial q_1
\partial q_2^2 } \frac{\partial \rho_H}
{\partial p_1 }+ \frac{\partial ^3 V}{\partial q_1 ^2\partial q_2 }
\frac{\partial \rho_H}{\partial p_2 }\right)
+ \frac{\alpha }{2}\frac{\partial ^2V}{\partial q_1
\partial q_2 }\left( {\frac{\partial ^2\rho_H }{\partial p_1
\partial q_2 } + \frac{\partial^2\rho_H }{\partial p_2 \partial q_1 }} \right)
\nn \\
&& + \frac{1}{4}\frac{\partial ^3V}{\partial q_1^2 \partial q_2
}\left[ \alpha ^2\left( {\frac{\partial ^3 \rho_H}{\partial p_1 \partial q_1 \partial q_2 }
+ \frac{1}{2}\frac{\partial ^3 \rho_H}{\partial p_2 \partial q_1^2 }} \right)
- \frac{\hbar^2}{2}\frac{\partial ^3\rho_H }{\partial p_1^2 \partial p_2 } \right]
\nn \\
&& + \frac{1}{4}\frac{\partial^3V}{\partial q_1 \partial q_2^2}
\left[ \alpha ^2\left( {\frac{\partial ^3 \rho_H}{\partial p_2 \partial q_1 \partial
q_2} + \frac{1}{2}\frac{\partial^3 \rho_H}{\partial p_1 \partial q_2^2 }} \right)
- \frac{\hbar^2}{2}\frac{\partial ^3\rho_H }{\partial p_1 \partial p_2^2 } \right]
\nn \\
&& + \frac{1}{16}\frac{\partial ^4V}{\partial q_1^2 \partial q_2^2
}\left[ \alpha ^3\left( {\frac{\partial ^4\rho_H}{\partial p_1 \partial q_1\partial
q_2^2 } + \frac{\partial ^4 \rho_H}{\partial p_2 \partial q_1^2 \partial q_2 }}\right)
 - \hbar ^2\alpha \left( {\frac{\partial ^4 \rho_H}{\partial p_1^2 \partial p_2
\partial q_2 } + \frac{\partial ^4 \rho_H}{\partial p_1 \partial p_2^2 \partial q_1 }}\right)\right].
\label{eq:Husimi_2D}
\ea
\end{widetext}
It is non easy to solve the Husimi equation of motion (\ref{eq:Husimi_2D}). Before we embark on this challenge, we first prove the energy conservation of the Husimi function in Sect.~\ref{sec:energy}, and then solve (\ref{eq:Husimi_2D}) by the test-particle method in Sect.~\ref{sec:TP}.

\subsection{Energy conservation\label{sec:energy}}

A coarse grained Hamiltonian, which describes energy conservation in the Husimi representation, was introduced by Takahashi \cite{Takahashi:1986a,Takahashi:1986b,Takahashi:1989ptps}, who identified  the quantum corrections to  the classical Hamiltonian in powers of $\hbar$ and then constructed a conserved Hamiltonian for the Husimi representation by adding these quantum corrections to the classical Hamiltonian. Explicit expressions for this coarse grained Hamiltonian were found for a few one-dimensional quantum systems \cite{Takahashi:1986a,Takahashi:1986b,Takahashi:1989ptps}. Here we propose a novel derivation of the conserved coarse grained Hamiltonian. Our approach, which involves no approximation, exploits the analytic properties of the transformation between the Wigner and Husimi distributions.

We now derive the coarse grained Hamiltonian for the two-dimensional Yang-Mills quantum mechanics model. The derivation for a one-dimensional quantum system is presented in Appendix~\ref{sec:energy_1D}. Our method can be easily extended to the derivation of the coarse grained Hamiltonian for higher-dimensional quantum systems with polynomial potentials.

The expectation value of a Hamiltonian in the Wigner representation is defined as:
\ba
\mathcal{E} [ \mathcal{H} W ] =\int_{-\infty}^{\infty} d\Gamma_{\mathbf{q},\mathbf{p}}
\;\mathcal{H} ( \mathbf{q}, \mathbf{p} ) W(t;\mathbf{q}, \mathbf{p}),
\label{eq:E_ave_W}
\ea
where $\mathcal{H}$ is the Hamiltonian, $W$ is the Wigner function defined in (\ref{eq:Wigner}), and
\be
d\Gamma_{\mathbf{q},\mathbf{p}} = \frac{d^{2}\mathbf{q}\, d^{2}\mathbf{p}}{(2\pi\hbar)^2}
\label{eq:dGamma}
\ee
is the four-dimensional phase space measure. In quantum mechanics, the energy of the system is calculated as $\langle \hat{\mathcal{H}} \rangle = \mathrm{tr} ( \hat\rho\hat{\mathcal{H}} )$. Starting from the Liouville equation (\ref{eq:Liouville}) it is straightforward to show, that $\partial \langle \hat{\mathcal{H}} \rangle / \partial t =0$. It is also easily shown \cite{Ballentine:1998book} that $\langle \hat{\mathcal{H}} \rangle=\mathcal{E} [ \mathcal{H} W ]$. Therefore, $\mathcal{E} [ \mathcal{H} W ]$ is a constant of motion under the time evolution of the Wigner distribution. We now apply the  convolution theorem to invert the transformation in (\ref{eq:Husimi-def-2}) and obtain:
\ba
\mathcal{E} [ \mathcal{H} W ] =\int_{-\infty}^{\infty} d\Gamma_{\mathbf{q},\mathbf{p}}
\;\mathcal{H}_H( \mathbf{q}, \mathbf{p}) \rho_{H}(t;\mathbf{q}, \mathbf{p}),
\label{eq:EHW-2}
\ea
where
\ba
\mathcal{H}_H ( \mathbf{q}, \mathbf{p} ) &=& \frac{1}{16\pi^4} \int_{-\infty}^{\infty} d^{2}\mathbf{q}'d^{2}\mathbf{p}' \;\mathcal{H}( \mathbf{q}',\mathbf{\mathbf{p}}')
\nn\\
&& \times \int_{-\infty}^{\infty}d^{2}\mathbf{u} \, d^{2}\mathbf{v}  \exp\left[\frac{\alpha}{4}
\mathbf{u}^2+ \frac{\hbar^2}{4\alpha} \mathbf{v}^2 \right.
\nn\\
&&\left. -i\mathbf{u} \cdot (\mathbf{q}'-\mathbf{q})-i\mathbf{v}\cdot(\mathbf{p}'-\mathbf{p})\right] ,
\label{eq:Hamiltonian-Husimi-1}
\ea
and $\mathbf{u}$ and $\mathbf{v}$ are the Fourier conjugate variables to $\mathbf{q}$ and $\mathbf{p}$, respectively. The expression of $\mathcal{H}_H$  in (\ref{eq:Hamiltonian-Husimi-1}) is not mathematically
well-defined because it involves exponentially growing Gaussian functions. However, $\mathcal{H}_H$
can be evaluated by analytic continuation. Let $\xi=-\alpha/4$ and  $\eta=-\hbar^2/(4\alpha)$. Then, we
evaluate the last two integrals in (\ref{eq:Hamiltonian-Husimi-1}) in the analytic region where $\xi>0$ and $\eta>0$ and obtain:
\ba
\mathcal{H}_H (\mathbf{q},\mathbf{p})= \frac{1}{16\pi^2 \xi \, \eta}  \int_{-\infty}^{\infty} d^{2}\mathbf{q}'d^{2}\mathbf{p}' \; \mathcal{H}(\mathbf{q}',\mathbf{p}') \nn\\ \times\exp \left[- \frac{(\mathbf{q}'-\mathbf{q})^2}{4 \xi}- \frac{(\mathbf{p}'-\mathbf{p})^2}{4 \eta} \right].
\label{eq:Hamiltonian_Husumi-2}
\ea
Again, we evaluate the integrals in (\ref{eq:Hamiltonian_Husumi-2}) in the analytic region where $\xi>0$ and $\eta>0$,
and then we substitute  $\xi=-\alpha/4$ and $\eta=-\hbar^2/(4\alpha)$ into its expression, thereby resulting in  a real and finite function $\mathcal{H}_{H} (\mathbf{q},\mathbf{p})$.  For example, by substituting (\ref{eq:YMH}) into (\ref{eq:Hamiltonian_Husumi-2}) and evaluating (\ref{eq:Hamiltonian_Husumi-2}) according to the above procedure, we obtain:
\ba
\mathcal{H}_H (\mathbf{q},\mathbf{p})& =&\frac{1}{2m} \left( p_1^2 + p_2^2 \right) + \frac{1}{2} g^2 q_1^2 q_2^2 \nn\\
&&-\frac{1}{4}  g^2 \alpha   \left(q_1^2 + q_2^2 \right)
\nn \\
&& +\frac{1}{8}g^2 \alpha^2-\frac{\hbar^2}{2m\alpha}.
\label{eq:Hamiltonian_Husumi-3}
\ea
The analytic function $\mathcal{H}_H (\mathbf{q},\mathbf{p})$ in (\ref{eq:Hamiltonian_Husumi-3}) is the
coarse grained Hamiltonian for the Yang-Mills quantum system whose conventional Hamiltonian is defined in (\ref{eq:YMH}). We now define the expectation value of the energy in the Husimi representation as:
\ba
\mathcal{E} [ \mathcal{H}_H \rho_H ] =\int_{-\infty}^{\infty} d\Gamma_{\mathbf{q},\mathbf{p}}
\;\mathcal{H}_H ( \mathbf{q}, \mathbf{p}) \rho_{H}
(t;\mathbf{q}, \mathbf{p}),
\label{eq:EHrho}
\ea
where $\mathcal{H}_H ( \mathbf{q}, \mathbf{p})$ is the coarse grained Hamiltonian defined in (\ref{eq:Hamiltonian_Husumi-3}). Using eqs.~(\ref{eq:YMH}, \ref{eq:Husimi_2D}, \ref{eq:EHrho}), it is straightforward to prove by explicit calculation that
\ba
\frac{\partial \mathcal{E} [ \mathcal{H}_H \rho_H ]}{\partial t} = 0.
\label{eq:energy_conservation}
\ea
Thus, $\mathcal{E} [ \mathcal{H}_H \rho_H ]$ is a constant of motion for the Husimi equation of motion (\ref{eq:Husimi_2D}) and can be identified as the total energy of the system. In Sect.~\ref{sec:Husimi_plots}, we verify numerically that $\mathcal{ E} [ \mathcal{H}_H \rho_H  ]$ is a constant of motion.

\section{Test particle method \label{sec:TP}}

The numerical solution of the Husimi equation of motion for
one-dimensional quantum systems has been investigated, e.g., in
\cite{Trahan:2003jcp,Lopez:2006jcp}. Because our goal is to apply
the Husimi representation to quantum systems in two or more
dimensions, we need a method that is capable of providing solutions
to the Husimi equation of motion for higher-dimensional systems. As
a practical approach to this problem, we here adopt the
test-particle method. This method was previously applied by Heller
\cite{Heller:1981jcp}, who assumed that the wave function is a
superposition of frozen Gaussian wave packets. The test-particle method was also
used to describe the time evolution of the Husimi function of one-dimensional quantum
systems by L\'opez, Martens and Donoso \cite{Lopez:2006jcp}.
Manipulating the Husimi equation of motion algebra\"ically, these
authors obtained the equations of motion for the test particles. The
equations of motion for test particles obtained in this manner
exhibit a nonlinear dependence on the Husimi distribution. However,
we note that the true equation of motion for the Husimi
distribution is a linear partial differential equation, which
encodes the superposition principle for quantum states. The
nonlinear dependence of the equations of motion for the
test-particles representing the Husimi distribution in
\cite{Lopez:2006jcp} implies a violation of this principle. We note
that the superposition principle is crucial to our investigation. To
study the entropy production of the Yang-Mills quantum system and
the approach to thermal equilibrium, we need to consider highly
excited states of the system, whose energies form a quasi-continuum.
Thus, the time evolution of the system is described by the
superposition of eigenstates with almost the same energy. When the
superposition principle is violated, we cannot expect to describe
the time evolution of such states correctly.

Therefore, we here apply the test-particle method in a way that respects the superposition principle. Instead of adopting the strategy proposed in \cite{Lopez:2006jcp}, we obtain the equations of motion for the test particles by taking the first few moments on the Husimi equation of motion. This approach preserves the superposition principle for solutions of the Husimi equation of motion. In Sect.~\ref{sec:QDEs}, we derive the equations of motion for the test particles, obtain the uncertainty relation for Husimi distribution, and prove that the energy conservation holds for each individual test particle. In Sect.~\ref{sec:initial_conditions}, we describe the method by which we choose the initial conditions for the Husimi equation of motion. In Sect.~\ref{sec:fixed-width}, we discuss additional approximations that we use for the Gaussian test functions.

\subsection{Equations of motion for the test particles \label{sec:QDEs}}

Now we briefly describe the test-particle method. Our goal is to solve the Husimi equation of motion in (\ref{eq:Husimi_2D}) and obtain the time dependence of the Husimi distribution. As stated before, the Husimi distribution is a density distribution on the phase space, and it is positive semi-definite for all times. Therefore, we can approximate the time-dependent Husimi distribution by the superposition of a sufficiently large number $N$ of Gaussian functions, whose centers can be considered as the (time-dependent) positions and momenta of $N$ ``test particles''.

For these Gaussian functions, we assume that we can neglect
all correlations between $q_1$ and $q_2$, between $p_1$ and $p_2$,
between $q_1$ and $p_2$, and between $q_2$ and $p_1$. Under these
assumptions, the Husimi distribution can be written as
\begin{widetext}
\ba
\rho_H ( {t; \mathbf{q},\mathbf{p} } ) &=&
\frac{\hbar^2}{N} \sum\limits_{i=1}^{N} \sqrt {\tilde {N}^i (t)}
\exp \left[  - \frac{1}{2}c_{q_1 q_1 }^i (t)\left( {q_1 - \bar {q}_1^i (t)} \right)^2
-\frac{1}{2}c_{q_2 q_2 }^i (t) \left( {q_2 - \bar{q}_2^i (t)} \right)^2 \right]
\nn \\
&& \times \exp \left[ - \frac{1}{2}c_{p_1 p_1 }^i (t) \left( {p_1 - \bar {p}_1^i (t)} \right)^2
- \frac{1}{2}c_{p_2 p_2 }^i (t)\left( {p_2 - \bar{p}_2^i (t)} \right)^2 \right]
\nn \\
&& \times \exp \left[ - c_{q_1 p_1 }^i (t) \left( {q_1 - \bar{q}_1^i (t)} \right)
\left( {p_1 - \bar {p}_1^i (t)} \right)
- c_{q_2 p_2 }^i (t)\left( q_2 - \bar{q}_2^i (t) \right)
\left( p_2 - \bar {p}_2^i (t) \right) \right] .
\label{eq:ansatz}
\ea
\end{widetext}
In order to satisfy the normalization condition for the Husimi distribution:
\ba
\int_{-\infty}^{\infty} d\Gamma_{\mathbf{q},\mathbf{p}} \;
\rho_H( \mathbf{q},\mathbf{p} ;t )=1, \label{eq:rho_norm}
\ea
we normalize each Gaussian according to:
\ba
\tilde {N}^i (t) = \Delta_1^i (t) \Delta_2^i (t),
\ea
where we introduced the abbreviations:
\ba
\Delta_1^i (t) &=& \left[ {c_{q_1 q_1 }^i (t) c_{p_1 p_1 }^i (t)
- \left( {c_{q_1 p_1 }^i(t)} \right)^2} \right],  \label{eq:Delta_1}
\\
\Delta_2^i (t) &=& \left[ {c_{q_2 q_2 }^i (t)c_{p_2 p_2 }^i (t) -
\left( {c_{q_2 p_2 }^i (t)} \right)^2} \right]. \label{eq:Delta_2}
\ea We require that $\tilde {N}^i (t)>0$ for all times. The
assumption of setting $c^i_{q_1 q_2}(t) = c^i_{p_1 p_2}(t) =
c^i_{q_1 p_2}(t) = c^i_{q_2 p_1}(t)=0$ in (\ref{eq:ansatz}) is motivated
by the fact that $c^i_{q_1 p_1}(t)$ and $c^i_{q_2 p_2}(t)$ encode the dominant
correlations induced by the dynamics.
For purposes further down, we have examined numerically that even
when setting $c^i_{q_1 p_1}(t) = c^i_{q_2 p_2}(t) = 0$ for all
times, the correlations between $q_1$ and $p_1$ and between $q_2$
and $p_2$ are produced by the ensemble of Gaussians as time evolves,
by virtue of the contribution of a large number of test functions. Therefore, the
ansatz in (\ref{eq:ansatz}) is justified.

Owing to (\ref{eq:ansatz}), the solution to the Husimi equation of motion will depend on the chosen particle number $N$, and so will the Wehrl-Husimi entropy. In the limit $N \to \infty$ we expect both, the Husimi distribution and the Wehrl-Husimi entropy, to approach values that are independent of the test particle approximation scheme. We will confirm this expectation in Sect.~\ref{sec:number_dependence} by investigating the particle number dependence of our numerical result for the Wehrl-Husimi entropy.

The main task for us is to determine the optimal solutions for the time-dependent variables $\bar{q}_1^i(t)$, $\bar{q}_2^i(t)$, $\bar{p}_1^i(t)$, $\bar{p}_2^i(t)$, $c_{q_1 q_1}^i(t)$, $c_{q_2 q_2}^i(t)$, $c_{p_1 p_1}^i(t)$, $c_{p_2 p_2}^i(t)$, $c_{q_1 p_1}^i(t)$, and $c_{q_2 p_2}^i(t)$. In other words, instead of directly solving (\ref{eq:Husimi_2D}), we seek a system of the equations of motion for the ten time-dependent variables. This goal can be achieved by evaluating the moments on both sides of the Husimi equation of motion. The resulting equations constitute a system of ordinary differential equations for the ten time-dependent variables of each test particle labeled by $i=1,2,...,N$. Overall, we thus have to solve $10N$ equations of motion. These can be grouped into $N$ independent systems of ten coupled differential equations, each of which can be solved separately.

Generally, the moment of a function $f(t;\mathbf{q},\mathbf{p})$ with respect to a weight function $w(\mathbf{q},\mathbf{p})$ is defined as,
\ba
I_w[f] &=& \int_{-\infty}^{\infty} d\Gamma_{\mathbf{q},\mathbf{p}}   \,
 \left[ w (\mathbf{q},\mathbf{p} )  f (t; \mathbf{q},\mathbf{p}) \right].
\ea
Therefore, after we apply the ten moments $I_{q_1}$, $I_{q_2}$, $I_{p_1}$, $I_{p_2}$, $I_{q_1^2}$, $I_{q_2^2}$, $I_{p_1^2}$, $I_{p_2^2}$, $I_{q_1 p_1}$ and $I_{q_2 p_2}$ to the Husimi equation of motion (\ref{eq:Husimi_2D}), we obtain ten equations of motions for each test particle $i$ for the ten variables representing the location in phase space and width of each test particle.  In eqs.~(\ref{eq:I_x1}-\ref{eq:I_p2}), we present the equations obtained from the first moments $I_{q_1}$, $I_{q_2}$, $I_{p_1}$ and  $I_{p_2}$ of (\ref{eq:Husimi_2D}) associated with the location of the test particle. The equations for the evolution of the test particle widths, obtained from the second moments $I_{q_2^2}$, $I_{p_1^2}$, $I_{p_2^2}$, $I_{q_1 p_1}$ and $I_{q_2 p_2}$ of (\ref{eq:Husimi_2D}) are presented in eqs.~(\ref{eq:I_x1x1}-\ref{eq:I_x2p2}) of the Appendix~\ref{sec:Appendix:EOMs}.

The equations for the first moments of (\ref{eq:Husimi_2D})
are:
\ba
\dot{\bar {q}}_1^i (t) - \frac{1}{m}\bar {p}_1^i (t) &=& 0 ,
\label{eq:I_x1}
\\
\dot{\bar {q}}_2^i (t) - \frac{1}{m}\bar {p}_2^i (t) &=& 0 ,
\label{eq:I_x2}
\\
\dot{\bar {p}}_1^i (t) + \left. {\frac{\partial V}{\partial q_1 }}
\right|_{ \bar{\mathbf{q}}^i(t) }
\hspace{0.35\linewidth} &&
\nn \\
\qquad + \frac{1}{2}\left( {\frac{c_{p_2 p_2 }^i (t)}
{\Delta_2^i \left(t \right ) } - \frac{\alpha}{2} } \right)\left.
{\frac{\partial ^3V}{\partial q_1 \partial q_2^2 }}
\right|_{ \bar{\mathbf{q}}^i(t) } &=& 0,
\label{eq:I_p1}
\\
\dot{\bar {p}}_2^i (t) + \left. {\frac{\partial V}
{\partial q_2 }} \right|_{ \bar{\mathbf{q}}^i(t) }
\hspace{0.35\linewidth} &&
\nn \\
\qquad +  \frac{1}{2}\left( \frac{c_{p_1 p_1 }^i (t)}{
\Delta_1^i \left(t \right ) } - \frac{\alpha}{2} \right)\left.
\frac{\partial ^3 V }{\partial q_1^2 \partial q_2 }
\right|_{ \bar{\mathbf{q}}^i(t) } &=& 0,
\label{eq:I_p2}
\ea
where $\Delta_1^i (t)$ and $\Delta_2^i (t)$ are defined in (\ref{eq:Delta_1}) and (\ref{eq:Delta_2}), respectively. The subscript $\bar{\mathbf{q}}^i(t)$ in the partial derivatives of the potential energy $V (q_1,q_2)$ in (\ref{eq:I_p1}, \ref{eq:I_p2}) denotes that the partial derivatives are evaluated at $(q_1, q_2)=\bar{\mathbf{q}}^i(t)$, where
\ba
\bar{\mathbf{q}}^i \left( t \right)=\left( \bar{q}_1^i \left(t \right), \bar{q}_2^i \left(t \right) \right).
\label{eq:q_bar}
\ea
Instead of solving the Husimi equation of motion (\ref{eq:Husimi_2D}), we now solve (\ref{eq:I_x1}-\ref{eq:I_p2}) and (\ref{eq:I_x1x1}-\ref{eq:I_x2p2}) for each test particle $i=1,2,...,N$ and then construct the Husimi distribution by superposition. These test particle equations of motion can be solved numerically by applying the Runge-Kutta method when proper initial conditions are given. The method of choosing the initial conditions will be discussed in Sect.~\ref{sec:initial_conditions}.

To ensure the existence of the solutions, we need to confirm that eqs.~(\ref{eq:I_x1x1}-\ref{eq:I_x2p2}) are nonsingular. We write the system of differential equations (\ref{eq:I_x1x1}-\ref{eq:I_x2p2}) in the form $\mathbf{A}v=b$, where $v$ and $b$ are column vectors and
\ba
v = \left( \dot c_{q_1 q_1 }^i, \dot c_{p_1 p_1 }^i, \dot c_{q_1 p_1}^i,
\dot c_{q_2 q_2 }^i, \dot c_{p_2 p_2 }^i, \dot c_{q_2 p_2 }^i \right)^T.
\ea
The system of equations would be singular if $\det \mathbf{A} = 0$, which implies,
\ba
\Delta_1^i (t)  \Delta_2^i (t)=0 .
\label{eq:singularity}
\ea
This condition is equivalent to $\tilde {N}^i(t) = 0$. Equation~(\ref{eq:singularity}) violates the constraint that $\tilde{N}^i(t) > 0$; therefore, (\ref{eq:I_x1}-\ref{eq:I_p2}) and (\ref{eq:I_x1x1}-\ref{eq:I_x2p2}) are never singular.

The uncertainty relation for the Husimi distribution for one-dimensional quantum systems has been derived in, {\em e.~g.}, \cite{Ballentine:1998book}. Here we generalize their result to the case of two dimensions. The uncertainty relation for the Husimi distribution $\rho_H\left( t; q_1,q_2,p_1,p_2 \right)$ reads:
\ba
\left( \Delta q_j \right)_H \left( \Delta p_j \right)_H \geq \hbar,
\label{eq:uncertainty}
\ea
where
\ba
\left( \Delta q_j \right)_H^2 &=& \int_{-\infty}^{\infty} d\Gamma_{\mathbf{q},\mathbf{p}}
\left[   \left( q_j^2-\langle  q_j \rangle_H \right)^2  \right.
\nn \\
&& \qquad \left.  \times \rho_H(t; \mathbf{q} , \mathbf{p})   \right],
\\
\left( \Delta p_j \right)_H^2 &=& \int_{-\infty}^{\infty} d\Gamma_{\mathbf{q},\mathbf{p}}
\left[ \left( p_j^2-\langle  p_j \rangle_H \right)^2 \right.
\nn \\
&& \qquad \times \left. \rho_H(t; \mathbf{q} , \mathbf{p}) \right] ,
\ea
for $j=1,2$ with
\ba
\langle  q_j \rangle_H &=& \int_{-\infty}^{\infty}  d\Gamma_{\mathbf{q},\mathbf{p}}\,
q_j \, \rho_H(t; \mathbf{q} , \mathbf{p}),
\\
\langle  p_j \rangle_H &=& \int_{-\infty}^{\infty}  d\Gamma_{\mathbf{q},\mathbf{p}}\,
p_j \, \rho_H(t; \mathbf{q} , \mathbf{p}).
\ea
We emphasize that the uncertainty relation (\ref{eq:uncertainty}) does not serve as an additional constraint
when we solve the Husimi equation of motion (\ref{eq:Husimi_2D}). As long as the initial condition $\rho_H  \left( {0; q_1 ,q_2 , p_1 ,p_2  } \right)$ satisfies (\ref{eq:uncertainty}),  the solution to the Husimi equation of
motion satisfies the uncertainty relation (\ref{eq:uncertainty}) for all times. This results from the fact that the quantum effect is encoded in the Husimi equation of motion itself.

\subsection{Initial conditions \label{sec:initial_conditions}}

In order to solve the equations of motions (\ref{eq:I_x1}-\ref{eq:I_p2}, \ref{eq:I_x1x1}-\ref{eq:I_x2p2}), we need to assign initial conditions for the Husimi distribution at $t=0$. We next describe the method we use to assign the initial conditions, $\left\{ {\bar q_1^i(0),\bar q_2^i(0),\bar p_1^i(0),\bar p_2^i(0)} \right\}$ and the initial widths for each test particle $i$. Our goal is to assign initial conditions so that the initial Husimi distribution satisfies the four conditions at $t=0$: ({\em i}) $\rho_H (0;\mathbf{q}, \mathbf{p}) \ge 0$, ({\em ii}) the normalization condition in (\ref{eq:rho_norm}), ({\em iii}) the uncertainty relation in (\ref{eq:uncertainty}), and ({\em iv}) the relation between moments:
\ba
\int_{-\infty}^{\infty} d\Gamma_{\mathbf{q},\mathbf{p}}\, \rho_H(0; \mathbf{q} , \mathbf{p})
\ge \int_{-\infty}^{\infty} d\Gamma_{\mathbf{q},\mathbf{p}}\,  \left[ \rho_H(0; \mathbf{q} , \mathbf{p}) \right]^2.
\label{eq:cond_iii}
\ea
Our strategy is as follows. First of all, we formally write (\ref{eq:ansatz}) as:
\ba
{\rho_H }(t; \mathbf{q} , \mathbf{p})
& =& \frac{1}{N}\sum\limits_{i = 1}^N
K(\mathbf{q} - \bar{\mathbf q}^i(t), \mathbf{p} - \bar{\mathbf{p}}^i(t) ),
\label{eq:ansatz2}
\ea
where $K$ denotes the Gaussian function for each test particle. For $t=0,$ the Husimi distribution (\ref{eq:ansatz2}) can be expressed as
\ba
\rho_H (0; \mathbf{q}, \mathbf{p})  &=& \int_{-\infty}^{\infty} d\Gamma_{\mathbf{q}',\mathbf{p}'} \,
  K( \mathbf{q}-\mathbf{q}', \mathbf{p}-\mathbf{p}')
\nn \\
&& \qquad \times \phi( \mathbf{q}', \mathbf{p}'),
\label{eq:Husimi_0}
\ea
\noindent where $\phi$ denotes the distribution of the test particle locations in the phase space. We abbreviate the phase space variables for clarity: $\bm{\chi}= (q_1,q_2,p_1,p_2)$ and $\bm{\chi}'= (q'_1,q'_2,p'_1,p'_2)$. Owing to the four conditions ({\em i})--({\em iv}) stated above, we choose the Husimi distribution at $t=0$ to be a Gaussian distribution:
\ba
\rho_H(0; \bm{\chi})   &=& \hbar^2  \left( \prod_{a=1}^4\gamma _H^{a} \right)^{1/2}
\nn   \\
&& \times \exp \left[  - \frac{1}{2} \sum_{a=1}^4 \gamma _H^a \left( \chi^a- \mu_H^a \right)^2\right],
\label{eq:initial_Husimi}
\ea
where  $\gamma_H^a$ and $\mu_H^a$ for $a=1,\ldots,4$ are to be determined. In (\ref{eq:initial_Husimi}) we do not assume any correlation between position and momentum locations at $t=0$, implying that we initially set  $c_{x_1 p_1}^i(0)=c_{x_2 p_2}^i(0)=0$ for $i=1,\ldots,N$ in (\ref{eq:ansatz}).

The main idea of choosing initial conditions is that, according to (\ref{eq:Husimi_0}), we can represent the initial Husimi distribution (\ref{eq:initial_Husimi}) to be the sum of Gaussian test functions by randomly assigning $\left\{ {\bar q_1^i(0),\bar q_2^i(0),\bar p_1^i(0),\bar p_2^i(0)} \right\}$ for $i=1,...,N$ according to the  distribution $\phi$. Our remaining tasks are then to determine the parameters in (\ref{eq:initial_Husimi}) and to obtain the functional forms for $K$ and $\phi$. In (\ref{eq:initial_Husimi}), $\mu_H^a$ can be assigned freely by choice, but the $\gamma_H^a$ are subject to the conditions ({\em iii}) and ({\em iv}). Substituting (\ref{eq:initial_Husimi}) into the conditions ({\em iii}) and ({\em iv}), expressed by eqs.(\ref{eq:uncertainty}) and (\ref{eq:cond_iii}), respectively, we obtain from ({\em iii}):
\ba
\prod_{a=1}^4 \left( \gamma _H^{a}  \right)^{-1/2}  \ge \hbar ^2 ,
\label{eq:condition}
\ea
and from ({\em iv}):
\ba
\prod_{a=1}^4 \left( \gamma _H^{a}  \right)^{-1/2}   \ge \hbar^2/4 .
\ea
Since eq.~(\ref{eq:condition}) is the stronger of the two conditions, we adopt it as the constraint for the initial Husimi distribution. To represent $\rho_H (0, \bm{\chi})$ in (\ref{eq:initial_Husimi}), we chose the following functional forms  for $K$ and $\phi$ at $t=0$:
\ba
K(\bm{\chi}-\bm{\chi}')  &=& \hbar^2  \left( \prod_{a=1}^4\gamma _K^{a} \right)^{1/2}
\nn \\
&& \times \exp \left[   - \frac{1}{2} \sum_{a=1}^4 \gamma _K^a
  \left( \chi^a- \chi'^a \right)^2\right],  \label{eq:K_0}
\ea
and
\ba
\phi( \bm{\chi})  &=& \hbar^2 \left( \prod_{a=1}^4 \gamma _\phi^{a} \right)^{1/2}
\nn \\
&& \times \exp \left[   - \frac{1}{2} \sum_{a=1}^4 \gamma _\phi^a
  \left( \chi^a- \mu_{\phi}^a \right)^2\right].
\label{eq:phi_0}
\ea
This choice implies that we represent the initial Husimi distribution as the convolution of test particle Gaussian functions $K$ and a Gaussian distribution $\phi$ of  test particle locations in phase space. In (\ref{eq:ansatz2}) at $t=0$, $\rho_H$ is denoted as the sum of Gaussian functions, each of which may possess distinct widths.
However, when we choose to express (\ref{eq:ansatz2}) at $t=0$ in terms of the convolution of $K$ and $\phi$, we no longer have the flexibility to assign different widths for each individual Gaussian. Instead, for $K$ in (\ref{eq:Husimi_0}, \ref{eq:K_0}) we  should assign:
\ba
\gamma_K^1&=&c_{q_1 q_1 }(0),\qquad \gamma_K^2=c_{q_2 q_2 }(0),
\nn\\
\gamma_K^3 &= &c_{p_1 p_1}(0),\qquad \gamma_K^4=c_{p_2 p_2 }(0),
\label{eq:gamma_K}
\ea
where the suppression of the label $i$ implies that all test particles possess the same width at $t=0$.

It is advantageous to use the convolution of $K$ and $\phi$ in (\ref{eq:Husimi_0}) to represent $\rho_H$ because the parameters in (\ref{eq:initial_Husimi}, \ref{eq:K_0}, \ref{eq:phi_0}) can be related to satisfy the constraint imposed by the uncertainty condition, as described below. In (\ref{eq:phi_0}), $\mu_{\phi}^a$ denotes the location of the center of the distribution of loci of the test particles in the phase space. According to (\ref{eq:Husimi_0}, \ref{eq:initial_Husimi}, \ref{eq:K_0}, \ref{eq:phi_0}), it is clear that the center of the distribution of loci of test particles must coincide with the center of the initial Husimi distribution.  We thus must assign
\ba
\mu_{\phi}^a=\mu_H^a,
\label{eq:mu_relation}
\ea
where $\mu_H^a$ are selected by choice. Moreover, since the $\gamma_H^a$ are subject to the constraint (\ref{eq:condition}), we obtain relations between $\gamma_H^a$, $\gamma_K^a$ and $\gamma_{\phi}^a$, which allow us to determine $\gamma_K^a$ and $\gamma_{\phi}^a$. By applying the convolution theorem to  (\ref{eq:Husimi_0}), we obtain the following relations:
\ba
\frac{1}{\gamma_H^a} &=& \frac{1}{\gamma_K^a} + \frac{1}{\gamma_{\phi}^a},
\label{eq:relation}
\ea
for $a=1,\ldots,4$. Once we select the values of $\gamma_H^a$ based on (\ref{eq:condition}), we must determine $\gamma_K^a$ and $\gamma_{\phi}^a$ according to (\ref{eq:relation}). Furthermore, owing to (\ref{eq:gamma_K}), the choice of $\gamma_K^a$ is subject to the constraints
\ba
\gamma_K^a \ge \gamma_H^a \qquad {\rm for}~ a=1,\ldots,4 .
\label{eq:constraint_1}
\ea
Furthermore, $\gamma_K^a$ must be assigned in the domain where the solutions of  (\ref{eq:I_x1}-\ref{eq:I_p2}) and (\ref{eq:I_x1x1}-\ref{eq:I_x2p2}) are stable. We discuss our choice of initial conditions in more detail in Sect.~\ref{sec:Husimi_plots}.

The number $N$ of test particles plays a crucial role for the accuracy of numerical results. If we set $N=1$ in (\ref{eq:ansatz2}), we find that $\rho_H= K$,  and thus $\gamma_H^a=\gamma_K^a$. This special case is called the single-particle ansatz. In general, the single particle ansatz is insufficient as representation of ${\rho_H }\left( t;{{q_1},{q_2},{p_1},{p_2}} \right)$,  because the Husimi distribution will not retain a Gaussian shape for all times, even if we initialize it as a Gaussian at $t=0$.

As a specific example, we present and compare the solutions of the Husimi equation of motion in one dimension in Fig.~\ref{fig:1D_plots} in Appendix~\ref{sec:energy_1D}.  Figure~\ref{fig:1D_plots}  shows the difference between the solution $\rho_H (t;q,p)$ for the single particle ansatz [panels (a) and (b)] and for the many-particle ansatz [panels (c) and (d)], for the same Hamiltonian defined in eqs.~(\ref{eq:H_1D-1}). The initial conditions are also discussed in Appendix~\ref{sec:energy_1D}. From Fig.~\ref{fig:1D_plots}, it is obvious that the single-particle ansatz is insufficient in representing the solution $\rho_H (t;q,p)$  for $t>0$. We conclude that we need a sufficiently large test-particles number $N$ in (\ref{eq:ansatz2}) to represent the evolution of the Husimi distribution. We discuss the test-particle number dependence of our numerical results in Sect.~\ref{sec:numerical}.

\subsection{Fixed-width ansatz \label{sec:fixed-width}}

Once the initial conditions are obtained, the numerical solutions to eqs.~(\ref{eq:I_x1}-\ref{eq:I_p2},
\ref{eq:I_x1x1}-\ref{eq:I_x2p2}) can be obtained by the Runge-Kutta method. These equations can be dramatically simplified by fixing the Gaussian widths in our ansatz (\ref{eq:ansatz}) for the Husimi distribution.
The precise definition of the fixed-width ansatz reads as follows: For each particle $i$,
\ba
c_{q_1 q_1}^i (t) &=& c_{q_1 q_1} (0),\qquad c_{q_2 q_2}^i (t) = c_{q_2 q_2} (0),
\nn \\
c_{p_1 p_1}^i (t) &=& c_{p_1 p_1} (0),\qquad c_{p_2 p_2}^i (t) = c_{p_2 p_2} (0),
\nn \\
c_{q_1 p_1}^i (t) &=& c_{q_1 p_1} (0),\qquad c_{q_2 p_2}^i (t) = c_{q_2 p_2} (0),
\label{eq:fiq_width_definition}
\ea
where $c_{q_1 q_1}(0)$, $c_{q_2 q_2}(0)$, $c_{p_1 p_1}(0)$, $c_{p_2 p_2}(0)$, $c_{q_1 p_1}(0)$, and  $c_{q_2 p_2}(0)$ are chosen to be the same for all $i$.

In the variable-width ansatz, we solve the ordinary differential equations (\ref{eq:I_x1}-\ref{eq:I_p2}, \ref{eq:I_x1x1}-\ref{eq:I_x2p2}) simultaneously for each test particle $i$. In the fixed-width ansatz, we fix the values of  $c_{q_1 q_1}^i(t)$, $ c_{q_2 q_2}^i(t)$,  $c_{p_1 p_1}^i(t)$, $c_{p_2 p_2}^i(t)$, $c_{q_1 p_1}^i(t)$, and $c_{q_1 p_1}^i (t)$ to be constant for $t \geq 0$.  Therefore, in the fixed-width ansatz, eqs.~(\ref{eq:I_x1x1}-\ref{eq:I_x2p2}) cannot be satisfied, and eqs.~(\ref{eq:I_x1}-\ref{eq:I_p2}) are the only equations of motion for each test particle $i$.  We apply the fixed-width ansatz because (\ref{eq:I_x1}-\ref{eq:I_p2})  are obtained from the first moments of (\ref{eq:Husimi_2D}) and thus serve as the leading contribution to (\ref{eq:Husimi_2D}). From a physical viewpoint, equations (\ref{eq:I_x1}-\ref{eq:I_p2}) determine the "locations" of test particles in the phase space as functions of time, while eqs.~(\ref{eq:I_x1x1}-\ref{eq:I_x2p2}) govern the time-varying widths of each test-particle Gaussian. In Sect.~\ref{sec:numerical} we evaluate all of the numerical results  based on the fixed-with ansatz in (\ref{eq:fiq_width_definition}).

The conservation of energy is not only true for $\rho_H$, as shown in Sect.~\ref{sec:energy}, but also holds for each individual test particle. We now prove the conservation of  energy for each individual test particle in the fixed-width ansatz. The proof can be easily generalized  to the case of variable widths. In the fixed-width ansatz, the test-particle space is spanned by the test-particle positions and momenta $(\bar{\mathbf{q}} , \bar{\mathbf{p}})$. We define a function $\bar{\mathcal{H}}_H$ in the test-particle space as follows:
\ba
\bar{\mathcal{H}}_H \left(\bar{\mathbf{q}} , \bar{\mathbf{p}} \right)
& =& \int_{-\infty}^{\infty} d\Gamma_{\mathbf{q},\mathbf{p}}\,
\mathcal{H}_H \left(\mathbf{q}, \mathbf{p} \right)
K \left(\mathbf{q} - \bar{\mathbf q}, \mathbf{p} - \bar{\mathbf p} \right),
\nonumber\\
\label{eq:barH_H}
\ea
where $\mathcal{H}_H$ denotes the coarse-grained Hamiltonian defined in Sect.~\ref{sec:energy} and $K$ is defined in (\ref{eq:ansatz2}). We note that the functional form of $K$ is independent of the test-particle label $i$. With the help of (\ref{eq:energy_conservation}) and (\ref{eq:barH_H}), it is straightforward to show that
\ba
\frac{\partial \bar{\mathcal{H}}_H \left(\bar{\mathbf{q}}^i (t) , \bar{\mathbf{p}}^i (t)\right)}{\partial t} = 0,
\label{eq:energy_conservation_TP}
\ea
where $i=1, ..., N$. In view of (\ref{eq:energy_conservation_TP}), $ \bar{\mathcal{H}}_H \left(\bar{\mathbf{q}}^i (t) , \bar{\mathbf{p}}^i (t)\right) $ can be identified as the energy of an individual test particle $i$. Due to (\ref{eq:energy_conservation_TP}),  the histogram of test particle energies $ \bar{\mathcal{H}}_H \left(\bar{\mathbf{q}}^i (t) , \bar{\mathbf{p}}^i (t)\right) $ remains unaltered at all times. We apply this result to the numerical calculation in Sect.~\ref{sec:numerical}.

Before we end this Section, a general consideration is in order. In principle, any smooth, positive definite, normalizable function on the phase space can be represented to any desired precision by a sufficient number of sufficiently narrow Gaussian functions with fixed width. However, it is important to keep in mind that these conditions are not satisfied, in general, by the Wigner function or the classical phase space distribution of a chaotic dynamical system. The Wigner function is in general not positive definite, and the classical phase space distribution does not remain smooth for an arbitrary initial condition. The presence of exponentially contracting directions in phase space ensure that, over time, the classical phase space distribution will develop structure on exponentially small scales, which cannot be described by  superposition of fixed-width Gaussian functions.

The Husimi transform of the Wigner function cures both problems. It removes regions of negative values from the quantum phase space distribution, and its respect for the uncertainty relation ensures that the phase space distribution remains smooth on the scale set by $\hbar$ and the smearing parameter $\alpha$. As a result, the fixed-width Gaussian ansatz will always be able to represent the Husimi distribution and track its evolution faithfully over time, if a sufficiently large number of sufficiently narrow Gaussian test functions is employed. On one hand, the width of Gaussian test functions cannot be larger than the width of the initial Husimi distribution so that the Gaussian test functions can represent $\rho_H$ faithfully, as indicated in (\ref{eq:constraint_1}). On the other hand, the width of Gaussian test functions must not be too narrow in order to ensure that the solutions of (\ref{eq:I_x1}-\ref{eq:I_p2}) are stable.  We do not attempt to give a rigorous proof of these assertion here, but content ourselves with the heuristic argument presented above. We will explore the convergence of or numerical solution for the fixed-width ansatz for large values of $N$ at the end of the next Section.

\section{Numerical results\label{sec:numerical}}

We now present  our numerical results. Throughout our calculations, we have used the fixed-width ansatz as described in Sect.~\ref{sec:fixed-width}. In Sect.~\ref{sec:Husimi_plots}, we present the numerical results for the evolution of the Husimi distribution and the Wehrl-Husimi entropy of the Yang-Mills quantum system using $N=1000$ test particles. In Sect.~\ref{sec:GLE}, we obtain the Lyapunov exponents, the average Kolmogorov-Sina\"i entropy and the logarithmic breaking time for Yang-Mills quantum mechanics. In Sect.~\ref{sec:number_dependence}, we compare the Wehrl-Husimi entropies for $N=1000$ and $N=3000$ test particles and explore the test particle number dependence of the saturation value of the Wehrl-Husimi entropy. In Sect.~\ref{sec:canonical}, we obtain the partition function and entropy for the canonical ensemble. Then, in Sect.~\ref{sec:MC}, we evaluate the microcanonical distribution and entropy, and we compare the saturated Wehrl-Husimi entropy to the microcanonical and canonical entropies.

\begin{figure}[h]
\includegraphics[width=0.45\textwidth]{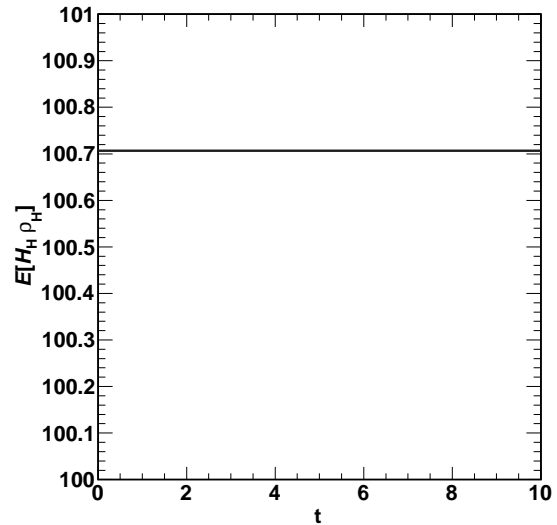}
\caption{Conservation of the coarse grained energy (\ref{eq:EHrho}) during time evolution of the Husimi distribution. This shows that a state with energy $\mathcal{E} \left[\mathcal{H}_H \rho_H \right] = 100.707$ for $t=0$ remains at the same energy for $t>0$, with relative precision better than $10^{-4}$ up to $t=10$. $\rho_H$ is obtained from (\ref{eq:ansatz2}) with $N=1000$ fixed-width test particles.}
\label{fig:energy}
\end{figure}

\begin{figure}
\includegraphics[width=0.45\textwidth]{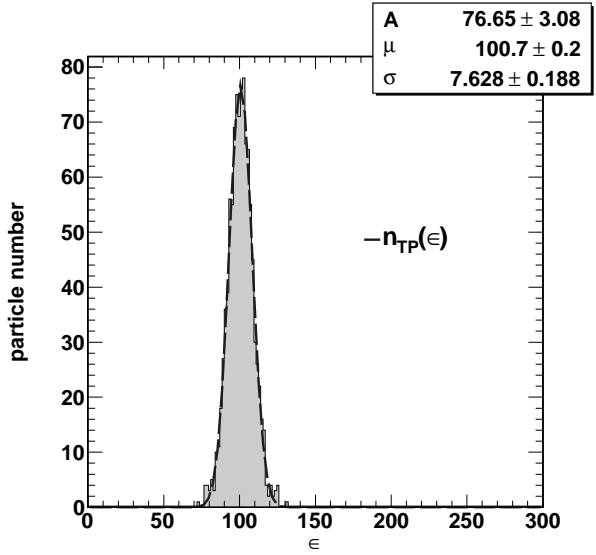}
\caption{Energy histogram for $N=1000$ test particles  at $t=0$. $\epsilon$ denotes the test-particle energy, which is defined in (\ref{eq:energy_def}), and the labels on the vertical axis denote test-particle numbers. A normal distribution $n_{\rm TP} (\epsilon)$ is used to fit the histogram. $A$, $\mu$ and $\sigma$ are the fit parameters for ${n_{\mathrm {TP}}}( \epsilon)$, which are defined in (\ref{eq:n_TP}). The values for the fit parameters are shown in the plot.}
\label{fig:fitting}
\end{figure}

\subsection{Husimi distribution and Wehrl-Husimi entropy \label{sec:Husimi_plots}}

For our numerical calculations, we fix the parameters $m=g=\alpha=\hbar=1$ in (\ref{eq:Husimi_2D}). Initially, we set the number of test particles to $N=1000$. We choose a minimum uncertainty initial Husimi distribution (\ref{eq:initial_Husimi}) by setting:
\ba
\gamma_H^a=1 \qquad {\rm for}~ a=1,\ldots,4,
\label{eq:H_0_widths}
\ea
which satisfies the constraint (\ref{eq:condition}). Besides, in (\ref{eq:initial_Husimi}) we choose
\ba
\mu_H^1=\mu _H^2 =0, \qquad  \mu_{H}^3 =\mu _H^4=10 .
\label{eq:mu_H}
\ea
Owing to (\ref{eq:mu_relation}, \ref{eq:mu_H}), we then have
\ba
\mu _\phi^1=\mu _\phi^2 =0, \qquad \mu^1_\phi=\mu^2_\phi=10.
\label{eq:mu_phi}
\ea

\noindent For a fixed-width ansatz,  the solutions of  (\ref{eq:I_x1}-\ref{eq:I_p2}) are stable under the following constraint:
\ba
\frac{c_{q_1 q_1}(0)+c_{q_2 q_2}(0)}{c_{q_1 q_1}(0)\, c_{q_2 q_2}(0)} \ge \alpha,  \label{eq:constraint_2}
\ea
which can be confirmed by a linear stability analysis. Besides, we set $c_{q_1 p_1}(0)=c_{q_2 p_2}(0)=0$ according to Sect.~\ref{sec:initial_conditions}. Thus, due to (\ref{eq:gamma_K}, \ref{eq:constraint_2}), our choices of $\gamma_K^1$ and $\gamma_K^2$ are constrained by:
\ba
\frac{\gamma_K^1 +\gamma_K^2 }{ \gamma_K^1  \gamma_K^2} \ge \alpha.  \label{eq:constraint_3}
\ea
In summary, our choice of $\gamma_K^a$ is restricted by the two constraints (\ref{eq:constraint_1}, \ref{eq:constraint_3}) together with the settings (\ref{eq:H_0_widths}) and $\alpha=1$. In view of the discussion in Sect.~\ref{sec:initial_conditions}, we satisfy these constraints by the choice
\ba
\gamma_K^a=\frac{3}{2}, \qquad \gamma_{\phi}^a =3,  \qquad (a=1,\ldots,4).
\label{eq:gamma_K_0}
\ea
As described in Sect.~\ref{sec:initial_conditions}, we randomly generate test particle locations $\left\{ {\bar q_1^i(0),\bar q_2^i(0),\bar p_1^i\left( 0 \right),\bar p_2^i(0)} \right\}$ for $i=1,...,N$ according to $\phi$ in (\ref{eq:phi_0}), with parameters given by (\ref{eq:mu_phi}, \ref{eq:gamma_K_0}). For the fixed-width ansatz with the initial conditions (\ref{eq:gamma_K_0}), we solve (\ref{eq:I_x1}-\ref{eq:I_p2}) for each test particle $i$ and repeat the procedure for $i=1,2,...,N$.

Using eqs.~(\ref{eq:Hamiltonian_Husumi-3}, \ref{eq:EHrho}) where $\rho_H$ is obtained from (\ref{eq:ansatz2}) with $N=1000$ fixed-width test particles, we verify numerically that $\mathcal{ E} \left[ \mathcal{H}_H \rho_H
\right ]$ is a constant of motion. This is illustrated in Fig.~\ref{fig:energy}, which shows that a state with initital energy $\mathcal{E} \left[ \mathcal{H}_H \rho_H \right] = 100.707$ remains at the same energy with relative precision better than $10^{-4}$ up to $t=10$. Since the initial "locations" of test particles in the phase space are generated randomly according to $\phi$ in (\ref{eq:phi_0}), different sets of $\left\{ {\bar{\mathbf{q}}^i(0),\bar{\mathbf{p}}^i\left( 0 \right)} \right\}$ generated by different runs of  the  computer program may result in  differences of $\mathcal{ E} \left[ \mathcal{H}_H \rho_H \right ]$ at $t=0$ of less than 0.5 percent. Thus, for any set of initial locations for $N=1000$ test particles, the energy of the state at $t=0$ is $\mathcal{E} \left[ \mathcal{H}_H \rho_H \right] = 100.6 \pm 0.5$.

The energies of individual test particles can be studied by the following method. We denote the test-particle energy
variable $\epsilon$ as
\ba \epsilon=\bar{\mathcal{H}}_H
\left(\bar{\mathbf{q}} , \bar{\mathbf{p}} \right),
\label{eq:energy_def}
\ea
where $\bar{\mathcal{H}}_H
\left(\bar{\mathbf{q}} , \bar{\mathbf{p}} \right)$ is defined in
(\ref{eq:barH_H}). Because we choose the fixed-width Gaussian $K$ with the parameters $\gamma_K^a$ in (\ref{eq:gamma_K_0}) and set $m=g=\alpha=\hbar=1$, we obtain
\ba
\bar{\mathcal{H}}_H
\left(\bar{q}_1, \bar{q}_2 , \bar{p}_1, \bar{p}_2 \right)
&=& \frac{1}{2} \left(\bar{p}_1^2+\bar{p}_2^2  \right)+\frac{1}{2} \bar{q}_1^2 \bar{q}_2^2
\nonumber \\
&+&\frac{1}{12} \left( \bar{q}_1^2 + \bar{q}_2^2 \right)+\frac{13}{72}.
\label{eq:barH_H-2}
\ea
The energy for an individual test particle is denoted as $i$ $ \epsilon_i =\bar{\mathcal{H}}_H \left(\bar{\mathbf{q}}^i (t) , \bar{\mathbf{p}}^i (t)\right) $. Owing to (\ref{eq:ansatz2}), the energy of the state is the average energy of the test particles:
\be
\mathcal{E} \left[ \mathcal{H}_H \rho_H \right ]= \frac{1}{N} \sum_{i=1}^{N} \epsilon_i,
\ee
provided  that $N$ is sufficiently large.   In Fig.~\ref{fig:fitting}, we plot the energy histogram at $t=0$ for $N=1000$ test particles, which we fit to a normal distribution:
\be
{n_{\mathrm {TP}}}\left( \epsilon  \right) = A
\exp \left[ { - \frac{1}{{2{\sigma ^2}}}{{\left( {\epsilon - \mu } \right)}^2}} \right] .
\label{eq:n_TP}
\ee
The values of the fit parameters $A$, $\mu$ and $\sigma$ are listed in Fig.~\ref{fig:fitting} for $N=1000$. We note that the histogram of test particle energies remains unaltered as time evolves, as shown in Sect.~\ref{sec:fixed-width}.

\begin{figure*}
\includegraphics[width=0.45\textwidth]{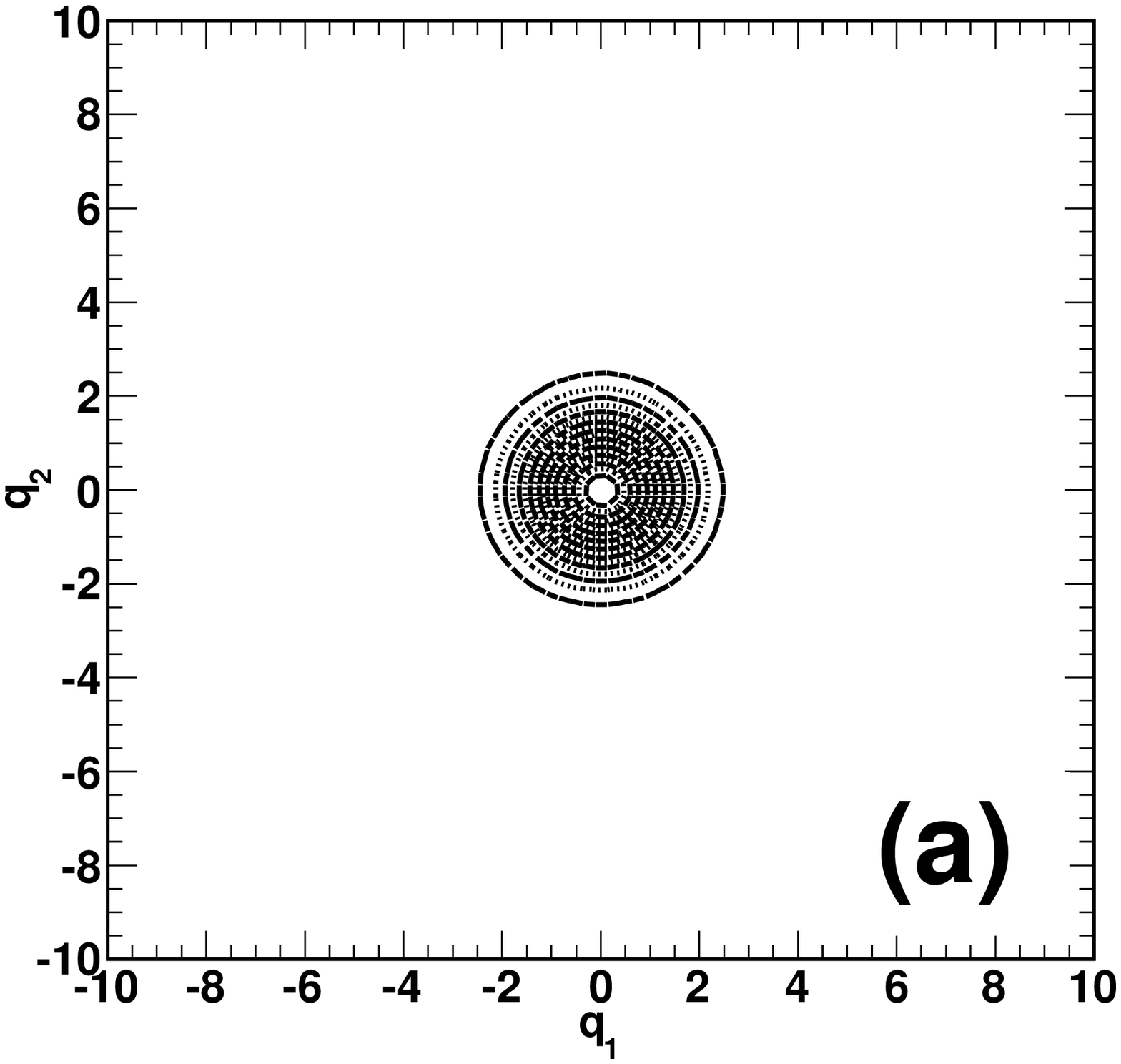}
\includegraphics[width=0.45\textwidth]{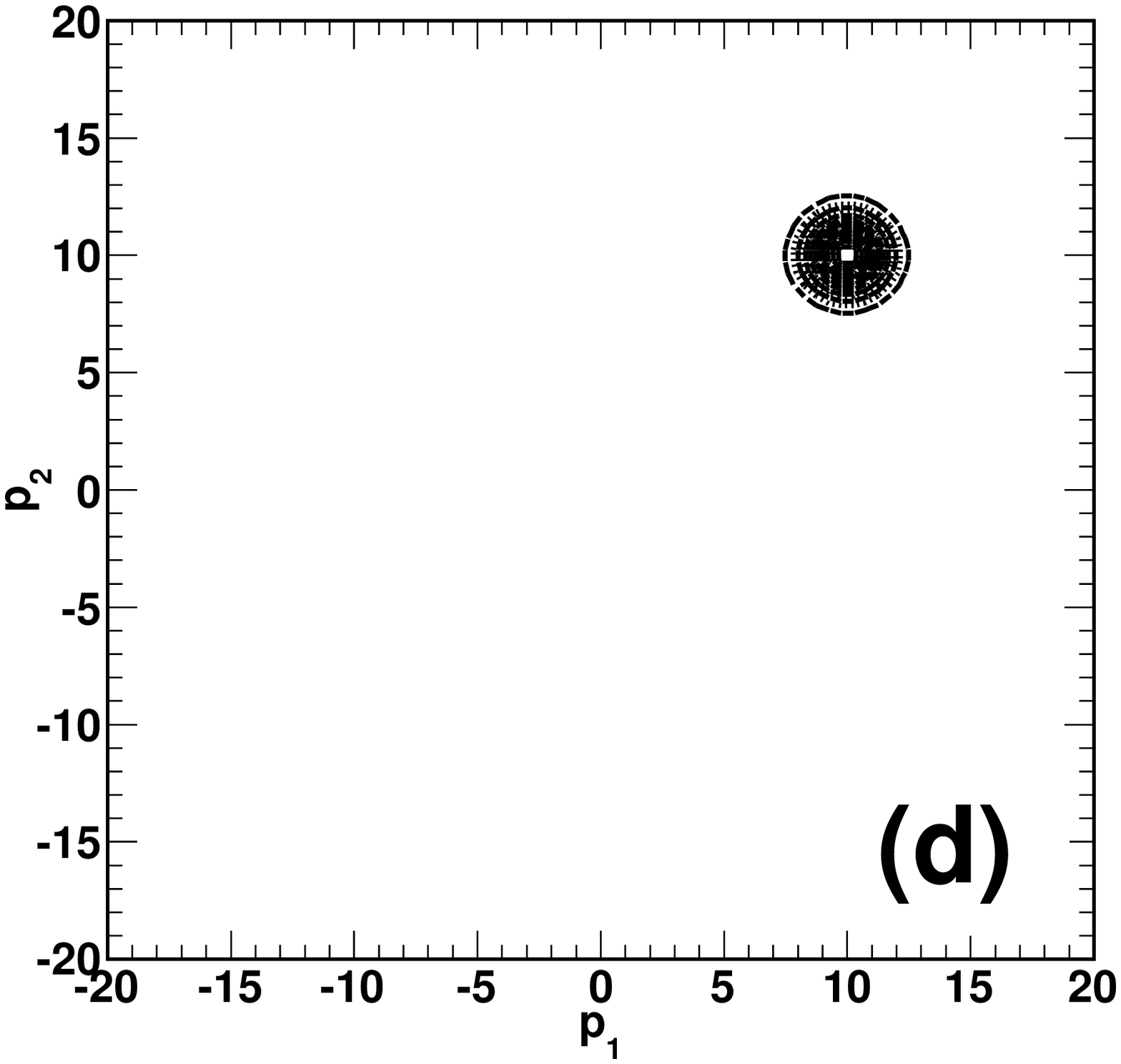}
\includegraphics[width=0.45\textwidth]{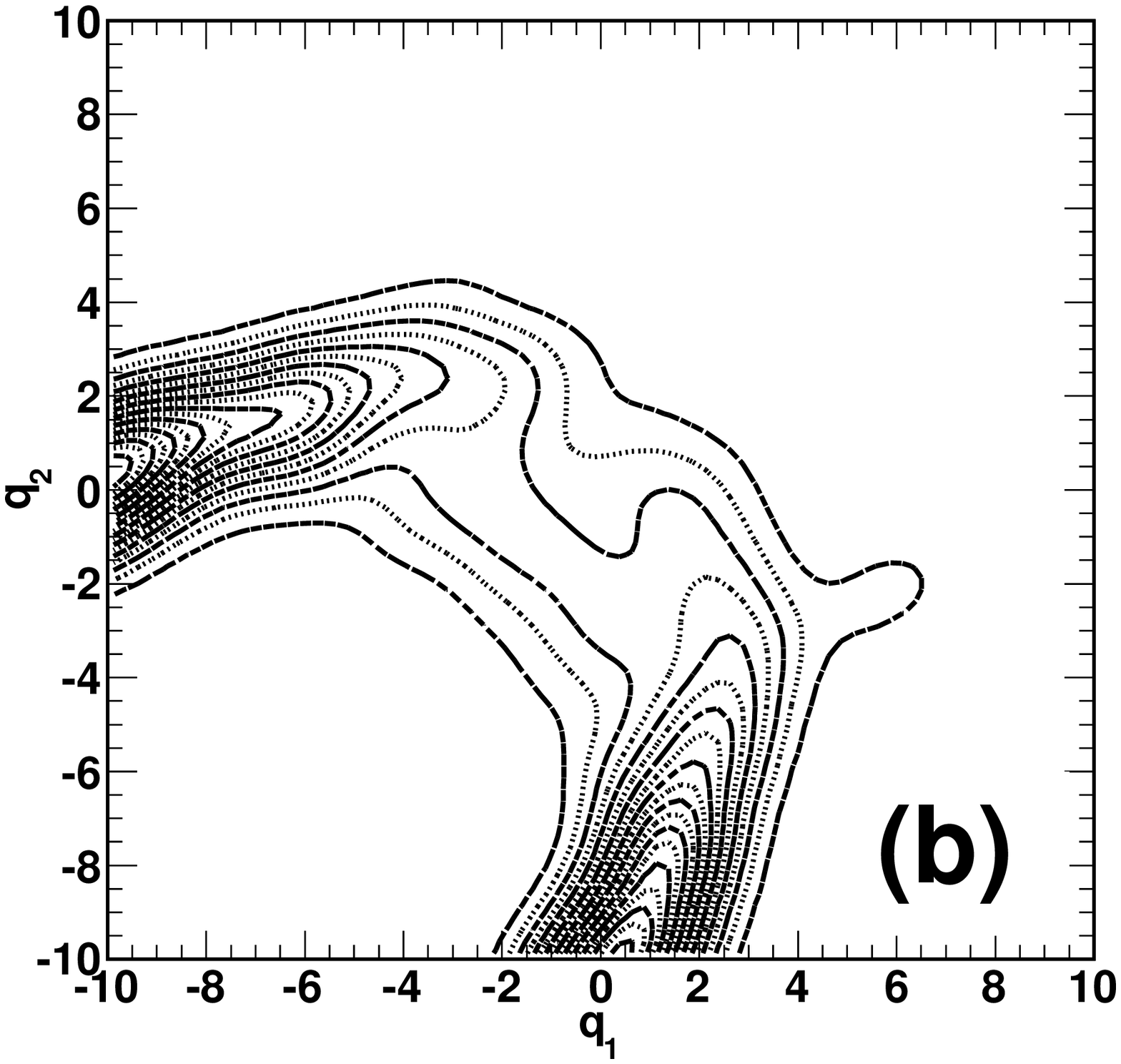}
\includegraphics[width=0.45\textwidth]{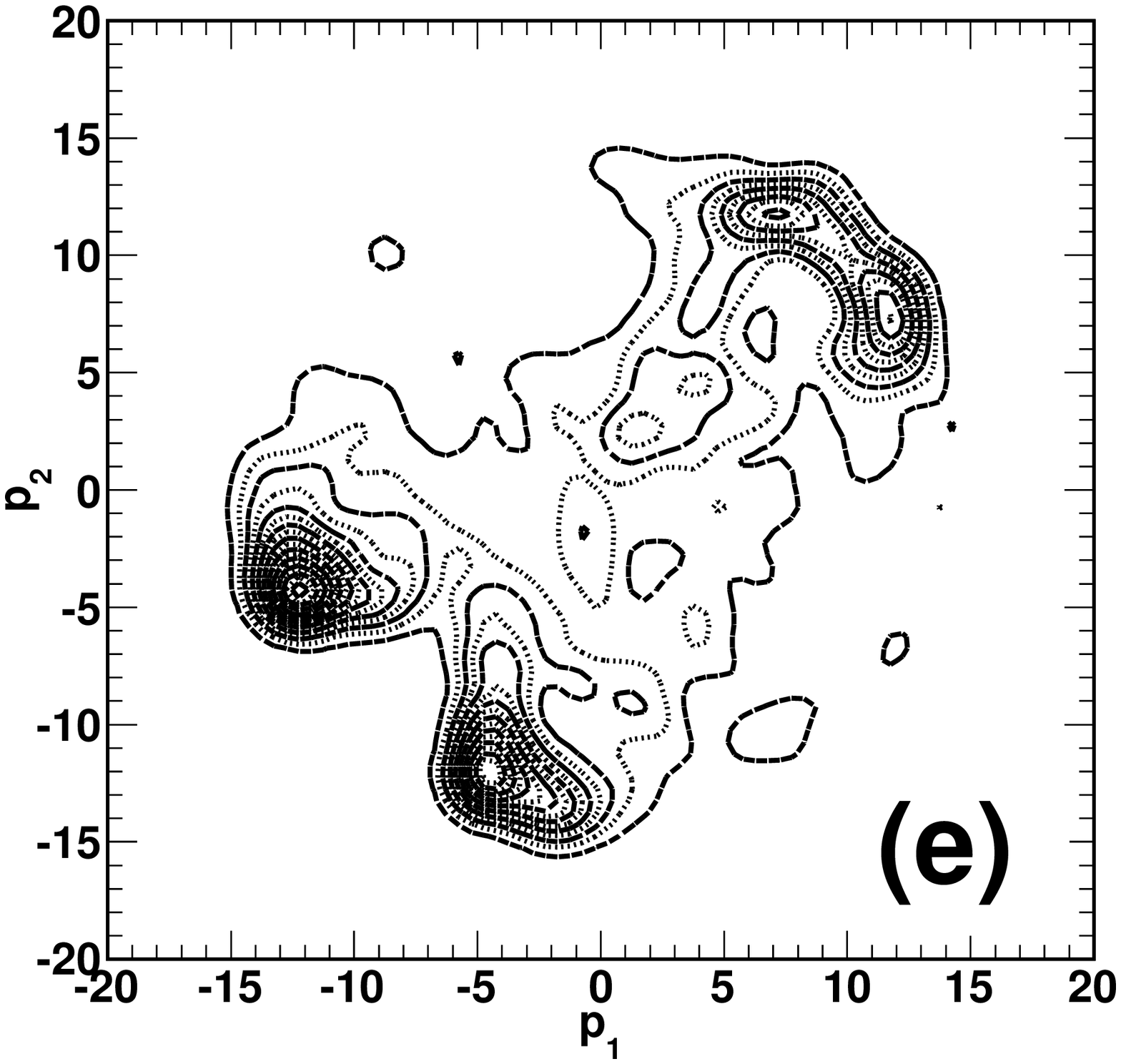}
\includegraphics[width=0.45\textwidth]{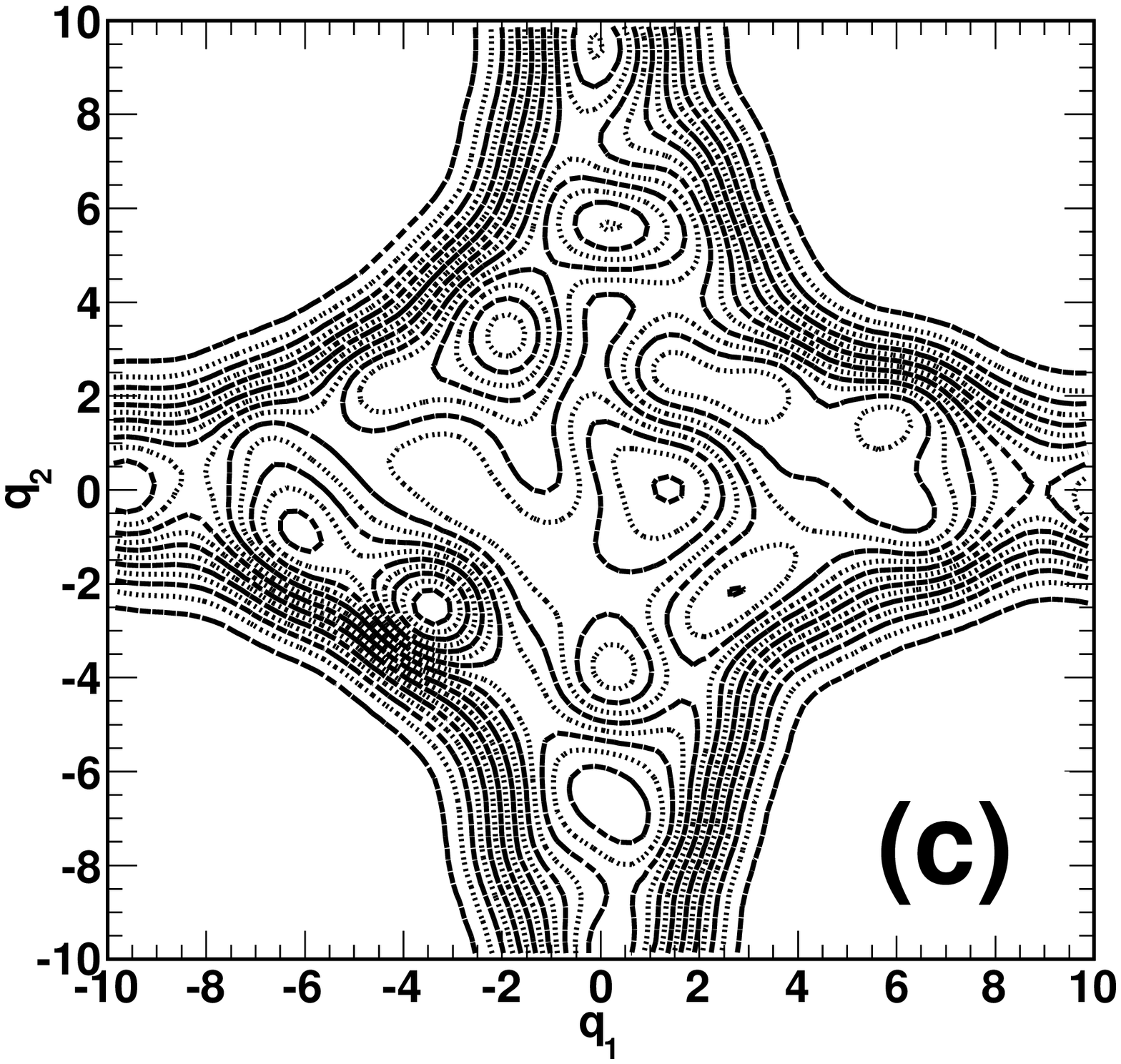}
\includegraphics[width=0.45\textwidth]{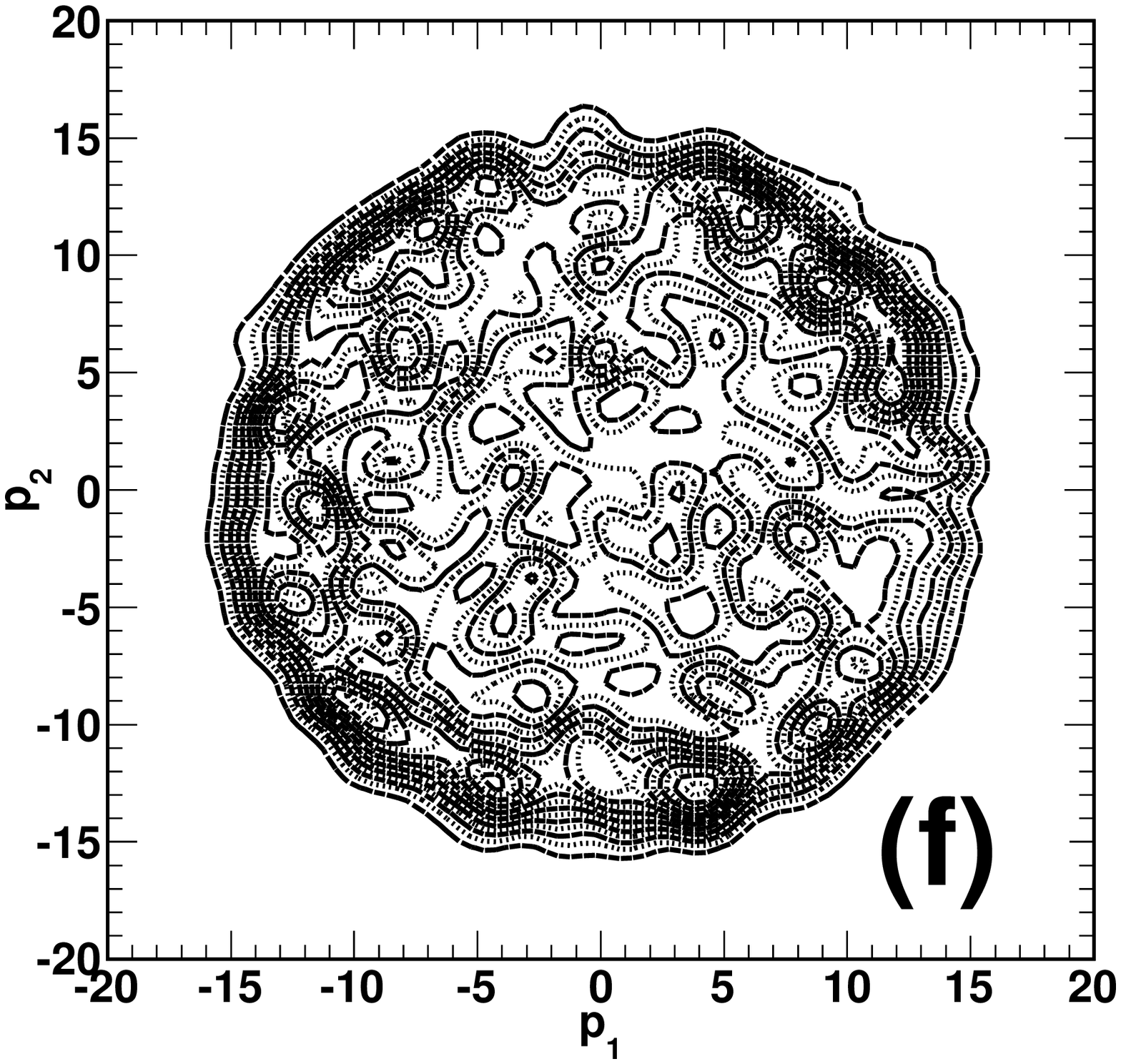}
\caption{Two-dimensional projections of the Husimi distribution on position space $F_q (t; q_1,q_2)$ at times (a) $t=0$, (b) $t=2$ and (c) $t=10$, and on momentum space $F_p (t; p_1,p_2)$ at times (d) $t=0$, (e) $t=2$ and (f) $t=10$. The number of test particles is $N=1000$.}
\label{fig:fiq_all}
\end{figure*}

To visualize the Husimi distribution as a function of time, it is useful to project the distribution either onto the two-dimensional position space $(q_1,q_2)$ or onto momentum space $(p_1,p_2)$ by integrating out the remaining two variables. To this end, we define the following two distribution functions:
\begin{widetext}
\ba
 F_q \left(t; q_1, q_2 \right) &=& \int_{-\infty}^{\infty} dp_1 dp_2
\,\rho_H \left( t; q_1,q_2,p_1,p_2 \right)
\nn \\
&=& \frac{ 2\pi\hbar^2}{N} \sum\limits_{i=1}^{N}
\sqrt{\frac{\Delta_1\Delta_2}{ c_{p_1 p_1}c_{p_2 p_2}} }
 \exp \left[  - \frac{ \Delta_1}{2c_{p_1 p_1}} \left( q_1 -\bar{q}_1^i (t)  \right) ^2
- \frac{\Delta_2}{2 c_{p_2 p_2}} \left( q_2 -\bar{q}_2^i (t)  \right) ^2  \right] ;
\label{eq:F_x}
\ea
\ba
F_p \left( t; p_1, p_2\right) &=& \int_{-\infty}^{\infty} dq_1 dq_2
\,\rho_H \left( t; q_1,q_2,p_1,p_2  \right)
\nn \\
&=& \frac{ 2\pi\hbar^2} {N}  \sum\limits_{i=1}^{N}
\sqrt{\frac{\Delta_1\Delta_2}{ c_{q_1 q_1}c_{q_2 q_2}} }
 \exp \left[  -  \frac{\Delta_1}{2c_{q_1 q_1}} \left( p_1 -\bar{p}_1^i (t)  \right) ^2
-\frac{\Delta_2}{2 c_{q_2 q_2}}  \left( p_2 -\bar{p}_2^i (t)  \right) ^2 \right] .
\label{eq:F_p}
\ea
\end{widetext}
We can conveniently visualize the evolution of the Husimi distribution $\rho_H(t;  q_1,q_2,p_1,p_2)$ by showing contour plots of the two-dimensional projections $F_q(t; q_1, q_2) $ and $F_p( t; p_1, p_2)$. Figure~\ref{fig:fiq_all} shows $F_q$ and $F_p$ side by side at times $t=0$, $t=2$, and $t=10$, respectively. At the initial time, $F_q(0; q_1, q_2)$ is chosen as a Gaussian distribution centered around the origin in position space, while $F_p(0; p_1, p_2)$ is a Gaussian function centered around $(p_1,p_2) = (10,10)$. The projected initial distributions are shown panels (a) and (d) of Fig.~\ref{fig:fiq_all}. As shown next in panels (b) and (e) of Fig.~\ref{fig:fiq_all}, $F_q$ and $F_p$ at $t=2$ are beginning split into distinct clusters. This behavior is caused by the fact that test particles bounce off the equipotential curves defined by $\epsilon=\bar{\mathcal{H}}_H \left(\bar{\mathbf{q}} , \mathbf{0} \right)$.

Closer inspection of the time evolution of $F_q(t; q_1, q_2)$ and $F_p( t; p_1, p_2)$ reveals that gross features of the Husimi distribution $\rho_H(t;\mathbf{q}, \mathbf{p})$ remain approximately unchanged for $t\ge6$. To wit, the panels (c) and (f) of Fig.~\ref{fig:fiq_all}, presenting $F_q$ and $F_p$ at $t=10$, show that the contours of $F_q(10; q_1, q_2)$ follow equipotential lines, while the contours of $F_p( 10; p_1, p_2) $ are shaped as concentric circles, {\em i.~e.}, lines of constant kinetic energy. The time evolution of $F_q$ demonstrates that   test particles starting from their initial positions localized around the origin in position space $(q_1,q_2)$ eventually spread all over the region enclosed by the equipotential curves defined by $\epsilon=\bar{\mathcal{H}}_H \left(\bar{\mathbf{q}} , \mathbf{0} \right)$. This behavior is a result of the fact that the Yang-Mills quantum system is chaotic implying a strong sensitivity of test particle trajectories on their initial conditions.

The Wehrl-Husimi entropy $S_{H}(t)$  defined in (\ref{eq:entropy}) is the coarse grained entropy of the quantum system. The numerical evaluation of the four-dimensional integral in the definition (\ref{eq:entropy}) of the entropy $S_H(t)$ is nontrivial because the upper (lower) limits of the integral in each dimension are infinite and
the width of each test particle Gaussian is tiny.  Therefore, we use the following method to evaluate the integrals
efficiently. For each discretized time step $t_k$, we find the largest absolute values of the test particle positions and momenta. Since each Gaussian is narrow and the Husimi distribution is nearly zero outside the regions of support of the test particles, we can assign $\pm(\max_i |\bar{q}_1^i(t_k)|+b) $ as the limits of integration over the variable $q_1$. We choose $b= 6 (\gamma_K^1)^{-1/2} $ to ensure that the Gaussians of all test particles are fully covered by the integration range within our numerical accuracy. Similar limits are assigned to the
integrations over $q_2$, $p_1$, and $p_2$, respectively. These integration limits ensure that the integrals run over the whole domain of phase space where the Husimi distribution has support. We verify the accuracy of Simpson's rule by evaluating the normalization for $\rho_{H}(t;\mathbf{q}, \mathbf{p})$ for various time $t$. We find that the numerical results coincide with (\ref{eq:rho_norm}) within errors of less than $0.3\%$. We then perform the numerical integration by Simpson's rule.

\begin{figure}
\includegraphics[width=0.45\textwidth]{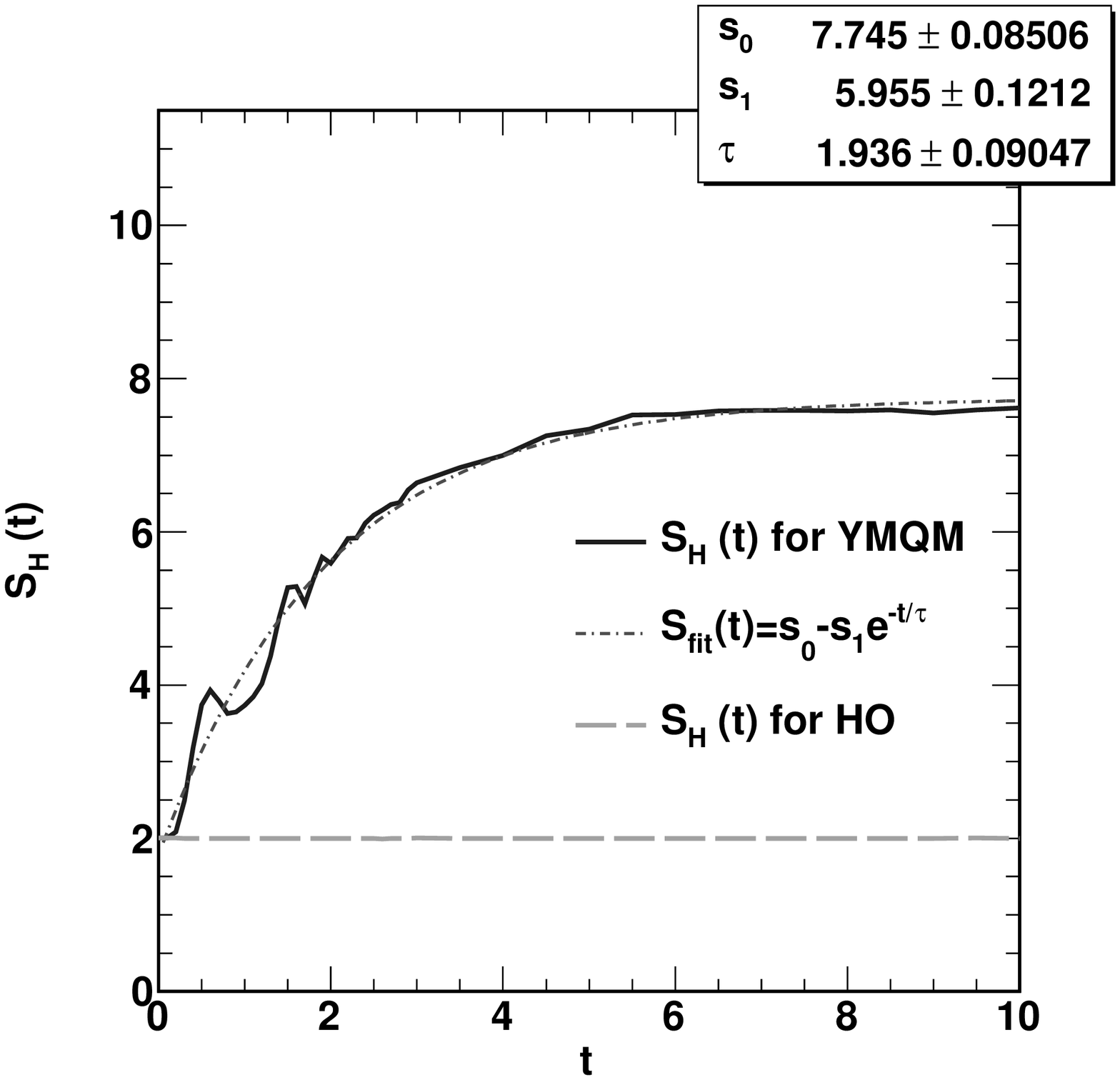}
\caption{The time evolution of the Wehrl-Husimi entropy
$S_H(t)$ for Yang-Mills quantum mechanics (YMQM), the fit function
$S_{\mathrm{fit}}(t)$ for the Wehrl-Husimi entropy, and $S_H(t)$ for
the harmonic oscillator (HO).  We set the same initial condition at
$t=0$ both for YMQM and HO. The figure shows that $S_H(t)$ for YMQM
starts from $S_H(0) \approx 2.0$ and saturates to $7.6$ for $t \ge
6.5$, while $S_H(t)$ for HO remains at $2.0$ for all times. The fit
parameters for $S_{\mathrm{fit}}(t)$ are listed in the figure.}
\label{fig:entropy}
\end{figure}

Our results for the Wehrl-Husimi entropy $S_H(t)$ for $N=1000$ test particles are shown in Fig.~\ref{fig:entropy}. We evaluate $S_H(t)$ for Yang-Mills quantum mechanics (YMQM) and for the harmonic oscillator (HO), for comparison. The Hamiltonian for YMQM is given in (\ref{eq:YMH}), while the Hamiltonian for HO is:
\ba
{\mathcal H} = \frac{1}{2m}\left( {p_1^2 + p_2^2 } \right) + \frac{1}{2}v^2(q_1^2 + q_2^2),
\label{eq:HO}
\ea
where we set $m=v=1$. We remind the reader that initially $\rho_H(0)$ is chosen as a minimum uncertainty distribution satisfying the constraints (\ref{eq:condition}, \ref{eq:H_0_widths}) with the total number of test particles $N=1000$. We assign the same initial condition both for YMQM and HO, and we compare the difference in their Wehrl-Husimi entropies as time evolves.  Figure~\ref{fig:entropy} shows that $S_H(0) \approx 2.0$, and ${ S_H }(0) \ge 2$ for $t\ge 0$ for YMQM, in agreement with the conjecture (\ref{eq:conjecture}). For late times, Fig.~\ref{fig:entropy} reveals that $S_H(t)$ for YMQM saturates to $7.7$ for $t \ge 6.5$. In order to find the characteristic time for the growth of the entropy, we fit $S_H(t)$ for YMQM to the parametric form:
\ba
S_{\mathrm{fit}} (t)=s_0-s_1 \exp (-t/\tau),
\label{eq:s_0}
\ea
where $s_0$, $s_1$ and $\tau$ are fit parameters. The fit shown as a dash-dotted line in Fig.~\ref{fig:entropy} corresponds to the parameters $s_0 \approx7.7$, $s_1 \approx 6.0$ and $\tau \approx 1.9$. On the other hand, $S_H(t)$ for HO starts from $S_H(0) \approx 2.0$ and then remains at $2.0$ for all times.

In Fig.~\ref{fig:entropy}, we note that the coarse grained entropy does not increase continuously as time evolves. This fact can be interpreted in the framework of Zwanzig's formalism for the time evolution of ``relevant'' density operator \cite{Jancel:1969book,Rau:1996ea}. In Zwanzig's formalism, one defines the relevant density operator as $\hat \rho_R (t)=\hat P \hat \rho(t)$, where $\hat{P}$ denotes the projection operator. The transition of the density operator $\hat\rho(t) \to \hat\rho_R(t)$ and of corresponding entropies $S[\hat\rho(t)] \to S[\hat \rho_R(t)]$ is referred to as generalized coarse graining \cite{Rau:1996ea,Zeh:2007vf}. By applying $\hat{P}$
to (\ref{eq:Liouville}), one obtains the equation for time evolution of $\hat \rho_R(t)$. The non-Markovian part of this equation reads:
\be
\frac{\partial \hat{\rho}_R (t)}{\partial t}=- \int_0^t dt'
\;\hat{G}(t')  \hat{\rho}_R (t-t'),
\label{rho_R_EOM}
\ee
where $\hat{G}$ denotes the so-called memory kernel \cite{Jancel:1969book,Rau:1996ea, Zeh:2007vf}. It can be shown that  $d S[\hat \rho_R(t)]/dt$ receives contributions from the non-Markovian term indicated in (\ref{rho_R_EOM}). Therefore, $S[\hat \rho_R(t)]$ in general does not increase monotonically as a function of time. The Husimi equation of motion in (\ref{eq:Husimi_general}) contains a similar memory effect. Therefore, in Fig.~\ref{fig:entropy} the coarse grained entropy $S_H(t)$ does not increase continuously as time evolves, and the second law of thermodynamics holds only in a time averaged sense \cite{Rau:1996ea}.

\subsection{Lyapunov exponents \label{sec:GLE}}

Since the classical system corresponding of YMQM is almost chaotic, we evaluate the average Kolmogorov-Sina\"i (KS) entropy for this system. For a two dimensional  system, the KS entropy is defined as:
\ba
h_{\mathrm{KS}}= \sum_{j=1}^{4} \lambda_j \, \theta( \lambda_j ), \label{eq:KS_entropy}
\ea
where $\lambda_j$'s are the Lyapunov exponents (LE). To obtain the full spectrum of the LEs, we utilize the following procedure. First, we divide a large time interval, from $t=0$ to $t=t_{\mathrm{max}}$, into a number of slices. Each time slice is labeled by its final time $t_k$, where $k=1,2,...,k_{\mathrm{max}}$. Let  $\bar{\bm{\chi}}^{i} (t)=( {\bar q_1^i(t),\bar q_2^i(t),\bar p_1^i(t),\bar p_2^i(t)} )$ denote the position of test particle $i$ in phase space. At $t=0$, we perform four orthogonal perturbations on the initial condition: $\bm{\pi}^i_j(0)=\bar{\bm{\chi}}^{i} (0) + \epsilon\, \hat{\mathbf{e}}_j$, for $j=1,...,4$, where $\hat{\mathbf{e}}_j$'s are orthonormal vectors and we set $\epsilon=10^{-4}$.   For each time slice $t\in[t_{k-1}, t_k]$, we solve eqs.~(\ref{eq:I_x1}-\ref{eq:I_p2}) and obtain one reference trajectory $\bar{\bm{\chi}}^{i} (t)$ and four modified trajectories $\bm{\pi}^i_j (t)$, where $j=1,...,4$. Define the four deviation vectors: $\bm{\delta}^i_j(t)= \bm{\pi}^i_j (t) - \bar{\bm{\chi}}^i (t)$. After obtaining the four  deviations $\bm{\delta}^i_j(t_k)$, we orthogonalize these four vectors and rescale their lengths back to $\epsilon$. We store the four rescaling factors $r^i_j (t_k)$ for each $j$ and $k$, and we repeat the above procedures for the representative test particles $i=1,...,N_{\mathrm{rep}}$, where $N_{\mathrm{rep}} \le N$. For the case of $N=1000$, we choose $N_{\mathrm{rep}}=100$. Besides, we set $t_k=2k$ and $t_{\mathrm{max}}=100$, and therefore $k_{\mathrm{max}}=50$. Finally, we obtain the full Lyapunov spectrum:
\ba
\lambda_j=\frac{1}{N_{\mathrm{rep}}} \sum_{i=1}^{N_{\mathrm{rep}}} \frac{1}{t_{\mathrm{max}}}  \ln \left[ \prod_{k=1}^{k_{\mathrm{max}}} r^i_j (t_k) \right],
\ea
where $j=1,...,4$. The numerical values of the LEs for YMQM are:
\ba
\lambda_1 &=& 1.216,  \qquad \lambda_2 = 2.344 \times 10^{-2}, \nonumber \\
\lambda_3 &=& -2.349 \times 10^{-2},  \qquad \lambda_4 = -1.223. \label{eq:GLEs}
\ea
If we take the classical limit $\hbar \to 0$ and $\alpha \to 0$ for (\ref{eq:Husimi_2D}) and repeat the above procedure, we obtain the LEs for the regular classical equations of motion without the quantum (Husimi) corrections:
\ba
\lambda_1^c &=& 1.283,  \qquad \lambda_2^c = 1.599 \times 10^{-2}, \nonumber \\
\lambda_3^c &=& -1.629 \times 10^{-2},  \qquad \lambda_4^c = -1.287.  \label{eq:GLEs_classical}
\ea
From (\ref{eq:GLEs}) and (\ref{eq:GLEs_classical}), we observe that classical solutions conserve the energy and the phase space better than the quantum solutions. By (\ref{eq:KS_entropy}, \ref{eq:GLEs}), we obtain the average KS entropy for YMQM: $h_{\mathrm{KS}} \approx 1.24$.

In addition, we calculate the logarithmic breaking time for YMQM, which is defined as \cite{Berman:1978pa, Zaslavsky:1981pr, Iomin:2001pre}:
\ba
{\tau _\hbar } \approx \frac{1}{\Lambda }\ln \left( {\frac{I}{\hbar }} \right),   \label{eq:LBT}
\ea
where $I$ is the characteristic action and $\Lambda$ is the characteristic Lyapunov exponent.  We set $\Lambda=h_{\mathrm{KS}}$ for YMQM. We utilize two methods for obtaining the action $I$. One of these is to obtain $I$ from the classical dynamical variables $(\mathbf{q}, \mathbf{p})$:
\ba
I = \oint_{C} {{\mathbf{p}} \cdot d{\mathbf{q}}}\;.  \label{eq:action_1}
\ea
The integration is taken over the curve $C$ constrained by $\mathcal{H}=E$, where $\mathcal{H}$ is defined in (\ref{eq:YMH}) and $E$ denotes the classical energy of the system.  If we consider the case where a classical particle moves along the line $q_1=q_2$ in the position space and is subject to the potential energy $\frac{1}{2} q_1^2 q_2^2$, we obtain the period of motion of this classical particle:
\ba
T = 4\int_0^{{q_{\mathrm{max}}}} {\frac{{dq}}{{\sqrt {E - {\textstyle{1 \over 2}}{q^4}} }}},   \label{eq:period}
\ea
where $q=q_1=q_2$ and $q_{\mathrm{max}}=(2E)^{1/4}$. In the following numerical calculation, we set $E=100$.
Considering the periodic motion of this particle, we obtain by (\ref{eq:LBT}-\ref{eq:period}) that $I=263$, $T=1.97$ and $\tau _{\hbar} \approx 4.5$.  Alternatively, we evaluate the action by the integrating along test-particle trajectories obtained by  (\ref{eq:I_x1}-\ref{eq:I_p2}), which are the Husimi (quantum) equations of motion
in the fixed-width ansatz. Thus the action is:
\ba
I =\frac{1}{N} \sum_{i=1}^{N} \int_0^T dt\; {{\bar{\mathbf{p}}^i}(t) \cdot {\dot{ \bar{\mathbf{q}}}^i}(t) },   \label{eq:action_2}
\ea
where $T$ defined in (\ref{eq:period}). In (\ref{eq:action_2}), we estimate the time interval by the period of a classical particle moving along $q_1=q_2$ in the position space and is subject to the potential energy $\frac{1}{2} q_1^2 q_2^2$. By (\ref{eq:LBT}, \ref{eq:action_2}), we obtain  $I=267$ and $\tau _{\hbar} \approx 4.5$ in excellent agreement with the result of the first method. Moreover, comparing $\tau _{\hbar}$ to $\tau$ defined in (\ref{eq:s_0}), we conclude that $\tau _{\hbar}$ and $\tau$ are in the same order of magnitude, and $\tau _{\hbar}>\tau$.

\subsection{Particle number dependence \label{sec:number_dependence}}

\begin{figure}
\includegraphics[width=0.45\textwidth]{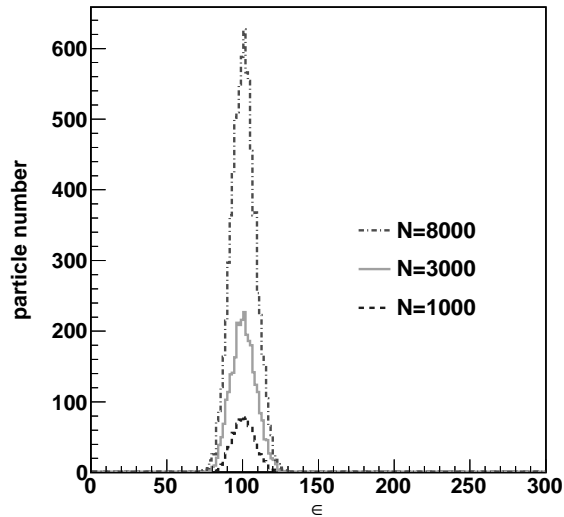}
\caption{Energy histograms of the test particles at $t=0$. The total numbers of test particles are $N=1000$, $N=3000$, and $N=8000$. $\epsilon$ denotes the test-particle energy, which is defined in (\ref{eq:energy_def}), and the labels on the vertical axis denote test-particle numbers. The initial locations of the test particles in the phase space are generated according to the normal distribution $\phi$ defined in (\ref{eq:phi_0}) with the parameters given in (\ref{eq:mu_phi}, \ref{eq:gamma_K_0}). In this plot, we show that $\mu$ and $\sigma$ are independent of $N,$ notwithstanding small fluctuations. By fitting the energy histograms for various choices of $N$, we obtain $\mu = 100.6$ and $\sigma = 8$, with fluctuations less than $0.5\%$ and $5\%$ respectively.}
\label{fig:histo-2}
\end{figure}

In Sect.~\ref{sec:Husimi_plots},  we studied the Husimi distribution and the Wehrl-Husimi entropy for Yang-Mills quantum system by using $N=1000$ test particles. We note that the results of the test particle method we used to obtain $S_H(t)$ depend on the number of test particles. The Husimi distribution $\rho_H (t;\mathbf{q},\mathbf{p}) $ depends on the particle number $N$ through the ansatz in (\ref{eq:ansatz2}), and so does the Wehrl-Husimi entropy $S_H(t)$.

Our main goal in this section is to quantify the dependence of the saturated Wehrl-Husimi entropy on the test particle number $N$. We proceed with this study by the following method. First, we plot the energy histograms for several different numbers of test particles (we choose $N=1000$, $N=3000$ and $N=8000$) in Fig.~\ref{fig:histo-2}. The distribution of the initial locations of the test particles in the phase space are generated according to the normal distribution $\phi$ defined in (\ref{eq:phi_0}), with the parameters given in (\ref{eq:mu_phi}, \ref{eq:gamma_K_0}). Figure~\ref{fig:histo-2} shows that the ranges of the test particle energies differ only slightly for $N=1000$, $N=3000$, and $N=8000$.
In other words, for the energy distribution $ {n_{\mathrm {TP}}}\left( \epsilon  \right)$ defined in (\ref{eq:n_TP}), the center $\mu$ and width $\sigma$ are independent of $N,$ notwithstanding small  fluctuations. By fitting the energy histograms for various choices of $N$, we obtain
\be
\mu = 100.6, \qquad \sigma = 8,
\label{eq:fit_parameters}
\ee
with fluctuations less than $0.5\%$ and $5\%$ respectively. We also define the normalized energy distribution of the test particles as
\ba
\bar{n}_{\mathrm{TP}}(\epsilon)
= \frac{n_{\mathrm{TP}}(\epsilon)}{ \int_0^{\infty} d\epsilon \; n_{\mathrm{TP}}(\epsilon)} .
\label{eq:norm_n_TP}
\ea
Thus we conclude that the energy histograms for all choices of $N$ correspond to a unique normalized energy distribution, $\bar{n}_{\mathrm{TP}}(\epsilon)$, which is unaltered by the time evolution and independent of $N$, provided that $N$ is sufficiently large.

\begin{figure}
\includegraphics[width=0.45\textwidth]{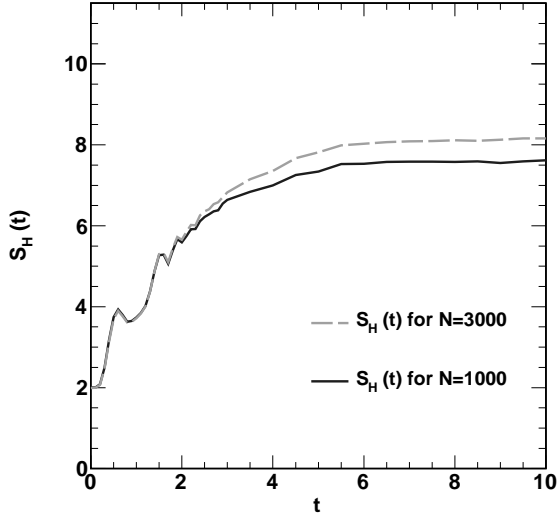}
\caption{The Wehrl-Husimi entropy $S_H(t)$ for $N=1000$ and $N=3000$ respectively. In both cases, the test particles are generated at $t=0$ by the same set of initial parameters in (\ref{eq:H_0_widths}--\ref{eq:mu_phi}, \ref{eq:gamma_K_0}). The Wehrl-Husimi entropies for both values of $N$ agree well for $t\le2$, but gradually diverge for $t>2$. $S_H(t)$ for $N=3000$ saturates to $8.1$, while  $S_H(t)$ for $N=1000$ saturates to $7.6$. The saturation level is reached in both cases for $t\ge6.5$.}
\label{fig:entropy-2}
\end{figure}

Next, we compute the Wehrl-Husimi entropy $S_H(t)$ for $N=3000$ under the same set of initial parameters (\ref{eq:H_0_widths}--\ref{eq:mu_phi}, \ref{eq:gamma_K_0}) we used in Sec.~\ref{sec:Husimi_plots} to calculate $S_H(t)$ for $N=1000$. We plot the Wehrl-Husimi entropy $S_H(t)$ for the two values of $N$ in Fig.~\ref{fig:entropy-2}. We observe that the Wehrl-Husimi entropy $S_H(t)$ for $N=1000$ and $N=3000$ agree well for $t\le2$, but gradually diverges when $t>2$. For both cases, the entropy begins to saturate at almost the same time, {\em viz.}, $t\ge6.5$. However, the saturation values are different: for $N=3000$, $S_H(t)$ saturates to 8.1, while for $N=1000$, $S_H(t)$ saturates to $7.6$.

\begin{figure}
\includegraphics[width=0.45\textwidth]{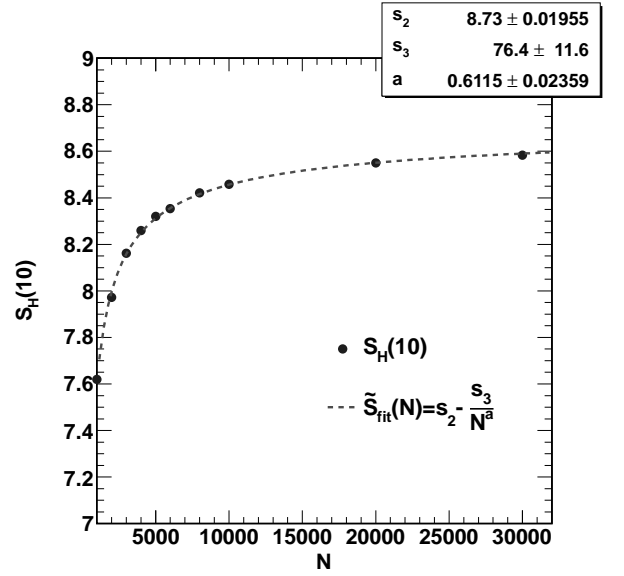}
\caption{$S_H(10)$ for several different test particle numbers $N$, indicated by the blue dots. We fit the curve by a fit function $\tilde{S}_{\mathrm{fit}}(N) $ defined in (\ref{eq:fit_SvsN}). The fit parameters are shown in the figure.}
\label{fig:SHN}
\end{figure}

Based on the above results, we decided to analyze the saturation values of $S_H(t)$ as a function of $N$. From Fig.~\ref{fig:entropy-2} we conclude that the saturation is reached for $t \ge 6.5$, independent of how large $N$ is. We thus use $S_H(10)$ as a proxy for the saturation value of $S_H(t)$. In Fig.~\ref{fig:SHN}, we plot $S_H(10)$ for several different test particle numbers $N$ and fit the curve by the function $\tilde{S}_{\mathrm{fit}}(N)$, defined as:
\be
 \tilde{S}_{\mathrm{fit}}(N)=s_2-\frac{s_3}{N^a},
\label{eq:fit_SvsN}
\ee
where $s_2$, $s_3$ and $a$ are parameters determined by the fit. We obtain:
\be
s_2=8.73, \;\;\; s_3=76.4, \;\;\; a=0.6115 .
\ee
If our hypothesis is correct that $S_H(10)$ represents the saturation value of $S_H(t)$ for any $N$, this
implies that the saturated value of $S_H(t)$ approaches $8.73$ for $N\to \infty$ for the initial conditions chosen for our numerical simulation.

\subsection{Canonical partition function and entropy \label{sec:canonical} }

We now consider the Yang-Mills Hamiltonian system in (\ref{eq:YMH}) for various classical ensembles. In the following numerical calculation, we use the same numerical parameters as those specified in Sect.~\ref{sec:Husimi_plots}. We begin by obtaining the canonical partition function and the canonical entropy for this system. We first determine the temperature of canonical ensemble of the Hamiltonian in (\ref{eq:YMH}) that would be reached if the system would approach thermal equilibrium. This temperature can be obtained by the following procedure.
First of all, the total energy of the system is defined in (\ref{eq:EHrho}) and was evaluated numerically to be  $\mathcal{E} \left[ \mathcal{H}_H \rho_H \right]= 100.6 \pm 0.5$, as shown in Sect.~\ref{sec:Husimi_plots}. On the other hand, the canonical ensemble average of the Hamiltonian $\mathcal H$
\ba
\langle \mathcal{H} \rangle _C = \frac{1}{Z} \int_{-\infty}^{\infty}
d\Gamma_{\mathbf{q},\mathbf{p}}\,
\mathcal{H}\, \exp\left( - \mathcal{H}/T \right),
\label{eq:enr_can}
\ea
where $T$ is the temperature and the partition function is defined as
\ba
Z = \int_{-\infty}^{\infty} d\Gamma_{\mathbf{q},\mathbf{p}}\,
\exp\left( - \mathcal{H}/T \right).
\label{eq:partition_function}
\ea
We then fix $\langle \mathcal{H} \rangle_C$ to the total energy of the system $\mathcal{E} \left[ \mathcal{H}_H \rho_H \right]$:
\ba
\mathcal{E} \left[ \mathcal{H}_H \rho_H \right]=\langle \mathcal{H} \rangle_C,
\label{eq:canonical}
\ea
from which we determine the temperature $T_{\chi}$ of the equivalent canonical ensemble.

When we try to evaluate (\ref{eq:canonical}) by substituting the Hamiltonian of the Yang-Mills system (\ref{eq:YMH}) into (\ref{eq:enr_can}), we encounter a problem associated with the classical limit of the quantum system. The integrals over $q_1$ and $q_2$ exhibit a logarithmic divergence owing to the special form of the potential $V(q_1,q_2)$, which vanishes along the axes $q_1=0$ and $q_2=0$. A classical particle can therefore escape toward infinity in the hyperbolic channels along the $q_1,q_1$ axes \cite{Matinyan:2005cm}. In contrast, the escape of a quantum mechanical particle to infinity is forbidden by quantum fluctuations. The channels grow narrower as the particle moves away from the origin, and more and more energy is required to localize the particle in the direction orthogonal to the channel. The uncertainty relation thus provides for effectively finite boundary conditions; as a result, the energy levels of the quantum system are discretized \cite{Simon:1983jy}.

Matinyan and M\"uller \cite{Matinyan:2005cm,Matinyan:2005hp} showed that this quantum effect could be accounted for in the semi-classical limit by adding a harmonic term to the Hamiltonian:
\ba
{\mathcal H} = \frac{1}{2m}\left( {p_1^2 + p_2^2 } \right) + \frac{1}{2}g^2q_1^2 q_2^2
+\frac{g^2\hbar^2}{2mT} \left( q_1^2 + q_2^2 \right),
\label{eq:YMH-intermediate}
\ea
where the last term encodes the quantum correction. Thus, instead of inserting the classical Hamiltonian into (\ref{eq:YMH}), we apply the Hamiltonian with quantum corrections:
\ba
{\mathcal H} = \frac{1}{2m}\left( {p_1^2 + p_2^2 } \right) + \frac{1}{2}g^2q_1^2 q_2^2
+\frac{1}{2} m\omega^2\left( q_1^2+q_2^2\right),
\label{eq:YMH-2}
\ea
where
\ba
\omega^2 =\frac{\hbar^2 g^2}{2m^2 T} .
\label{eq:omega_eff}
\ea
The additional term arises from the commutator of the kinetic and potential energy in the semiclassical expansion of the partition function \cite{Matinyan:2005hp}. After setting $m=g=\hbar=1$ we can now solve eq.~(\ref{eq:canonical}) for the equivalent temperature $T_\chi$. We obtain $T_{\chi} \approx 67.1$ and $\omega \approx 0.0863$.

Starting from the Hamiltonian (\ref{eq:YMH-2}), we obtain the partition function for the canonical ensemble \cite{Tomsovic:1991jpa, Matinyan:2002pd}:
\ba
Z\left( \omega, T \right)=\frac{{m{T^{3/2}}}}{{\sqrt {2\pi } g{\hbar
^2}}}\exp \left( {\frac{{{m^2}{\omega ^4}}}{{4{g^2}T}}}
\right){K_0}\left( {\frac{{{m^2}{\omega ^4}}}{{4{g^2}T}}} \right),
\label{partition_function_2}
\ea
where $K_0(z)$ denotes the modified Bessel function of the second kind. Since  the free energy in the canonical ensemble theory is $F= - {T\ln Z} $ and the entropy is given by $S_C = -\partial F/\partial T$, the entropy of our system in the canonical ensemble is:
\ba
S_C( \omega, T) &=& \frac{3}{2} + \frac{m^2\omega^4}{4g^2 T}
\left( \frac{K_1\left( \frac{m^2\omega^4}{4g^2T} \right)}
{K_0\left(\frac{m^2\omega^4}{4g^2 T} \right)} - 1 \right)
\nn \\
&+& \ln \left[ \frac{m T^{3/2}}{\sqrt{2\pi} g\hbar^2} \exp \left( \frac{m^2\omega^4}{4g^2 T} \right)
K_0\left(\frac{m^2\omega^4}{4g^2 T} \right) \right].
\label{eq:entropy_canonical}
\ea
The partition function $Z$ diverges for $\omega = 0$, and so does the canonical entropy $S_{C}$.  Both divergences are cured by the quantum correction to the Hamiltonian (\ref{eq:YMH-2}). In view of the discussion above, we obtain the canonical entropy as $S_C(\omega,T_\chi) \approx 9.70$.

\subsection{Microcanonical distribution and entropy  \label{sec:MC}}

In this section, we compare the late-time Husimi distribution to the microcanonical distribution. Since the Yang-Mills quantum system is an isolated system, we anticipate that the Husimi distribution after equilibration would approach the microcanonical distribution.

We obtain the appropriate microcanonical distribution by the following procedure. First, we construct the microcanonical distribution in the test particle space by
\ba
\bar{\rho}_{\mathrm{MC}}\left(  \bar{\mathbf{q}} , \bar{\mathbf{p}}   \right)
= \frac{1}{\Xi} \int_{0}^{\infty} d\epsilon \; \delta\left[
\bar{\mathcal{H}}_H \left(\bar{\mathbf{q}} , \bar{\mathbf{p}} \right)
-\epsilon\right]  {\bar{n}_{\mathrm {TP}}}\left( \epsilon  \right),
\label{eq:bar_rho_MC}
\ea
where $\bar{\mathcal{H}}_H \left(\bar{\mathbf{q}} , \bar{\mathbf{p}} \right)$ is defined in (\ref{eq:barH_H}), $\epsilon$ is defined in (\ref{eq:energy_def}), $\bar{n}_{\mathrm{TP}} (\epsilon )$ is defined in (\ref{eq:norm_n_TP}), and $\Xi $ is the normalization constant. We note that the initial energy distribution for our system is not strictly a delta function $\delta[  \bar{\mathcal{H}}_H (\bar{\mathbf{q}} , \bar{\mathbf{p}} ) -\epsilon]$, because we generated the test particle positions in phase space randomly according to the distribution $\phi$ defined in eq.~(\ref{eq:phi_0}). Therefore, $\bar{\rho}_{\mathrm{MC}}\left(  \bar{\mathbf{q}} , \bar{\mathbf{p}}   \right)$ must be defined  as $\delta[  \bar{\mathcal{H}}_H (\bar{\mathbf{q}} , \bar{\mathbf{p}} ) -\epsilon]$ folded with the energy distribution of test particles shown in (\ref{eq:bar_rho_MC}). According to (\ref{eq:energy_conservation_TP}), the energy is conserved for each test particle individually, and thus ${\bar{n}_{\mathrm {TP}}}\left( \epsilon  \right)$ remains unchanged as time evolves.  Using (\ref{eq:n_TP}), (\ref{eq:norm_n_TP}) and (\ref{eq:bar_rho_MC}), we easily obtain
\ba
\bar{\rho}_{\mathrm{MC}}\left(  \bar{\mathbf{q}} , \bar{\mathbf{p}}   \right)
= \frac{1}{\Xi'} \exp \left[
-\frac{1}{2\sigma^2}\left(\bar{\mathcal{H}}_H \left(\bar{\mathbf{q}} , \bar{\mathbf{p}} \right) - \mu \right)^2\right],
\label{eq:bar_rho_MC_2}
\ea
where $\mu$ and $\sigma$ are input from (\ref{eq:fit_parameters}), $\Xi'$ is the redefined normalization constant and $\bar{\mathcal{H}}_H \left(\bar{\mathbf{q}} , \bar{\mathbf{p}} \right)$ is obtained from (\ref{eq:barH_H-2}). In the test particle space, $\bar{\rho}_{\mathrm{MC}}$ is normalized as:
\ba
\int_{ - \infty }^\infty  d\Gamma_{\bar{\mathbf{q}},\bar{\mathbf{p}}}\,
\bar{\rho}_{\mathrm{MC}}({\bar {\mathbf {q} }},\bar {\mathbf {p} }) = 1.
\label{eq:norm_bar_rho_MC}
\ea

\begin{figure}
\includegraphics[width=0.45\textwidth]{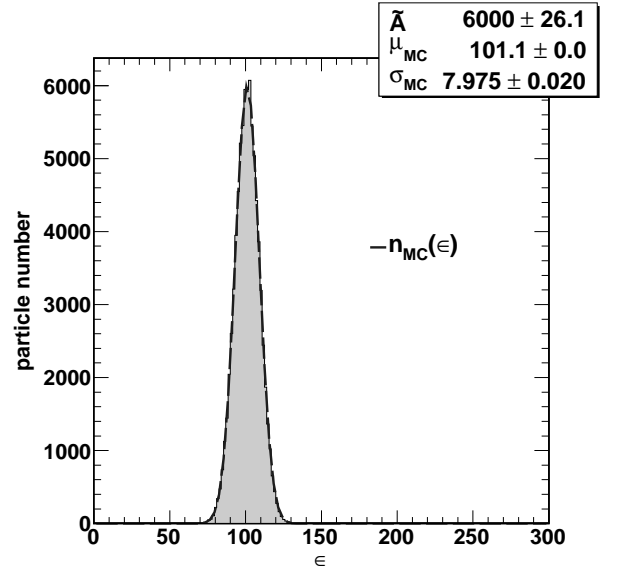}
\caption{  Energy histogram of test functions for $\bar{\rho}_{\mathrm{MC}}\left(  \bar{\mathbf{q}} , \bar{\mathbf{p}} \right)$, which is defined in (\ref{eq:bar_rho_MC_2}). The test functions are generated by Metropolis-Hastings algorithm, and the total number of test functions is $M=8\times10^4$. $\epsilon$
denotes the test-particle energy, which is defined in (\ref{eq:energy_def}), and the labels on the vertical axis denote test particle numbers. A normal distribution $n_\mathrm{MC} (\epsilon)$ is used to fit this histogram. $\tilde{A}$, $\mu_{\mathrm {MC}}$ and $\sigma_{\mathrm {MC}}$ are the fit parameters for ${n_{\mathrm
{MC}}}( \epsilon)$, which are defined in (\ref{eq:n_TP}). The values for the fit parameters are shown in the plot.}
\label{fig:fitting_MC_energy}
\end{figure}

\begin{figure}
\includegraphics[width=0.45\textwidth]{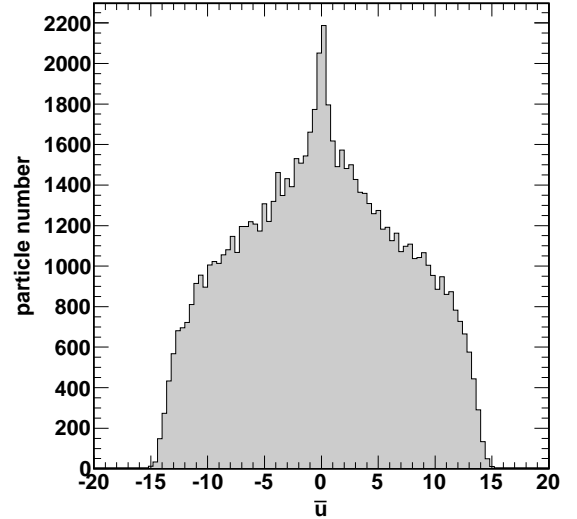}
\caption{   $\bar{u}$-histogram of test functions for $\bar{\rho}_{\mathrm{MC}}\left(  \bar{\mathbf{q}} , \bar{\mathbf{p}}   \right)$, which is defined in (\ref{eq:bar_rho_MC_2}). The test functions are generated by Metropolis-Hasting Algorithm, and the total number of test functions is $M=8\times10^4$. $\bar{u}$ is defined as $\bar{u}=\bar{q}_1 \bar{q}_2$, the labels on the vertical axes denote test-particle numbers.}
\label{fig:fitting_MC_u}
\end{figure}

\noindent To obtain the microcanonical distribution in the phase space $\rho_{\mathrm{MC}} \left(\mathbf{q} , \mathbf{p} \right)$, we convolute $\bar{\rho}_{\mathrm{MC}}$ with test particle Gaussian $K$, which yields:
\ba
&&\rho_{\mathrm{MC}} \left(\mathbf{q} , \mathbf{p} \right)   \nonumber \\
&&= \int_{-\infty}^{\infty} d\Gamma_{\bar{\mathbf{q}},\bar{\mathbf{p}}}
\bar{\rho}_{\mathrm{MC}} \left(\bar{\mathbf{q}} , \bar{\mathbf{p}} \right)
K \left(\mathbf{q} - \bar{\mathbf q}, \mathbf{p} - \bar{\mathbf p} \right),
\label{eq:rho_MC}
\ea
where $\bar{\rho}_{\mathrm{MC}}$ is defined in (\ref{eq:bar_rho_MC_2}) and $K$ is defined in (\ref{eq:K_0}).
The microcanonical entropy is then obtained as:
\ba
S_\mathrm{MC} =  - \int_{-\infty}^{\infty} d\Gamma_{\mathbf{q},\mathbf{p}} \,
{ \rho}_\mathrm{MC}(\mathbf{q},\mathbf{p}) \ln {\rho}_\mathrm{MC}(\mathbf{q},\mathbf{p}).
\label{eq:S_mc}
\ea

Before we proceed, we briefly comment on the reason why $\rho_{\mathrm{MC}} \left(\mathbf{q} , \mathbf{p} \right)$ should be constructed by (\ref{eq:rho_MC}). In statistical physics, the microcanonical distribution of an isolated system of energy $E$ is conventionally obtained by $\rho_{\mathrm{MC}}=\delta(\mathcal{H}-E)/\Omega$, where $\Omega$ is the total number of microstates that satisfies the constraint $\mathcal{H}=E$. If we substitute this conventional definition of $\rho_{\mathrm{MC}}$ into (\ref{eq:S_mc}), it is straightforward to show that $S_{\mathrm{MC}}$ is not well defined. However, if one approximates $\delta(\mathcal{H}-E)$ by an Gaussian distribution centered on $E$ with a finite width $\sigma_g$, $S_{\mathrm{MC}}$ becomes well defined and is a function of both, $E$ and $\sigma_g$. Therefore, $\rho_{\mathrm{MC}} (\mathbf{x}, \mathbf{p})$ in (\ref{eq:rho_MC}) is defined in a way that  encodes the coarse grained energy of the system, the width of energy distribution and the widths for the test particle Gaussians, all of which must be equivalent to those specified in our choice of the initial Husimi distribution $\rho_H (0;\mathbf{x}, \mathbf{p})$ in Sect.~\ref{sec:initial_conditions} and \ref{sec:Husimi_plots}.

\begin{figure*}
\includegraphics[width=0.45\textwidth]{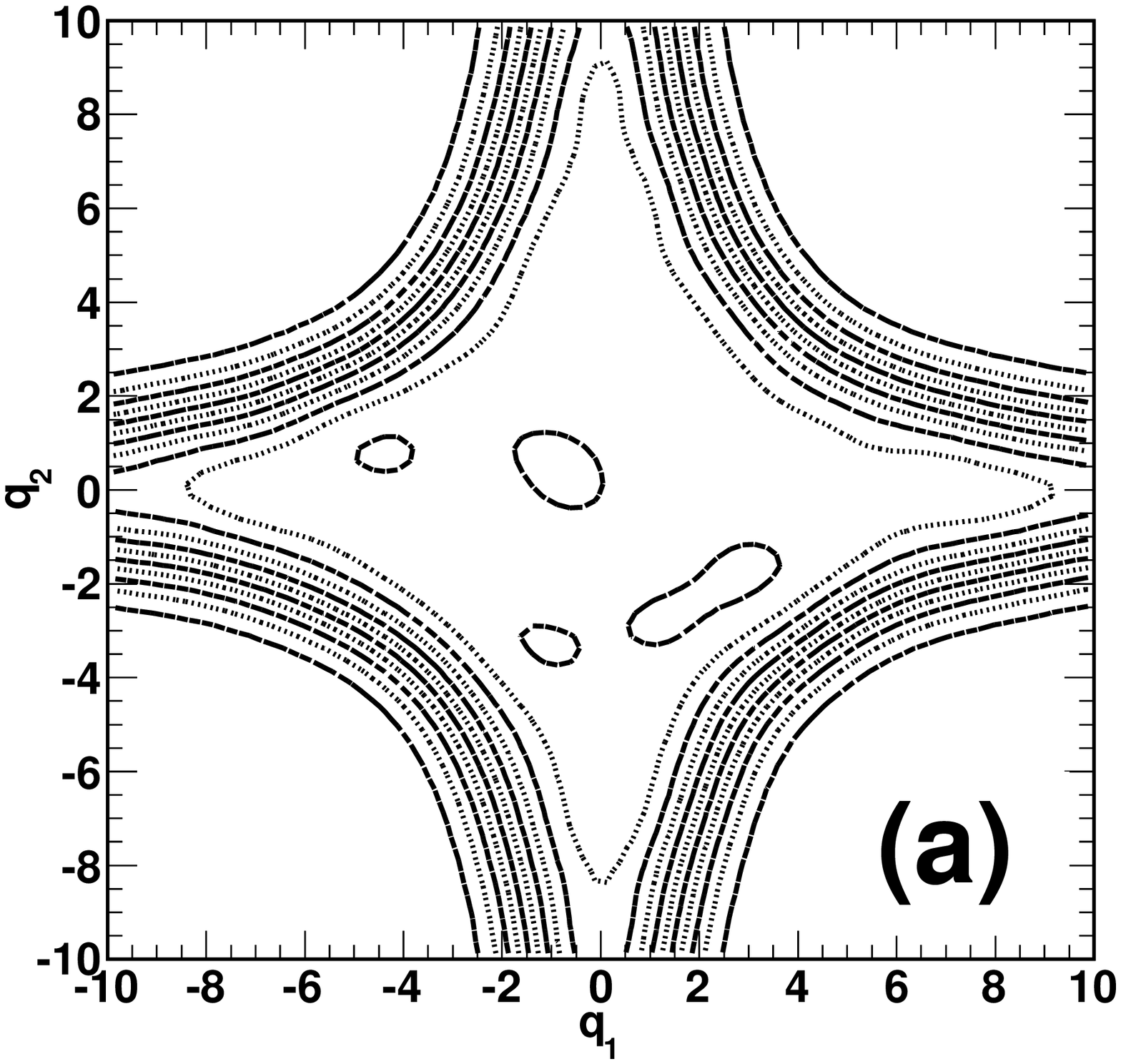}
\includegraphics[width=0.45\textwidth]{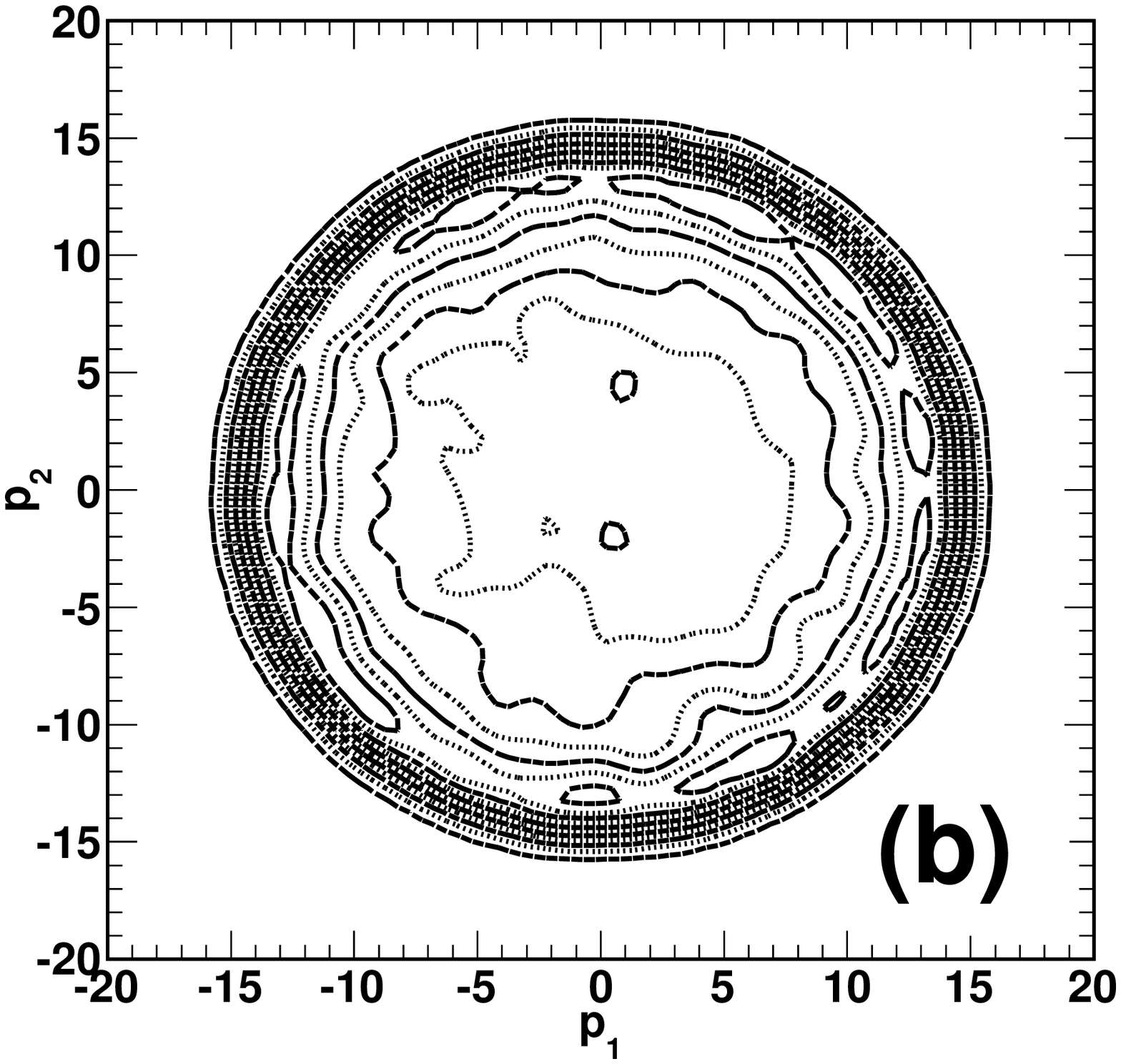}
\caption{The position and momentum projections of the microcanonical distribution function (a) $F^{\rm MC}_q(q_1, q_2)$  and (b) $F^{\rm MC}_p( p_1, p_2)$, defined in eqs.~(\ref{eq:F_q_mc}, \ref{eq:F_p_mc}). The test functions are generated by Metropolis-Hastings algorithm, and the total number of test functions is $M=8\times10^4$.}
\label{fig:MC}
\end{figure*}

Owing to the complexity of (\ref{eq:bar_rho_MC_2}) and the multidimensional integrals (\ref{eq:rho_MC}) and (\ref{eq:S_mc}), we adopt an alternative approach to evaluate $\rho_{\mathrm{MC}} \left(\mathbf{q} , \mathbf{p} \right)$, instead of directly evaluating eq.~(\ref{eq:rho_MC}). Our approach is briefly described as follows. Since $\bar{\rho}_{\mathrm{MC}}\left(  \bar{\mathbf{q}} , \bar{\mathbf{p}}   \right)$ in (\ref{eq:bar_rho_MC_2}) is a non-negative function and normalized by (\ref{eq:norm_bar_rho_MC}), we  generate a sufficiently large number of test functions in $(\bar{\mathbf{q}} , \bar{\mathbf{p}})$-space according to the distribution $\bar{\rho}_{\mathrm{MC}}\left(  \bar{\mathbf{q}} , \bar{\mathbf{p}}   \right)$. Thus $\bar{\rho}_{\mathrm{MC}}\left(  \bar{\mathbf{q}} , \bar{\mathbf{p}}   \right)$ can be represented as a sum of these test functions:
\be
\bar\rho _\mathrm{MC} ( \bar{\mathbf{q}},\bar{\mathbf{p}} )
= \frac{1}{M} \sum_{s = 1}^M \left[ \delta ( \bar{\mathbf{q}} - \bar{\mathbf{q}}^s )
\delta ( \bar{\mathbf{p}} - \bar{\mathbf{p}}^s ) \right],
\label{eq:bar_rho_MC_3}
\ee
where $(\bar{\mathbf{q}}^s,\bar{\mathbf{p}}^s)$ denotes the locations of the test functions, and $M$ is the total number of test functions. We generate $(\bar{\mathbf{q}}^s,\bar{\mathbf{p}}^s)$ by the Metropolis-Hastings algorithm using $5 \times 10^6$ iterations. After  excluding the first $10^5$ iterations, we randomly select, for instance, $M=8 \times 10^{4}$ points  $(\bar{\mathbf{q}}^s,\bar{\mathbf{p}}^s)$ from the remaining $4.9 \times 10^6$ iterations. In view of the shapes of the position and momentum projections of $\bar{\rho}_{\mathrm{MC}} \left(\bar{\mathbf{q}} , \bar{\mathbf{p}} \right)$, we make the following change of coordinates: $\bar{u}=\bar{q}_1 \bar{q}_2$ and $\bar{v}=\tan^{-1}(\bar{q}_2)$.
To ensure that the locations of the test functions are ergodic  in $(\bar{\mathbf{q}} , \bar{\mathbf{p}})$-space, we impose periodic boundary conditions to the random walks in the Metropolis-Hastings algorithm. For instance, when setting $\mu=100.6$ and $\sigma=8$ in (\ref{eq:bar_rho_MC_2}), we can map the entire domain in each dimension periodically to the region: $|\bar{u}|\le16$, $|\bar{v}|\le (\pi/2-10^{-5})$, $|\bar{p}_1| \le 16.5$ and $|\bar{p}_2| \le 16.5$. In this case, the acceptance rate is about $22\%$.

To verify the validity of the resulting microcanonical distribution, we plot the energy histogram of the test functions and compare it to the energy histogram of the test particles used to represent the Husimi distribution. In Fig.~\ref{fig:fitting_MC_energy}, we plot the energy of for the test functions for the microcanonical distribution. According to (\ref{eq:energy_def}), $\epsilon_s =\bar{\mathcal{H}}_H \left(\bar{\mathbf{q}}^s,\bar{\mathbf{p}}^s \right) $ denotes the energy for the test function $s$, for $s=1,...,M$. We fit the energy histogram for the test functions for $\bar{\rho}_{\mathrm{MC}} \left(\bar{\mathbf{q}} , \bar{\mathbf{p}} \right)$ by the normal distribution \be
{n_{\mathrm {MC}}}\left( \epsilon  \right) = \tilde{A}
\exp \left[ { - \frac{1}{{2{\sigma_{\mathrm {MC}}^2}}}{{\left( {\epsilon - \mu_{\mathrm {MC}} }
\right)}^2}} \right] .
\label{eq:n_MC}
\ee
The values of the fit parameters $\tilde{A}$, $\mu_{\mathrm {MC}}$ and $\sigma_{\mathrm {MC}}$ are listed in Fig.~\ref{fig:fitting_MC_energy} for $M=8 \times 10^4$. We obtain:
\ba
\mu_{\mathrm {MC}}=101.1, \qquad \sigma_{\mathrm {MC}}=7.975.   \label{eq:fit_parameters_MC}
\ea
We define the normalized energy distribution for test functions as
\ba
\bar{n}_{\mathrm{MC}}(\epsilon)= \frac{n_{\mathrm{MC}}(\epsilon)}{ \int_0^{\infty} d\epsilon \; n_{\mathrm{MC}}(\epsilon)}.  \label{eq:norm_n_MC}
\ea
Comparing (\ref{eq:fit_parameters}) to (\ref{eq:fit_parameters_MC}), we obtain $\mu_{\mathrm {MC}} \approx \mu$ and $\sigma_{\mathrm {MC}} \approx \sigma$, with the errors less than $0.5\%$. Therefore, we conclude that $\bar{n}_{\mathrm{MC}}(\epsilon)$ in (\ref{eq:norm_n_MC}) is practically identical to $\bar{n}_{\mathrm{TP}}(\epsilon)$ in (\ref{eq:norm_n_TP}), with the errors of less than $0.5\%$. Furthermore, in Fig.~\ref{fig:fitting_MC_u} we plot the $\bar{u}$-histogram of the test functions for  $\bar{\rho}_{\mathrm{MC}} \left(\bar{\mathbf{q}} , \bar{\mathbf{p}} \right)$, where $\bar{u}=\bar{q}_1 \bar{q}_2$. Figure~\ref{fig:fitting_MC_u} shows that the distribution of test functions is symmetric in the $\bar{u}$ coordinate.

Substituting (\ref{eq:bar_rho_MC_3}) to (\ref{eq:rho_MC}), we obtain:
\ba
\rho_{\mathrm{MC}} \left(\mathbf{q} , \mathbf{p} \right)
& =& \frac{1}{M}\sum\limits_{s = 1}^M
K(\mathbf{q} - \bar{\mathbf q}^s, \mathbf{p} - \bar{\mathbf p}^s ),
\label{eq:rho_MC_2}
\ea
where $K$ is defined in (\ref{eq:K_0}) and we choose $\gamma_K^a=3/2$ in (\ref{eq:gamma_K_0}). Clearly, $\rho_{\mathrm{MC}}$ is normalized by:
\ba
\int_{ - \infty }^\infty  d\Gamma_{\mathbf{q},\mathbf{p}}\,  \rho_{\mathrm{MC}}( \mathbf {q} ,\mathbf {p} ) = 1.   \label{eq:norm_rho_MC}
\ea
We visualize $\rho_{\mathrm{MC}} \left(\mathbf{q} , \mathbf{p} \right) $ in (\ref{eq:rho_MC_2}) by projecting on the $(q_1,q_2)$ and $(p_1,p_2)$ subspaces, respectively:
\ba
F_q^{\rm MC}( {{q_1},{q_2}} ) &=& \int_{ - \infty }^\infty  {d{p_1}d{p_2}} \,
{ \rho _\mathrm{MC}}( {{q_1},{q_2},{p_1},{p_2}} ),
\label{eq:F_q_mc}
\\
F_p^{\rm MC}( {{p_1},{p_2}} ) &=& \int_{ - \infty }^\infty {d{q_1}d{q_2}} \,
{ \rho _\mathrm{MC}}( {{q_1},{q_2},{p_1},{p_2}} ).
\label{eq:F_p_mc}
\ea
We plot $F^{\rm MC}_q(q_1, q_2)$ and $F^{\rm MC}_p( p_1, p_2 )$ in Fig.~\ref{fig:MC}. When we compare Fig.~\ref{fig:fiq_all} to Fig.~\ref{fig:MC}, we find that $F_q$ and $F_p$ at time $t=10$ are very similar in shape to $F^{\rm MC}_q$ and $F^{\rm MC}_p$, respectively. Contour lines of both $F_q(t=10)$ and $F^{\rm MC}_q$ follow equipotential curves, while the contour lines of both $F_p(t=10)$ and $F^{\rm MC}_p$ are shaped as concentric circles.

\begin{figure}
\includegraphics[width=0.45\textwidth]{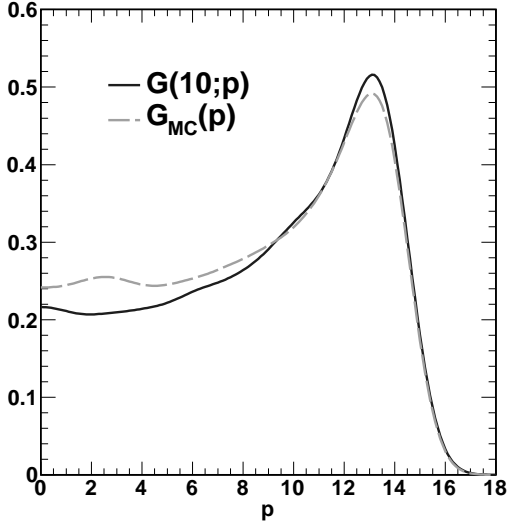}
\caption{Comparison of $G(t;p)$ at $t=10$ and $G_{\mathrm{MC}}(p)$. We define $G(t;p)$ and $G_{\mathrm{MC}}(p)$ in (\ref{eq:G_p}) and (\ref{eq:G_mc_p}) respectively. $G(10;p)$ is obtained from the momentum projection of $\rho_H(10;\mathbf{q}, \mathbf{p})$ composed of $N=10^4$ test particles, while $G_{\mathrm{MC}}(p)$ is obtained from the momentum projection of $\rho_\mathrm{MC}(\mathbf{q}, \mathbf{p})$ composed of $M=2\times10^4$ test functions.    }
\label{fig:G_Gmc}
\end{figure}

To quantify the similarities between $\rho_H(t;\mathbf{q}, \mathbf{p})$ at late times and $\rho_\mathrm{MC}(\mathbf{q}, \mathbf{p})$, we compare their momentum projections. By switching to polar coordinates $p_1=p\cos\theta$ and $p_2=p\sin\theta$, we define the following two projections:
\ba
G(t;p)&=&\int_{0}^{2\pi} d\theta \;F_p \left( t; p\cos\theta, p \sin \theta\right), \label{eq:G_p}  \\
G_{\mathrm{MC}}(p)&=&\int_{0}^{2\pi} d\theta \;F_p^{\mathrm{MC}} \left(  p\cos\theta, p \sin \theta\right), \label{eq:G_mc_p}
\ea
where $F_p$ and $F_p^{\mathrm{MC}}$ are defined in (\ref{eq:F_p}) and (\ref{eq:F_p_mc}) respectively. In Fig.~\ref{fig:G_Gmc}, we plot $G(10;p)$ and $G_{\mathrm{MC}}(p)$ in comparison. $G(10;p)$ is obtained from the momentum projection of $\rho_H(10;\mathbf{q}, \mathbf{p})$ composed of $N=10^4$ test particles, and $G_{\mathrm{MC}}(p)$ is obtained from the momentum projection of $\rho_\mathrm{MC}(\mathbf{q}, \mathbf{p})$ composed of $M=2\times10^4$ test functions. The figure shows that $G(10;p)$ and $G_{\mathrm{MC}}(p)$ have similar values for all $p$, and the largest deviation occurs at low $p$.  $G(10;p)$ and $G_{\mathrm{MC}}(p)$ at low $p$ receive contributions from the test functions located at the narrow ``channels'' along the coordinate axes in the position projections of $\rho_H$ and $\rho_{\mathrm{MC}}$, respectively. Since the number of test functions, $N$ and $M$, are finite, one expects larger fluctuations of the contributions from these narrow ``channels'', which explains the observed deviation at small $p$. Overall, the close similarity between $G(10;p)$ and $G_{\mathrm{MC}}(p)$ suggests that $\rho_H(t;\mathbf{q}, \mathbf{p})$ asymptotically approaches the microcanonical density distribution $\rho_\mathrm{MC}(\mathbf{q}, \mathbf{p})$.

\begin{figure}
\includegraphics[width=0.45\textwidth]{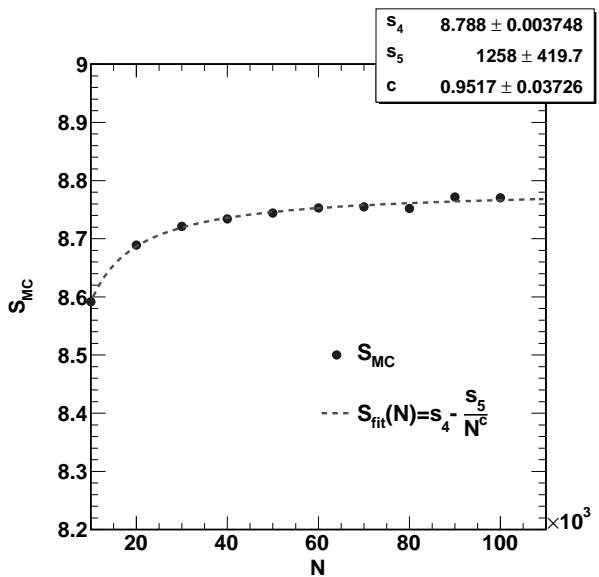}
\caption{The microcanonical entropy $S_{\mathrm{MC}}$ as a function of $M$, indicated by the blue dots. $S_{\mathrm{MC}}$ is defined in (\ref{eq:S_mc}). $M$ denotes the total number of test functions, as revealed in (\ref{eq:bar_rho_MC_3}) and (\ref{eq:rho_MC_2}). We set $\mu=100.6$ and $\sigma=8$ in (\ref{eq:bar_rho_MC_2}). Besides, we fit the curve by a fit function $\bar{S}_{\mathrm{fit}}(M) $ defined in (\ref{eq:fit_SmcvsM}). The fit parameters are shown in the figure.}
\label{fig:Smc_M}
\end{figure}

Finally, we obtain the microcanonical entropy $S_\mathrm{MC}$ by substituting (\ref{eq:rho_MC_2}) into (\ref{eq:S_mc}). We evaluated $S_\mathrm{MC}$ with the help of Simpson's rule and by applying the same integration techniques as those described in Sect.~\ref{sec:Husimi_plots}.  We verified the numerical precision of our approach by evaluating the normalization for $\rho_\mathrm{MC}(\mathbf{q}, \mathbf{p})$ for various choices of $M$ and found that the numerical result coincides with (\ref{eq:norm_rho_MC}) within errors of less than $0.6\%$. In addition to the errors associated with the use of Simpson's rule, $S_\mathrm{MC}$ possesses an additional error, typically less than $0.5\%$, which arises from the Monte-Carlo calculation of ${\bar \rho _\mathrm{MC}}(\bar{\mathbf{q}}, \bar{\mathbf{p}})$ in (\ref{eq:bar_rho_MC_3}). In Fig.~\ref{fig:Smc_M}, we plot $S_\mathrm{MC}$ for several different test function numbers $M$. We fit the data by the function
\be
\bar{S}_{\mathrm{fit}}(M)=s_4-\frac{s_5}{M^c}.
\label{eq:fit_SmcvsM}
\ee
The parameters determined by the fit are:
\be
s_4=8.788, \;\;\; s_5=1258, \;\;\; c=0.9517 .
\ee
We thus conclude that $S_{\mathrm MC} \approx 8.79$ is the microcanonical entropy for our chosen initial conditions.

In Sect.~\ref{sec:Husimi_plots}, we obtained the value $S_H(t=10) \to 8.73$ in the limit $N\to \infty$ for the initial conditions chosen for our numerical simulation. Under the same initial conditions, we found $S_\mathrm{MC} \to 8.79$ when $M \to \infty$. We conclude that the saturation value of the Wehrl-Husimi entropy coincides with the microcanonical entropy within errors, estimated at $1\%$.  Apart from numerical errors, the difference between the two entropy values may also be accounted for by the fact that at $t=10$ the system may not yet be completely equilibrated. Since $S_\mathrm{MC} <S_C$, we also conclude that the Yang-Mills quantum system is equilibrated microcanonically but not thermalized. The system does not have enough degrees of freedom to render the microcanonical and the canonical ensemble approximately identical.

\begin{figure}
\includegraphics[width=0.45\textwidth]{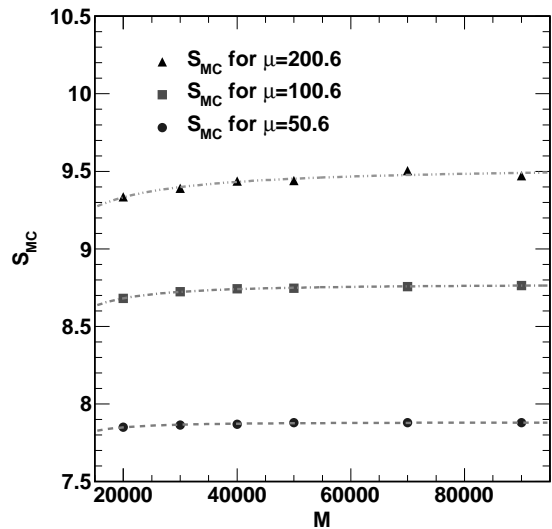}
\caption{The microcanonical entropy $S_{\mathrm{MC}}$ as a function of $M$ for the coarse grained energies $\mu=50.6$, $100.6$, and $200.6$. The corresponding widths $\sigma$, defined in (\ref{eq:bar_rho_MC_2}), for these energies are $\sigma=5.8$, $8.0$, and $11.5$. We fitted these points by the function $\bar{S}_{\mathrm{fit}}(M) $ defined in (\ref{eq:fit_SmcvsM}), and use the fit parameters to determine the asymptotic values of ${S}_{\mathrm{MC}}$ for $M\to \infty$, which are ${S}_{\mathrm{MC}}=7.88$, $8.77$, and $9.54$ (from bottom to top).}
\label{fig:Smc_E}
\end{figure}

In the above calculation, we studied the microcanonical distribution $S_{\rm{MC}}$ for the Yang-Mills quantum mechanics model at the coarse grained energy $\mu=\mathcal{E}[\mathcal{H}_H\rho_H] \approx 100.6$. We now briefly comment on how $S_{\rm{MC}}$ depends on the coarse grained energy of the system. In
Appendix~\ref{sec:scaling}, we  show that while the Yang-Mills Hamiltonian $\mathcal{H}$ possesses a scale invariance, the scale invariance of $\mathcal{H}_H$ is partially broken when we demand that the smearing function in (\ref{eq:Husimi-def-2}) should retains its minimal uncertainty. The reason is that, for any coarse grained average energy $\mu$, the relation $\xi \eta = \hbar^2/4$ constrains our ability to rescale $\xi$ and $\eta$ in (\ref{eq:Hamiltonian_Husumi-2}). Alternatively, we observe that the additional terms in the expression for $\bar{\mathcal{H}}_H \left(\bar{\mathbf{q}} , \bar{\mathbf{p}} \right)$ break the scaling symmetry of the original Yang-Mills Hamiltonian.

Despite the fact that the scaling properties of $\mathcal{H}_H$ are partially broken, we can examine numerically how $S_{\mathrm{MC}}$ changes when $\mu$ scales as $\mu \to \lambda_s^4 \mu$, where $\lambda_s$ is the scaling parameter. In analogy to (\ref{eq:S_mc_prime_scaling}), we parametrize the change in the microcanonical entropy as
\be
S_{\rm{MC}}(\mu) \to S_{\rm{MC}}(\mu) + r \ln\lambda_s,
\label{eq:S_mc_scaling}
\ee
where $r$ is a constant to be determined numerically. In order to find the value of $r$, we calculated $S_{\rm{MC}}$ by numerically evaluating (\ref{eq:S_mc}) for various choices of $\mu$ in (\ref{eq:bar_rho_MC_2}). In Fig.~\ref{fig:Smc_E}, we show $S_{\rm{MC}}$ as a function of $M$ for $\mu=50.6$, $\mu=100.6$, and
$\mu=200.6$, respectively. The corresponding widths $\sigma$, defined in (\ref{eq:bar_rho_MC_2}), for these energies are $\sigma=5.8$, $8.0$, and $11.5$, respectively.  In Fig.~\ref{fig:Smc_E}, we fitted these curves by
$\bar{S}_{\mathrm{fit}}(M) $ defined in (\ref{eq:fit_SmcvsM}). The fit parameters again determine the asymptotic values of ${S}_{\mathrm{MC}}$ for $M\to \infty$. The results are ${S}_{\mathrm{MC}}=7.88$, $8.77$, and $9.54$, respectively. From these results we can deduce the value $r=5.0\pm 0.2$.

In Appendix~\ref{sec:scaling} we show that the scale invariant Yang-Mills Hamiltonian ${\cal H}$ implies the value $r'=6$, where $r'$ is defined in (\ref{eq:S_mc_prime_scaling}). The difference between $r$ and $r'$ is attributed to the following reason: Since we demand the Gaussian smearing function in (\ref{eq:Husimi-def-2}) retains its minimal uncertainty encoded in the relation $\xi \eta = \hbar^2/4$, we are breaking the scaling symmetry of the Husimi Hamiltonian ${\cal H}_H$, as discussed in Appendix~\ref{sec:scaling}. This argument suggests that $S_{\rm{MC}}(\mu)$ changes less strongly under a scale transformation than na\"ively expected. Comparing the numerical value for $r$ with the analytical value for $r'$, we indeed obtain $r<r'$, which  confirms our expectation.

\section{Conclusions}

We have developed a general method to solve the Husimi equation of motion for two-dimensional quantum mechanical systems. We proposed a new method for obtaining the coarse grained Hamiltonian whose expectation value serves as a constant of motion for the time evolution of Husimi distribution. We solved the Husimi equation of motion by the Gaussian test particle method, using fixed-width Gaussians. Having obtained the Husimi distribution, we evaluated the Wehrl-Husimi entropy as a function of time for the Yang-Mills quantum system.

By comparing the Wehrl-Husimi entropy $S_H(t)$ obtained from different particle numbers, $N=1000$ and $N=3000$, we found that the values of $S_H(t)$ agree for $t<2$, and saturation is reached in both cases after $t \ge 6.5$. However, $S_H(t)$ for $N=3000$ saturates to a higher value than for $N=1000$. This result suggests that for a larger number of test particles the Husimi distribution is more evenly distributed in the phase space, and thus a larger value of $N$ results in a higher saturation value of the Wehrl-Husimi entropy. By evaluating $S_H(10)$ for various different $N$'s, we concluded that $S_H(10) \to 8.73$ for $N \to \infty$ for our chosen initial conditions.

In order to address the question of equilibration, we studied the Yang-Mills Hamiltonian system in the canonical and microcanonical ensembles. The canonical entropy for the system is $S_C \approx 9.70$. We obtained the microcanonical distribution by generating $M$ test functions. We observed that the saturated Husimi distribution closely resembles the microcanonical distribution. Moreover, we obtained the microcanonical entropy $S_{\mathrm{MC}} \to 8.79$ as $M \to \infty$ for the same choice of initial conditions.
Therefore, comparing the saturation value of the Wehrl-Husimi entropy to the microcanonical and canonical entropies, we conclude that $(S_H)_{max} \approx S_\mathrm{MC} <S_C$. This implies that, at late times, the Yang-Mills quantum system is microcanonically equilibrated but not thermalized.

It is straightforward to generalize the method introduced here to solve the Husimi equation of motion in three or more dimensions. However, for higher dimensions, the evaluation of the Wehrl-Husimi entropy becomes even more challenging owing to the increasing numbers of integrals. A new method will then be needed for the reliable evaluation of entropy.

\begin{acknowledgments}
We thank Christopher Coleman-Smith, Joshua W. Powell, Young-Ho Song and Steven Tomsovic for helpful discussions. This work was supported in part by the U.S. Department of Energy under grant DE-FG02-05ER41367.
\end{acknowledgments}

\appendix
\begin{widetext}

\section{Equations of motion for the test particles \label{sec:Appendix:EOMs}}

In Sect.~\ref{sec:TP} we obtained the equations of motion for the ten variables, $\bar{q}_1^i$, $\bar{q}_2^i$, $\bar{p}_1^i$, $\bar{p}_2^i$, $c_{q_1 q_1}^i$, $c_{q_2 q_2}^i$, $c_{p_1 p_1}^i$, $c_{p_2 p_2}^i$, $c_{q_1 p_1}^i$ and $c_{q_2 p_2}^i$, where $i$ labels the test particle. In (\ref{eq:I_x1}--\ref{eq:I_p2}), we listed the equations obtained from the first moments  $I_{q_1}$, $I_{q_2}$, $I_{p_1}$ and  $I_{p_2}$ of eq.~(\ref{eq:Husimi_2D}). The equations obtained from the second moments $I_{q_1^2}$, $I_{q_2^2}$, $I_{p_1^2}$, $I_{p_2^2}$, $I_{q_1 p_1}$, and $I_{q_2 p_2}$ of (\ref{eq:Husimi_2D}) are listed below:
\ba
&& \left[ 2\dot {c}_{q_1 p_1 }^i (t)c_{q_1 p_1 }^i(t)c_{p_1 p_1 }^i (t)  - \dot {c}_{q_1 q_1}^i (t)
\left( {c_{p_1 p_1 }^i (t)} \right)^2
- \dot {c}_{p_1 p_1 }^i (t)\left( {c_{q_1 p_1}^i (t)} \right)^2 \right]
+ \frac{2}{m}c_{q_1 p_1}^i (t) \Delta_1^i (t)  = 0,
\label{eq:I_x1x1}
\\
&& \left[ 2\dot {c}_{q_2 p_2 }^i (t)c_{q_2 p_2 }^i(t)c_{p_2 p_2 }^i (t) - \dot {c}_{q_2 q_2}^i (t)
\left( {c_{p_2 p_2 }^i (t)} \right)^2
- \dot {c}_{p_2 p_2 }^i (t)\left( {c_{q_2 p_2}^i (t)} \right)^2 \right]
+ \frac{2}{m}c_{q_2 p_2 }^i(t) \Delta_2^i (t) = 0,
\label{eq:I_x2x2}
\ea
\ba
&& \left[ 2\dot {c}_{q_1 p_1 }^i(t) c_{q_1 p_1 }^i(t) c_{q_1 q_1 }^i(t) -
\dot {c}_{q_1 q_1 }^i (t)\left( {c_{q_1 p_1 }^i \left(t \right)} \right)^2
- \dot {c}_{p_1 p_1 }^i (t) \left( c_{q_1 q_1}^i(t) \right)^2 \right]
\nn \\
&& \hspace{0.2\textwidth}
- \left. \left[ 2 \frac{\partial ^2V}{\partial q_1^2 }
\right|_{ \bar{\mathbf{q}}^i(t) }  + \left( \frac{c_{p_2 p_2 }^i (t)}{\Delta_2^i (t)}
- \frac{\alpha}{2} \right) \left. \frac{\partial ^4V}{\partial q_1^2\partial q_2^2}
\right|_{{\bf\bar{q}}^i(t)}  \right]
c_{q_1 p_1 }^i (t) \Delta_1^i(t)  = 0,
\label{eq:I_p1p1}
\\
&& \left[ 2\dot {c}_{q_2 p_2 }^i (t)c_{q_2 p_2 }^i
(t)c_{q_2 q_2 }^i (t) - \dot {c}_{q_2 q_2}^i (t)\left( {c_{q_2 p_2 }^i (t)} \right)^2
- \dot {c}_{p_2 p_2 }^i (t)\left( {c_{q_2 q_2}^i (t)} \right)^2\right]
\nn \\
&& \hspace{0.2\textwidth}
- \left[  \left. 2 \frac{\partial ^2V}{\partial q_2^2} \right|_{{\bf \bar {q}}^i(t)}
+ \left( \frac{c_{p_1 p_1 }^i (t)}{\Delta_1^i (t)} - \frac{\alpha}{2} \right)
\left. \frac{\partial ^4V}{\partial q_1^2\partial q_2^2}
\right|_{ \bar{\mathbf{q}}^i(t) } \right]
c_{q_2 p_2 }^i (t) \Delta_2^i(t) = 0,
\label{eq:I_p2p2}
\ea
\ba
&& \left[ \dot {c}_{q_1 q_1 }^i (t)c_{p_1 p_1 }^i
(t)c_{q_1 p_1 }^i (t) + \dot {c}_{p_1 p_1}^i (t)
c_{q_1 q_1 }^i (t)c_{q_1 p_1 }^i(t)
- \dot {c}_{q_1 p_1 }^i (t)\left( {c_{q_1 q_1}^i (t)c_{p_1 p_1 }^i (t)
+ \left( {c_{q_1p_1 }^i(t)} \right)^2} \right) \right]
\nn \\
&& \hspace{0.2\textwidth}
+ \left[ \frac{\hbar ^2}{2m\alpha } - \frac{1}{m}\left(
{\frac{c_{q_1 q_1 }^i (t)}{\Delta_1^i (t)}} \right)
+ \left( {\frac{c_{p_1 p_1 }^i (t)}{\Delta_1^i(t)} - \frac{1}{2}\alpha } \right)
\left. {\frac{\partial ^2V}{\partial q_1^2 }} \right|_{ \bar{\mathbf{q}}^i(t) }  \right.
\nn \\
&& \hspace{0.2\textwidth} \left.
+ \frac{1}{2} \left({\frac{c_{p_1 p_1 }^i (t)}{\Delta_1^i (t)
} - \frac{\alpha}{2} } \right) \left({\frac{c_{p_2 p_2 }^i (t)}{\Delta_2^i(t)
} - \frac{\alpha}{2} } \right) \left. {\frac{\partial ^4V}{\partial q_1^2
\partial q_2^2 }} \right|_{ \bar{\mathbf{q}}^i(t)  } \right]
\left( \Delta_1^i(t) \right)^2= 0,
\label{eq:I_x1p1}
\ea
\ba
&& \left[ \dot {c}_{q_2 q_2 }^i (t)c_{p_2 p_2 }^i
(t)c_{q_2 p_2 }^i (t) + \dot {c}_{p_2 p_2}^i (t)
c_{q_2 q_2 }^i (t)c_{q_2 p_2 }^i(t)
- \dot {c}_{q_2 p_2 }^i (t)\left( {c_{q_2 q_2}^i (t)
c_{p_2 p_2 }^i (t) + \left( {c_{q_2
p_2 }^i(t)} \right)^2} \right) \right]
\nn \\
&& \hspace{0.2\textwidth}
+ \left[ \frac{\hbar ^2}{2m\alpha } - \frac{1}{m}\left(
{\frac{c_{q_2 q_2 }^i (t)}{\Delta_2^i (t)}} \right)
+ \left( {\frac{c_{p_2 p_2 }^i (t)}{\Delta_2^i(t)} - \frac{1}{2}\alpha } \right)
\left. {\frac{\partial ^2V}{\partial q_2^2 }} \right|_{ \bar{\mathbf{q}}^i(t)} \right.
\nn \\
&& \hspace{0.2\textwidth} \left.
+ \frac{1}{2}\left( \frac{c_{p_1 p_1 }^i (t)}{\Delta_1^i (t)} - \frac{\alpha}{2} \right)
\left({\frac{c_{p_2 p_2 }^i (t)}{\Delta_2^i (t)} - \frac{\alpha}{2} } \right)
\left. \frac{\partial ^4V}{\partial q_1^2 \partial q_2^2} \right|_{ \bar{\mathbf{q}}^i(t) }  \right]
\left( \Delta_2^i (t) \right)^2 =0,
\label{eq:I_x2p2}
\ea
where $i=1,2,...,N$, and $\Delta_1^i \left(t \right )$, $\Delta_2^i\left(t \right )$ and $\bar{\mathbf{q}}^i(t)$ are defined in
(\ref{eq:Delta_1}), (\ref{eq:Delta_2}) and (\ref{eq:q_bar}), respectively.

\section{Husimi equation of motion in one dimension \label{sec:energy_1D}}

\begin{figure*}
\includegraphics[width=0.45\textwidth]{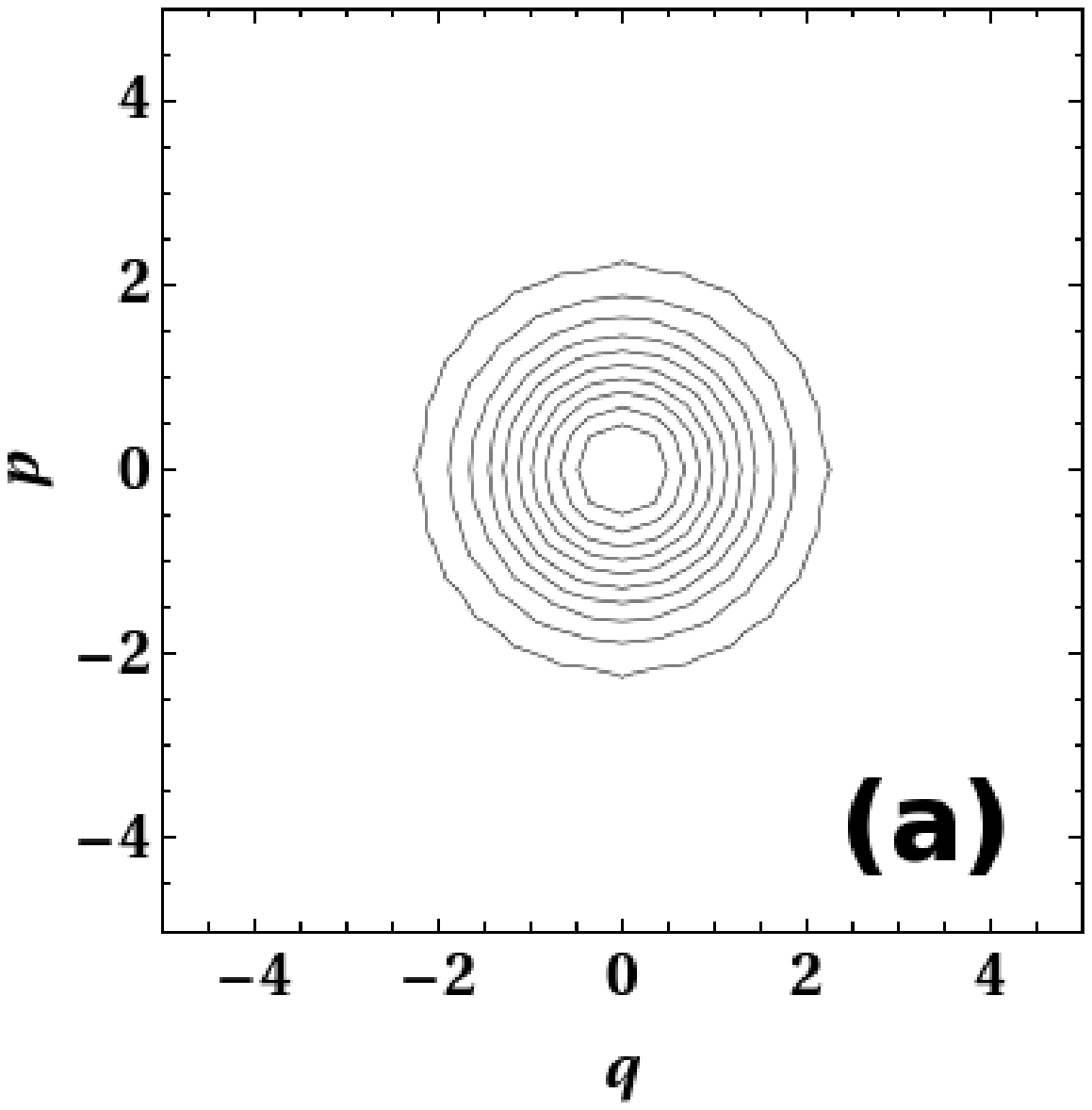}
\includegraphics[width=0.45\textwidth]{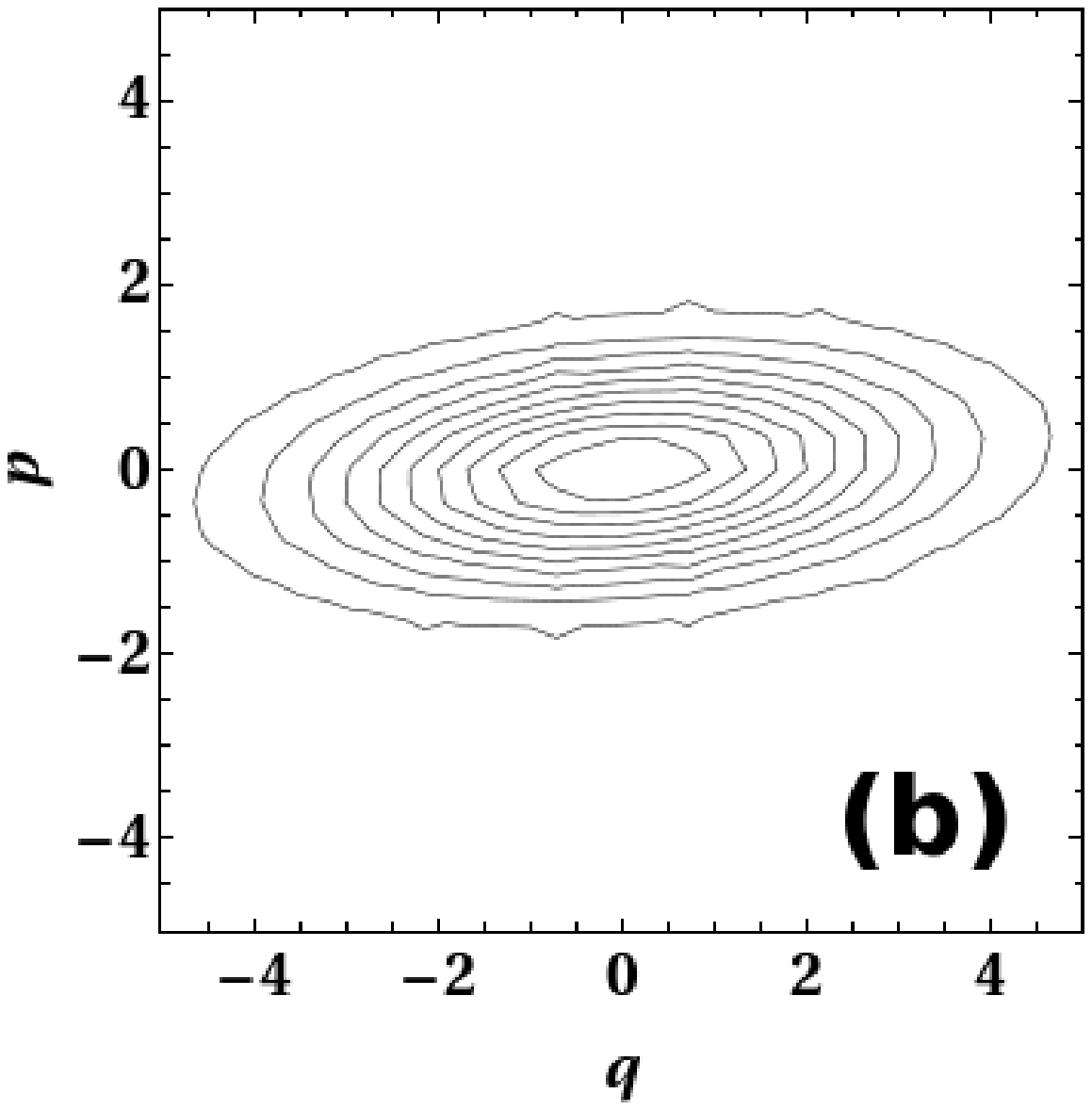}
\includegraphics[width=0.45\textwidth]{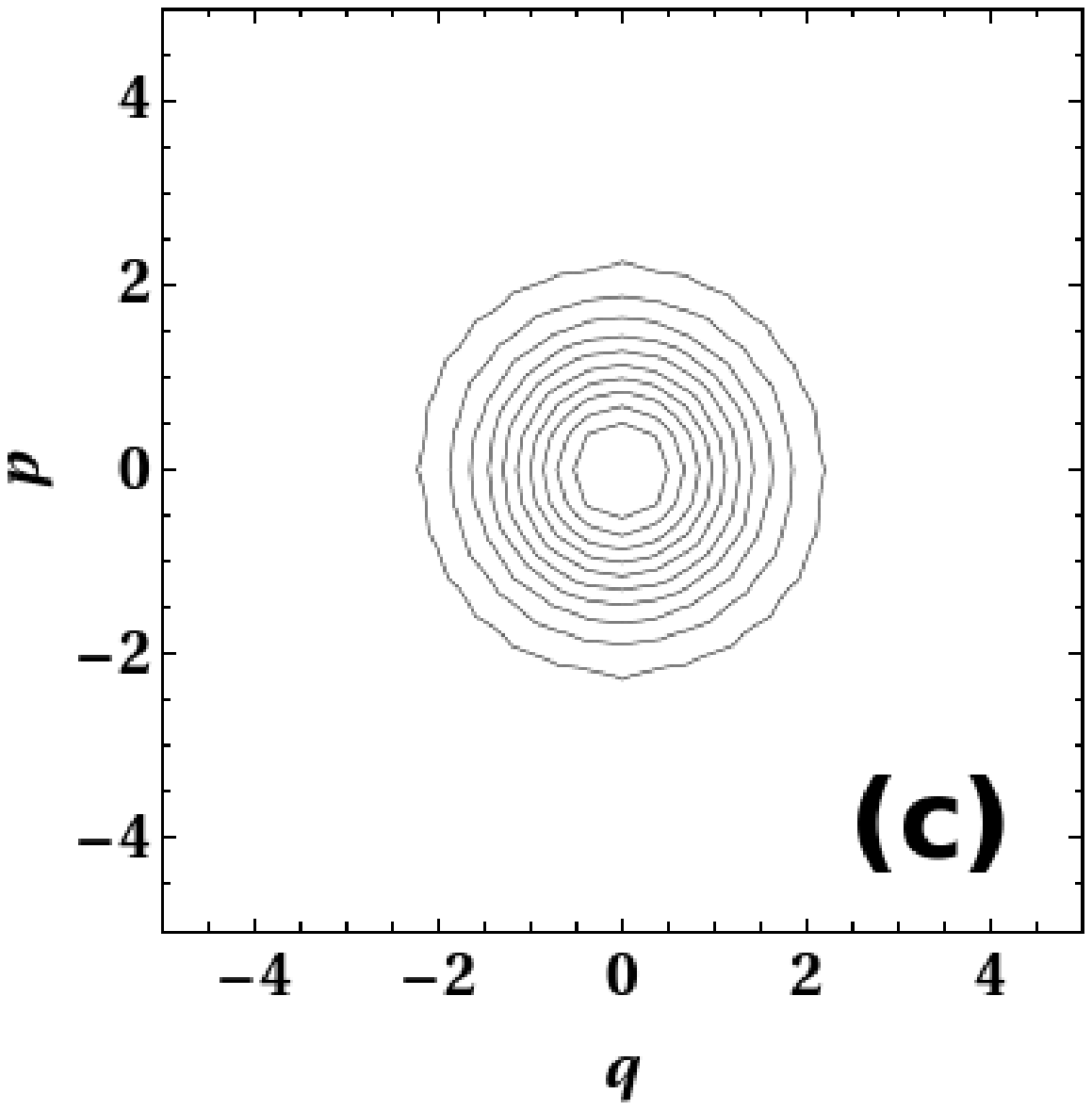}
\includegraphics[width=0.45\textwidth]{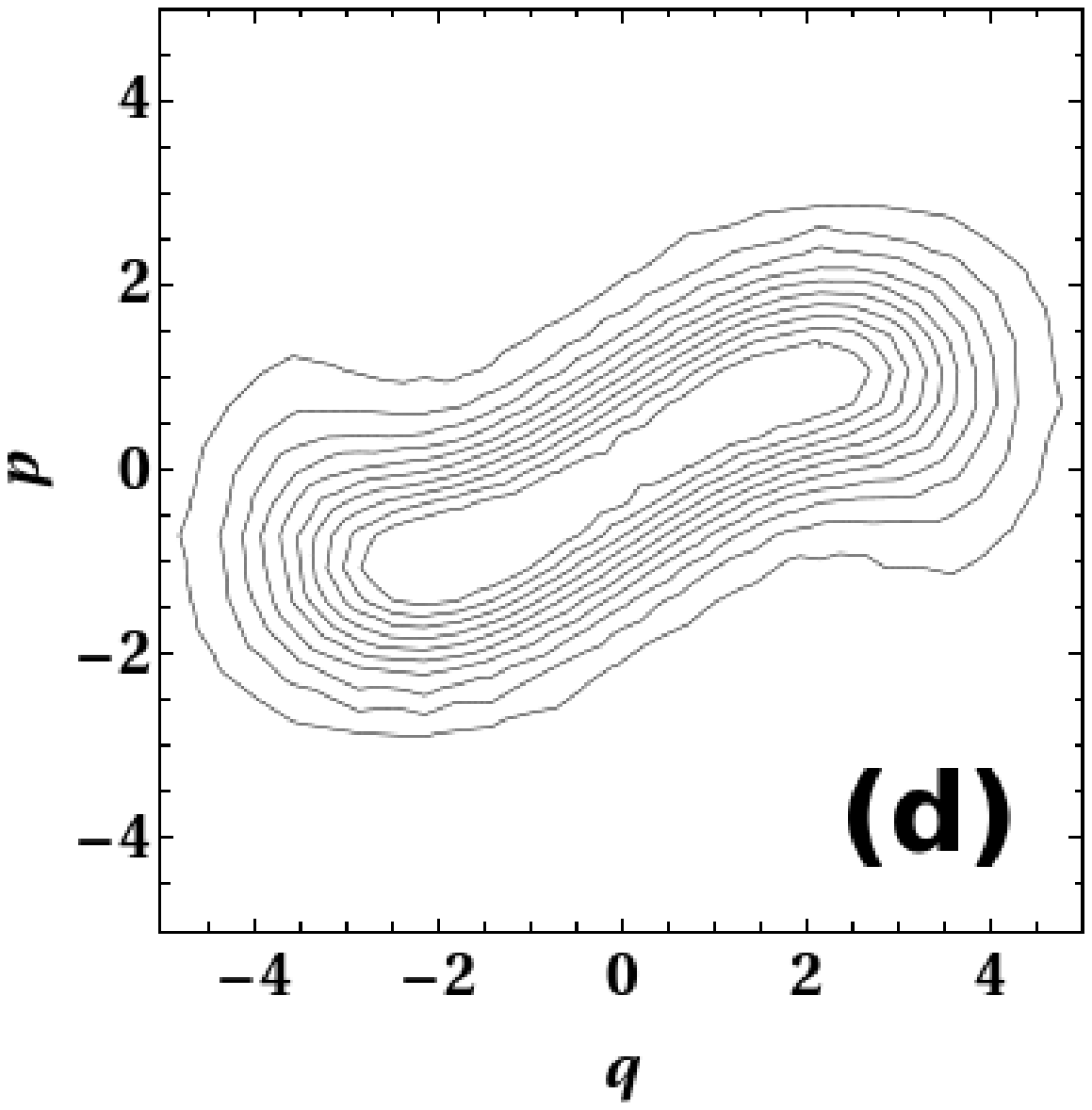}
\caption{ Solutions of the Husimi equation of motion in one dimension. The Hamiltonian is defined in (\ref{eq:H_1D-1}) of Appendix~\ref{sec:energy_1D}. The parameters as chosen as $\kappa=1$ and $\zeta=1$. The initial conditions are discussed in Appendix~\ref{sec:energy_1D}. Panels (a) and (b) show $\rho_H(t;x,p)$ for a single test particle, at time (a) $t=0$ and (b) $t=2$. Panels (c) and (d) show $\rho_H(t;q,p)$ or the many test particles, at times (c) $t=0$ and (d) t=2. It is obvious that for $t>0$ this single particle ansatz is insufficient to represent the solution.}
\label{fig:1D_plots}
\end{figure*}

The Husimi equation of motion for one-dimensional quantum systems was derived in \cite{OConnell:1981pla}. For the potential energy $V(q)$ being a $\mathcal{C}^{\infty}$-differentiable function of $q$, the Husimi equation of motion in one dimension is:
\ba
 \frac{\partial \rho_H }{\partial t} &=& -\frac{1}{m}  {\left( {p + \frac{\hbar ^2}{2\alpha }
 \frac{\partial }{\partial p }}\right)} \frac{\partial \rho_H }{\partial q }
+ \sum\limits_{\lambda , \mu , \kappa }   \left[ {\frac{\left( {i\hbar } \right)^{\lambda  - 1}}{2^{\lambda
+ \mu  - 1}}}  {\frac{\alpha ^{\mu  - \kappa  }}{\lambda  !\kappa  !\left(
{\mu  - 2\kappa  } \right)!}}  \frac{\partial ^{\lambda + \mu  } V \left(x \right)}{\partial
q^{\lambda  + \mu  } }    \frac{\partial ^{\lambda  }}{\partial
p^{\lambda  } }\frac{\partial ^{\mu  - 2\kappa  }}{\partial
q^{\mu  - 2\kappa  } }\rho_H   \right] ,
\label{eq:Husimi_gen}
\ea
\end{widetext}
\noindent where $\lambda $, $\mu $ and $\kappa $ are summed over all non-negative integers subject to the constraints that $\lambda$ is odd and $\mu  - 2\kappa \ge 0$.

We discuss the energy conservation for the one-dimensional Hamiltonian. As a specific example, we start from the following one-dimensional Hamiltonian in the Wigner representation.
\ba
{\cal H}(x,p) = \frac{{{p^2}}}{{2m}} - \frac{\kappa }{2}{q^2} + \frac{\zeta }{{24}}{q^4},
\label{eq:H_1D-1}
\ea
where $\lambda$ and $\zeta$ are positive-valued parameters. We derive the corresponding one-dimensional coarse grained Hamiltonian as follows. The Husimi distribution for a one-dimensional quantum system can be obtained from the Wigner distribution by:
\ba
\rho_{H} (t; q,p) &=& \frac{1}{\pi \hbar} \int_{-\infty}^{\infty} dq'dp' \;
e^{-(q'-q)^2/\alpha -\alpha(p'-p)^2 /\hbar^2}
\nn\\
&& \qquad\qquad \times W(t; q', p').
\label{eq:rho_1D}
\ea
Starting from (\ref{eq:rho_1D}) and proceeding like in Sect.~\ref{sec:energy}, we obtain an expression similar to that of (\ref{eq:Hamiltonian-Husimi-1}), which reads
\begin{widetext}
\ba
\mathcal{H}_H \left( q, p \right) &=& \frac{1}{(2\pi)^2}\int_{-\infty}^{\infty} dx'dp' \;
\mathcal{H} \left( q',p'\right) \int_{-\infty}^{\infty} dudv
\exp\left[\frac{\alpha}{4} u^2+ \frac{\hbar^2}{4\alpha} v^2 -iu(q'-q)-iv(p'-p)\right] .
\label{eq:HH-1D}
\ea
\end{widetext}
Here $u$ and $v$ are Fourier conjugate variables to $q$ and $p$ respectively. Similar to the calculation in Sect.~\ref{sec:energy}, we set $\xi=-\alpha/4$ and $\eta=-\hbar^2/(4\alpha)$.  We evaluate the integrals in (\ref{eq:HH-1D}) in the analytic region where $\xi>0$ and $\eta>0$, and then substitute  $\xi=-\alpha/4$ and $\eta=-\hbar^2/(4\alpha)$ into the resulting analytical expression. In this manner, we obtain the coarse grained Hamiltonian:
\ba
{{\cal H}_H}(q,p) &=& \frac{{{p^2}}}{{2m}} - \frac{1}{2}\left( {\kappa  + \frac{{\alpha \zeta }}{4}} \right){q^2}
+ \frac{\zeta }{{24}}{q^4}
\nn \\
&& - \frac{{{\hbar ^2}}}{{4m\alpha }} + \frac{1}{{32}}\alpha (\alpha \zeta  + 8\kappa ).
\label{HH-1D-2}
\ea
Proceeding similarly as in Sect.~\ref{sec:energy}, we use eqs.~(\ref{eq:Husimi_gen}) and (\ref{HH-1D-2}) to prove that $\mathcal{E} \left[ \mathcal{H}_H \rho_H \right]$ is a constant of motion for the Husimi equation of motion in one dimension. Thus  $\mathcal{E} \left[ \mathcal{H}_H \rho_H \right]$ should be identified as the total energy corresponding to the Hamiltonian (\ref{eq:H_1D-1}).

Next, we solve the Husimi equation of motion (\ref{eq:Husimi_gen}) by using the test-particle method described in Sect.~\ref{sec:TP}. We begin by writing the Husimi distribution as:
\ba
\rho_H ( {t; q, p } ) &=&
\frac{\hbar^2}{N} \sum\limits_{i=1}^{N} \sqrt {\Delta^i (t)}
\exp \left[  - \frac{1}{2}c_{q q }^i (t)\left( {q - \bar {q}^i (t)} \right)^2  \right]
\nn \\
& \times & \exp \left[ - \frac{1}{2}c_{p p }^i (t) \left( {p - \bar {p}^i (t)} \right)^2
\right]
\nn \\
& \times & \exp \left[ - c_{q p }^i (t) \left( {q - \bar{q}^i (t)} \right)
\left( {p - \bar {p}^i (t)} \right)  \right] ,
\label{eq:ansatz_1D}
\ea
where $i=1,...,N,$ and we define
\ba
\Delta^i (t) &=& \left[ {c_{q q }^i (t) c_{p p }^i (t)
- \left( {c_{q p }^i(t)} \right)^2} \right].  \label{eq:Delta}
\ea
The moment of a function $f(t;q,p)$ with respect to a weight function $w(q,p)$ is defined as:
\ba
I_w[f] &=& \int \frac {d q \, dp}{ 2\pi\hbar}   \,
 \left[ w (q,p )  f (t; q,p) \right].
\ea
Applying the five moments $I_{q}$,  $I_{p}$, $I_{q^2}$, $I_{p^2}$ and $I_{q p}$ to the Husimi equation of motion (\ref{eq:Husimi_gen}), we obtain five equations of motions for each test particle $i$ for the five variables representing the location in phase space and width of each test particle.

These equations are:
\ba
\dot{\bar {q}}^i (t) - \frac{1}{m}\bar {p}^i (t) &=& 0 ,
\label{eq:I_x}
\\
\dot{\bar {p}}^i (t) + \left. {\frac{\partial V}{\partial q }}
\right|_{\bar{q}^i (t)}
\hspace{0.35\linewidth} &&
\nn \\
\qquad + \frac{1}{2}\left( {\frac{c_{p p }^i (t)}
{\Delta^i \left(t \right ) } - \frac{\alpha}{2} } \right)\left.
{\frac{\partial ^3V}{ \partial q^3 }}
\right|_{\bar{q}^i (t)} &=& 0,
\label{eq:I_p}
\ea

\begin{widetext}
\ba
&& \left[ 2\dot {c}_{q p }^i (t)c_{q p }^i(t)c_{p p }^i (t)  - \dot {c}_{q q}^i (t)
\left( {c_{p p }^i (t)} \right)^2
- \dot {c}_{p p }^i (t)\left( {c_{q p}^i (t)} \right)^2 \right]
+ \frac{2}{m}c_{q p}^i (t) \Delta^i (t)  = 0,
\label{eq:I_xx}
\ea
\ba
&& \left[ 2\dot {c}_{q p }^i c_{q p }^i c_{q q }^i -
\dot {c}_{q q }^i (t)\left( {c_{q p }^i \left(t \right)} \right)^2
- \dot {c}_{p p }^i (t) \left( c_{q q}^i(t) \right)^2 \right]
\nn \\
&& \hspace{0.2\textwidth}
- \left. \left[ 2 \frac{\partial ^2V}{\partial q^2 }
\right|_{\bar{q}^i (t)}  + \left( \frac{c_{p p }^i (t)}{\Delta^i (t)}
- \frac{\alpha}{2} \right) \left. \frac{\partial ^4V}{\partial q^4}
\right|_{\bar{q}^i (t)}  \right]
c_{q p }^i (t) \Delta^i(t)  = 0,
\label{eq:I_pp}
\ea
\ba
&& \left[ \dot {c}_{q q }^i (t)c_{p p }^i
(t)c_{q p }^i (t) + \dot {c}_{p p}^i (t)
c_{q q }^i (t)c_{q p }^i(t)
- \dot {c}_{q p }^i (t)\left( {c_{q q}^i (t)c_{p p }^i (t)
+ \left( {c_{q p }^i(t)} \right)^2} \right) \right]
\nn \\
&& \hspace{0.2\textwidth}
+ \left[ \frac{\hbar ^2}{2m\alpha } - \frac{1}{m}\left(
{\frac{c_{q q }^i (t)}{\Delta^i (t)}} \right)
+ \left( {\frac{c_{p p }^i (t)}{\Delta^i(t)} - \frac{1}{2}\alpha } \right)
\left. {\frac{\partial ^2V}{\partial q^2 }} \right|_{\bar{q}^i (t)}  \right.
\nn \\
&& \hspace{0.2\textwidth} \left.
+ \frac{1}{2} \left({\frac{c_{p p }^i (t)}{\Delta^i (t)
} - \frac{\alpha}{2} } \right)^2  \left. {\frac{\partial ^4V}{\partial q^4
 }} \right|_{\bar{q}^i (t)} \right]
\left( \Delta^i(t) \right)^2= 0,
\label{eq:I_xp}
\ea
\end{widetext}
\noindent  where $i=1,...,N$.  By solving (\ref{eq:I_x})-(\ref{eq:I_xp}) simultaneously for $i=1,...,N$, we obtain  $\bar{q}^i$, $\bar{p}^i$, $c_{xx}^i$, $c_{pp}^i$ and $c_{xp}^i$ as functions of time.

Finally, we solve these $5N$ equations of motions
for  the Hamiltonian system in (\ref{eq:H_1D-1}), with $\kappa=\zeta=1$. For choosing the initial conditions, we adopt the method similar to that introduced in Sect.~\ref{sec:initial_conditions}. Here, we briefly outline the ideas without showing the details.
We choose the initial conditions setting the initial Husimi distribution to be:
\ba
\rho_H (0; q, p)  &=& \int_{-\infty}^{\infty}  \frac{ d{q}' d {p}' }{ 2\pi\hbar } \,
  K( {q}-{q}', {p}-{p}')
\nn \\
&& \qquad \times \phi( {q}', {p}'),
\label{eq:Husimi_0_1D}
\ea
where $K$ and $\phi$ are defined in Sect.~\ref{sec:initial_conditions}. We express $\rho_H$, $K$ and $\phi$ in the forms of (\ref{eq:initial_Husimi}), (\ref{eq:K_0}) and (\ref{eq:phi_0}) respectively, with the redefined variables $\bm{\chi}=(q,p)$ and $\bm{\chi'}=(q',p')$ and the redefined indices $a=1,2$ for $\chi^a$, $\chi'^a$, $\mu_H^a$,  $\mu_\phi^a$, $\gamma_H^a$,  $\gamma_K^a$ and $\gamma_\phi^a$.  By convolution theorem, we obtain that:
\ba
\frac{1}{\gamma_H^a} &=& \frac{1}{\gamma_K^a} + \frac{1}{\gamma_{\phi}^a},
\label{eq:relation_1D}
\ea
for $a=1,2$.
At $t=0$, we choose $\gamma_H^a=1$. In the many-particle ansatz, we choose $N=1000$, $\gamma_K^a=3/2$ and $\gamma_\phi^a=3$.
And we choose $\mu_H^a=\mu_\phi^a=0$.  In the single particle ansatz, $\rho_H$ remains a single Gaussian for all times, and thus we choose $\gamma_H^a=1$ and $\mu_H^a=0$.
We plot $\rho_H (t;q,p)$ both for the single-particle and many-particle ansatz in Fig.~\ref{fig:1D_plots}.
We discuss the meaning of these results in Sect.~\ref{sec:initial_conditions}.

\section{Effects of coarse graining on the scale invariance of the Yang-Mills Hamiltonian  \label{sec:scaling}}

In this section, we discuss the effects of coarse graining on the scale invariance of the Yang-Mills Hamiltonian. We begin by constructing an alternative microcanonical distribution $\rho'_{\mathrm{MC}}$ in terms of the conventional Hamiltonian $\mathcal{H}$ in (\ref{eq:YMH}) and the conventional energy $E$, and we obtain the scaling of the  microcanonical entropy $S'_{\mathrm{MC}}$ with respect to that of  $E$. Furthermore, we show that, while $\mathcal{H}$ is scale invariant,  the scale invariance of the coarse grained Hamiltonian $\mathcal{H}_H$ is partially broken, due to the requirement that the smearing Gaussian function in the Husimi transformation (\ref{eq:Husimi-def-2}) should retain its minimal quantum mechanical uncertainty.

For the conventional Hamiltonian in (\ref{eq:YMH}), we construct an alternative microcanonical distribution $\rho'_{\mathrm{MC}}$ as:
\ba
\rho'_{\mathrm{MC}} =\frac{1}{\Omega}\exp\left(-\frac{\mathcal{H}-E}{2\sigma_g^2}\right).
\label{eq:rho_mc_prime}
\ea
As discussed in  Sect.~\ref{sec:MC}, approximating $\delta(\mathcal{H}-E)$ by a Gaussian distribution is a way to construct a microcanonical distribution that leads to a well-defined entropy. Define $\lambda_s$ as a scaling parameter. As the position and momentum scales as $\mathbf{q} \to \lambda_s \mathbf{q}$ and $\mathbf{p} \to \lambda_s^2 \mathbf{p}$ respectively, it is straightforward to show that $\mathcal{H} \to \lambda_s^4 \mathcal{H}$ and thus $E \to \lambda_s^4 E$.  The normalization condition:
\ba
\int d\Gamma_{\mathbf{q},\mathbf{p}} \,\rho'_{\mathrm{MC}}( \mathbf{q},\mathbf{p} )=1
\ea
must be scale invariant. Owing to the scaling $\Gamma_{\mathbf{q},\mathbf{p}} \to \lambda_s^6 \Gamma_{\mathbf{q},\mathbf{p}}$ we obtain $\Omega \to \lambda_s^2 \Omega$ and $\sigma_g \to \lambda_s^4 \sigma_g$. The microcanonical canonical entropy $S'_{\mathrm{MC}}$ is defined as:
\ba
S'_\mathrm{MC} =  - \int d\Gamma_{\mathbf{q},\mathbf{p}} \, \rho'_\mathrm{MC}(\mathbf{q},\mathbf{p}) \ln \rho'_\mathrm{MC}(\mathbf{q},\mathbf{p}),
\label{eq:S_mc_prime}
\ea
where $\rho'_\mathrm{MC}$ is given in (\ref{eq:rho_mc_prime}). The scaling of $S'_{\mathrm{MC}}$ follows from the scaling of $\mathcal{H}$ and $E$:
\ba
S'_{\mathrm{MC}}(E) \to S'_{\mathrm{MC}}(E) + r' \ln \lambda_s,
\label{eq:S_mc_prime_scaling}
\ea
where  $r'=6$.

The coarse grained Hamiltonian $\mathcal{H}_H (\mathbf{q},\mathbf{p})$ given in (\ref{eq:Hamiltonian_Husumi-3}) is obtained from $\mathcal{H} (\mathbf{q},\mathbf{p}) $ by the transformation (\ref{eq:Hamiltonian_Husumi-2}).  We now examine how $\mathcal{H}_H (\mathbf{q},\mathbf{p})$ scales when the positions and momenta scale as $\mathbf{q} \to \lambda_s \mathbf{q}$ and $\mathbf{p} \to \lambda_s^2 \mathbf{p}$, respectively. The uncertainty relation of a quantum state reads:
\ba
\Delta q_i \Delta p_j \ge \frac{\hbar}{2}\, \delta_{ij} ,
\label{eq:uncertainty_qs}
\ea
where $i,j=1,2$. We note the difference by a factor of 2 between (\ref{eq:uncertainty_qs}) and (\ref{eq:uncertainty}), which was pointed out in \cite{Ballentine:1998book}. From (\ref{eq:Hamiltonian_Husumi-2})
and (\ref{eq:uncertainty_qs}), it is straightforward to show that, when $\mathbf{q} \to \lambda_s \mathbf{q}$ and $\mathbf{p} \to \lambda_s^2 \mathbf{p}$, $\mathcal{H}_H$ will scale as $\mathcal{H}_H \to \lambda_s^4 \mathcal{H}_H$ only if the smearing parameters $\xi$ and $\eta$ scale as $\xi \to \lambda_s^2 \xi$ and $\eta \to \lambda_s^4 \eta$, respectively.  In addition, the constraint $\lambda_s \ge 1$ is imposed by the uncertainty relation (\ref{eq:uncertainty_qs}).

The Husimi distribution is defined as  a minimally smeared Wigner function, as can be seen from (\ref{eq:Husimi-def-2}). For the smearing Gaussian with minimal uncertainty, we have $\Delta q_j \Delta p_j = \hbar/2$ for $j=1,2$, and thus $\xi \eta = \hbar^2/4$. Therefore, we do not have the flexibility to scale the parameters $\xi$ and $\eta$ in the required way, if we demand that the smearing Gaussian in (\ref{eq:Husimi-def-2}) should retains its minimal uncertainty. As a consequence, the scaling symmetry of $\mathcal{H}_H$ is partially broken. We compare the different scaling behavior of $S'_{\mathrm{MC}}(E)$ and $S_{\mathrm{MC}}(\mu)$ at the end of Sect.~\ref{sec:MC}.


\end{document}